\documentclass[preprint,12pt,3p]{elsarticle}

\usepackage{amsmath}    %H141205 - For subequation
\usepackage{amssymb}
\usepackage{navigator}
\usepackage{changepage}		%H160129 - Adjust table to fit page
\usepackage[pdftex]{color}
\usepackage{caption}
\usepackage{subcaption}
\usepackage[nolists,tablesfirst,nomarkers]{endfloat}
 \biboptions{sort&compress}

\journal{Elsevier}

\begin{document}

\begin{frontmatter}

%\title{Sample article to present \texttt{elsarticle} class\tnoteref{label0}}
%\tnotetext[label0]{This is only an example}

%\title{Isogeometric analysis of functionally graded microplates for size-dependent problem}
\title{Isogeometric analysis for functionally graded microplates based on modified couple stress theory}

%\author[label1]{Xuan-Hoang Nguyen\fnref{label3}}
%\fntext[label3]{I also want to inform about\ldots}
%\fntext[label4]{Small city}
%\ead{author.one@mail.com}
%\ead[url]{author-one-homepage.com}

%\author[label1]{First Author}
%\author[label1]{Second Author}
%\author[label1]{Third Author}
%\author[label1]{Fourth Author}

\author[label1]{Hoang X. Nguyen}
\ead{xuan.h.nguyen@northumbria.ac.uk}

\author[label2]{Tuan N. Nguyen}
%\ead{address2@mail.com}

\author[label3]{M. Abdel-Wahab}
%\ead{address4@mail.com}

\author[label4]{S.P.A. Bordas}
%\ead{address5@mail.com}

\author[label5]{H. Nguyen-Xuan}
\ead{ngx.hung@hutech.edu.vn}

%\author[label4]{Fourth Author\corref{cor1}}
\author[label1]{Thuc P. Vo \corref{cor1}}
\ead{thuc.vo@northumbria.ac.uk}

%\address[label1]{Faculty of Engineering and Environment, Northumbria University, Newcastle upon Tyne NE1 8ST, United Kingdom}

\address[label1]{Department of Mechanical and Construction Engineering, Northumbria University, Newcastle upon Tyne NE1 8ST, United Kingdom}

\address[label2]{Department of Computational Engineering, Vietnamese-German University, Binh Duong New City, Vietnam}

\address[label3]{Soete Laboratory, Faculty of Engineering and Architecture, Ghent University, 9000 Ghent, Belgium}

\address[label4]{Faculte des Sciences, de la Technologie et de la Communication, University of Luxembourg, Luxembourg}

\address[label5]{Center for Interdisciplinary Research in Technology (CIRTech), Ho Chi Minh University of Technology (HUTECH), Ho Chi Minh City 700000, Vietnam}
%
%\fntext[add1]{Footnote address 1}
%\fntext[add2]{Footnote address 2}

\cortext[cor1]{Corresponding author}

\begin{abstract}
Analysis of static bending, free vibration and buckling behaviours of functionally graded microplates is investigated in this study. The main idea is to use the isogeometric analysis in associated with novel four-variable refined plate theory and quasi-3D theory. More importantly, the modified couple stress theory with only one material length scale parameter is employed to effectively capture the size-dependent effects within the microplates. Meanwhile, the quasi-3D theory which is constructed from a novel seventh-order shear deformation refined plate theory with four unknowns is able to consider both shear deformations and thickness stretching effect without requiring shear correction factors. The NURBS-based isogeometric analysis is integrated to exactly describe the geometry and approximately calculate the unknown fields with higher-order derivative and continuity requirements. The convergence and verification show the validity and efficiency of this proposed computational approach in comparison with those existing in the literature. It is further applied to study the static bending, free vibration and buckling responses of rectangular and circular functionally graded microplates with various types of boundary conditions. A number of investigations are also conducted to illustrate the effects of the material length scale, material index, and length-to-thickness ratios on the responses of the microplates. 

\end{abstract}

\begin{keyword}
%% keywords here, in the form: keyword \sep keyword
Functionally graded microplates \sep Modified couple stress theory \sep Refined plate theory \sep Quasi-3D theory \sep Isogemetric analysis.
%% MSC codes here, in the form: \MSC code \sep code
%% or \MSC[2008] code \sep code (2000 is the default)
\end{keyword}

\end{frontmatter}

%%
%% Start line numbering here if you want
%%
% \linenumbers

%% main text
%% main text
\section{Introduction}
\label{sec:Intro} 
Functionally graded materials (FGMs), which are initially proposed by group of Japanese aerospace researchers, are composite materials formed of two or more constituent phases in which material properties vary smoothly from one surface to the other. Consequently, FGMs are able to avoid high interlaminar shear stresses, stress concentration and delamination phenomenon which are often cited as shortcomings of laminated composite materials. A FGM consisting of ceramic and metal possesses higher thermal resistance and better ductility which are inherited from the ceramic and metal phases, respectively. Owning these striking features, FGMs are applicable to various fields of engineering including aerospace, nuclear power, chemistry and biology engineering. The FGMs are also widely studied within various types of structure such as beams \cite{trinh_fundamental_2015, vo_finite_2014, vo_static_2013, asghari_size_dependent_2010}, plates \cite{nguyen_new_2014, thai_new_2013, Zenkour200667, reddy_analysis_2000}, and shells \cite{mantari_refined_2015, torabi_linear_2013, tornabene_free_2009}.

Recent advances in technology lead to new industrial fields in which small-scale elements are involved. Such elements have been applied in the micro- and nano-electro-mechanical system \cite{fu_tini_based_2004, lee_metallic_2006}, actuators \cite{baughman_carbon_1999}, space and bio-engineering \cite{lau_carbon_2008}. These applications encourage new research area that focuses on investigating and predicting the behaviours of such micro structures. A number of attempts have been conducted to analyse characteristic of small-scale structures including experiment and computer simulation \cite{frankland_molecular_2002, liew_study_2004}. Nevertheless, while the former is mechanical and economical challenging due to the extremely small size of the elements, the latter exponentially raises computational cost. Therefore, the mathematical and mechanical models which account for the small-scale effects are needed. It is widely known that the behaviour of small-scale structures is size-dependent. This is confirmed following a number of theoretical and experimental works of Fleck et al. \cite{fleck_strain_1994}, Stolken and Evans \cite{stolken_microbend_1998}, and Lam et al. \cite{Lam20031477}. Having an experimental observation from bending test of epoxy polymeric microbeams, Lam et al. \cite{Lam20031477} point out that the bending rigidity increases $2.4$ times as a result of the reduction of beam thickness from $115$ $\mu$m to $20$ $\mu$m. 

In order to take into account the size effects, based on the classical elasticity theories, a number theories have been developed including nonlocal elasticity theory \cite{eringen_nonlocal_1972}, strain gradient theory \cite{fleck_phenomenological_1993}, and modified couple stress theory \cite{yang_couple_2002}. It is worth noting that the classical elasticity is fundamentally founded by the introduction of the Hook's Law in which the force and the change in displacement are linearly related via the stiffness of the component where the forces are applied. This principal of physics governs the linearly elastic behaviours of materials. As physics and mechanics develop, stress and strain which are linearly connected via the elastic modulus are cited as the primary measures of classical elasticity rather than force and displacement counterparts. Aiming for better prediction of materials' responses, Mindlin and Tiersten \cite{mindlin_effects_1962} and Mindlin \cite{mindlin_second_1965} have initially developed higher-order theories of elasticity. Based on the employment of deformation metrics, those theories could be classified into two categories including strain gradient and couple stress theories. With regard to the strain gradient theories, this concept was first developed by Fleck and Hutchinson \cite{fleck_phenomenological_1993} in which the Mindlin's theory \cite{mindlin_second_1965} was extended. There are two components, which are classified using the second-order deformation tensor that include stretch gradient tensor and rotation gradient tensor \cite{ansari_size_dependent_2013}. Within the concept of couple stress theory, both strain and curvature jointly govern the strength of the solid. In addition, while the antisymmetric part of the second-order deformation gradients is served as rotation gradients, the symmetric part is neglected. Based on the initial ideas of couple stress theory, a number of attempts have been conducted to further develop the concepts that are applicable for size-dependent problems. Yang et al. \cite{yang_couple_2002} proposed the equilibrium of moment of couples which was an additional equilibrium relation that forces the couple stress tensor to be symmetric. Therefore, the deformation energy is only influenced by the symmetric part of rotation gradient and the symmetric part of displacement gradient. In addition, instead of using two material length scale parameters as need in the classical couple stress theory, this modified couple stress theory (MCST) requires only one material length scale parameter to construct the constitutive relation. Park and Gao \cite{park_variational_2007} utilised the principle of minimum total potential energy to develop a variational formulation of the modified couple stress theory. This method not only derives the equilibrium equations but also forms the determination of boundary conditions which are not available in the Yang's theory \cite{yang_couple_2002}. In recent years, the modified couple stress theory has been applied to study the various behaviours of small-scale structures. For beam analysis, static bending, buckling, and vibration analysis have been conducted using Euler-Bernoulli \cite{park_bernoullieuler_2006}, Timoshenko \cite{xia_nonlinear_2010, ke_size_2011, roque_study_2013, thai_size_dependent_2015}, and higher-order beam theories \cite{salamat_talab_static_2012}. The modified couple stress theory was also applied in the small-scale plate analysis in several ways, Tsiatas \cite{tsiatas_new_2009} initially employed MCST to investigate the static bending response of isotropic Kirchhoff microplates. Yin et al. \cite{Yin2010386} solely investigated the vibration behaviour of Kirchhoff microplates using the standard separation of variables to derive the closed-form solution for natural frequency. Bending and vibration behaviours of Mindlin microplates were studied by Ma et al. \cite{ma_non_classical_2011} in which the thickness stretching effect was also taken into account. With regard to the small-scale functionally graded (FG) structures, a number of investigations have been conducted for micro FG beams and plates. Simsek et al. \cite{simsek_static_2013} investigated the static bending of functionally graded microbeams based on the modified couple stress theory. Buckling behaviour of FG microbeams was studied by Nateghi et al. \cite{nateghi_size_2012}. Based on the modified couple stress theory, Reddy et al. \cite{reddy_nonlinear_2012_a, reddy_nonlinear_2012_b, kim_general_2015, reddy_nonlinear_2015} studied nonlinear behaviour of small-scale FG microplates with different geometry shapes. Thai and his colleagues utilised the Navier's approach to derive  solutions for FG microplates in which Kirchoff, Minlind and sinusoidal plate theories were used along with the MCST \cite{Thai20131636, Thai2013142, thai_size_dependent_2013}. Ke et al. \cite{ke_free_2012, ke_bending_2012} employed $p$-version Ritz method and different quadrature method to solve free vibration and bending, buckling problems of the rectangular and annular FG Mindlin microplates based on MCST, respectively. Using modified couple stress theory, a refined plate theory was utilised to predict the bending, buckling, and vibration behaviours of FG microplates by He et al. \cite{He2015107} following the Navier's approach. Most of these efforts followed analytical approach which was able to solve some specific types of boundary condition.

A large proportion of the studies in small-scale FG structures employ the classical plate theory (CPT) and the first-order shear deformation theory (FSDT). However, the classical plate theory (or the Kirchoff-Love theory), which neglects shear deformation, provides acceptable solutions for thin plates, i.e. length-to-thickness ratios are quite large, only. Meanwhile, the first-order shear deformation theory (or Reissner-Mindlin plate theory), which accounts for transverse shear effects, is applicable for both thin and moderately thick plates \cite{della_croce_finite_2004, nguyen_xuan_edge_based_2010, zhang_non_classical_2013}. The shortcomings of the FSDT include inaccurate distribution of transverse shear strain/stress and violation of traction free boundary conditions at top and bottom surfaces. For this reason, shear correction factors are required to adjust the transverse shear stress distribution. However, these factors vary from problems and one may find it difficult to choose an appropriate value. In order to avoid using shear correction factors, the third-order shear deformation theory (TSDT) \cite{reddy_simple_1984}, higher-order shear deformation theory (HSDT) \cite{nguyen_xuan_isogeometric_2013}, sinusoidal shear deformation theory (SSDT) \cite{arya_zigzag_2002}, and refined plate theories (RPT) \cite{senthilnathan_buckling_1987} have been developed yielding more accurate and reliable results. The refined plate theory is initially proposed by Senthilnathan et al. \cite{senthilnathan_buckling_1987} by employing four unknowns and is one variable less than that of TSDT. Shimpi et al. \cite{shimpi_refined_2002, shimpi_two_2006, shimpi_free_2006} then further developed the RPT for isotropic and orthotropic plates by using only two variables. However, HSDT and RPT require C$^1$-continuity of general displacements which cause significant challenge to derive second derivative of deflection in the platform of finite element analysis (FEA) where C$^0$ elements are frequently used. In order to overcome the continuity issues, some C$^0$ approximations \cite{shankara_c0element_1996} and Hermite interpolation functions with C$^1$ elements \cite{reddy_analysis_2000} which involve adding extra unknown variables of derivative of displacement are adopted. 

Recently, a new numerical method so-called Isogeometric Analysis (IGA) which is able to deal with C$^1$-continuity problem without using any additional variables or Hermite interpolation function has been introduced by Hughes and his co-workers \cite{Hughes20054135}. This method bridges the existing gap between the fields of finite element analysis and computer-aided design (CAD). The essential idea of the isogeometric analysis is that the basis functions which are employed to exactly describe the geometry domain will also be used for approximations of unknown fields. Since geometry is exactly modeled and the number of unknown terms is not increased, it is expected that IGA would yield more accurate results with lower computational cost for HSDT and RPT problems in the comparison with regular FEA \cite{NguyenXuan2014222}. One may find the guidance on computer implementation of IGA in the open literature \cite{ vuong_isogat_2010, de_falco_geopdes_2011, nguyen_phu_isogeometric_2015}.  IGA has been widely implemented in a number of linearly and non-linearly mechanical and thermal problems such as static, free vibration, and buckling of laminated composite and functionally graded plates with various plate theories including layerwise, FSDT, HSDT, and RPT \cite{wang_superconvergent_2015, beirao_da_veiga_isogeometric_2012, thai_static_2012, nguyen_xuan_isogeometric_2013, Tran2013368, NguyenXuan2014222, thai_isogeometric_2013, tran_isogeometric_2013, phung_van_isogeometric_2015, tran_geometrically_2015, thai_isogeometric_2015}. However, as far as author's knowledge, there is no publication dealing with analysis of small-scale plates based on modified couple stress theory and NURBS basis functions.
 
In this study, in order to fill the aforementioned gap, the bending, free vibration and buckling behaviours of FG microplates based on the modified couple stress theory and four-variable refined plate and quasi-3D theories is investigated in conjunction with the isogeometric approach. While the modified couple stress theory is employed to capture the small-scale effects, the displacement fields of those microplates are expressed based on a novel seventh-order RPT and quasi-3D theory. The mechanical behaviours of FG microplates are then solved by the isogeometric analysis in which NURBS functions are simultaneously used to exactly describe the geometry boundaries and construct the basis functions of the approximations.

The outline of this study is expressed as follow. The next section presents theories which are applicable for analysis of FG microplates including modified couple stress theory, RPT and quasi-3D theory. In addition, a brief note on functionally graded material will also be introduced in this section. Section \ref{sec:IGA} focuses on the isogeometric and NURBS-based formulation of quasi-3D theory. The numerical examples which cover static bending, free vibration and buckling analysis of rectangular and circular FG microplates with various boundary conditions will be provided in Section \ref{sec:Exp}. Finally, some conclusions will be drawn in Section \ref{sec:Conclusion}.

\section{A novel theory for FG microplates}
\label{sec:Theory}

In this section, a brief review on the formulation of modified couple stress theory with only one material length scale that accounts for size-dependent effects is firstly presented. It is followed by the theoretical definition of functionally graded material of which microplates are made. Displacement field of plates is then derived based on the four-variables refined plate theory and quasi-3D plate theory where a novel seventh-order shear deformation theory is proposed.

\subsection{Modified couple stress theory}
\label{sec:Theory:MCS} 
According to the modified couple stress theory which is proposed by Yang et al. \cite{yang_couple_2002}, the strain energy density $w$ for linear isotropic material is a quadratic function of generalized strains
\begin{equation}
w = \frac{1}{2}\lambda \left( {\text{tr}{\pmb{{\bf{\varepsilon}}}}} \right)^2  + \mu \left( {{\pmb{{\bf{\varepsilon}}}}:{\pmb{{\bf{\varepsilon}}}} + l^2 {\boldsymbol{\chi }}:{\boldsymbol{\chi }}} \right),
\end{equation}
where $\lambda$ and $\mu$ are Lame's constants, $\mu$ is also known as shear modulus which is often denoted as $G$, $l$ represents material length scale parameter and the strain tensor $\pmb{{\bf{\varepsilon}}}$ and symmetric curvature tensor $\boldsymbol{\chi}$ are defined by
\begin{subequations}
\begin{align}
 {\pmb{{\bf{\varepsilon}}}} &= \frac{1}{2}\left[ {\nabla {\bf{u}} + \left( {\nabla {\bf{u}}} \right)^T } \right], \label{eqn:disp_strain_def}\\ 
 {\boldsymbol{\chi }} &= \frac{1}{2}\left[ {\nabla {\boldsymbol{\theta }} + \left( {\nabla {\boldsymbol{\theta }}} \right)^T } \right],
 \label{eqn:curvature_rotation}
\end{align}
\end{subequations}
where $\bf{u}$ is displacement vector and the rotation vector $\boldsymbol{\theta}$ is given by
\begin{equation}
\label{eqn:rotation_displ}
{\boldsymbol{\theta }} = \frac{1}{2}\text{curl}\left( {\bf{u}} \right).
\end{equation}
The strain energy $U$ restored in a deformed elastic body is then defined as
\begin{equation}
U =\int\limits_V {wdV}  = \int\limits_V {\left( {{\boldsymbol{\sigma }}:{\pmb{{\bf{\varepsilon}}}} + {\bf{m}}:{\boldsymbol{\chi }}} \right)dV},
\end{equation}
where $\boldsymbol{\sigma}$ and $\bf{m}$ are the symmetric stress tensor and the deviatoric part of the couple stress tensor, respectively. These components, $\boldsymbol{\sigma}$ and $\bf{m}$, which are conjugated to the deformation measures $\pmb{{\bf{\varepsilon}}}$ and $\boldsymbol{\chi}$, respectively, are given as
\begin{subequations}
\label{eqn:mcs:couple_stress}
\begin{align}
{\boldsymbol{\sigma }} &= \lambda \left( {\text{tr}{\pmb{{\bf{\varepsilon}}}}} \right){\bf{I}} + 2\mu {\pmb{{\bf{\varepsilon}}}}, \\ 
{\bf{m}} &= 2\mu l^2 {\boldsymbol{\chi }},
\end{align}
\end{subequations}
where $\bf{I}$ is the identity matrix. Apparently, Eq. (\ref{eqn:mcs:couple_stress}) shows there is only one material length scale needed and the deviatoric couple stress tensor $\bf{m}$ is also symmetric, from which the modified couple stress theory is formed.

\subsection{Functionally graded material}
\label{sec:Theory:FGM}

\begin{figure}
	\centering
	\includegraphics[width=1.0\textwidth]{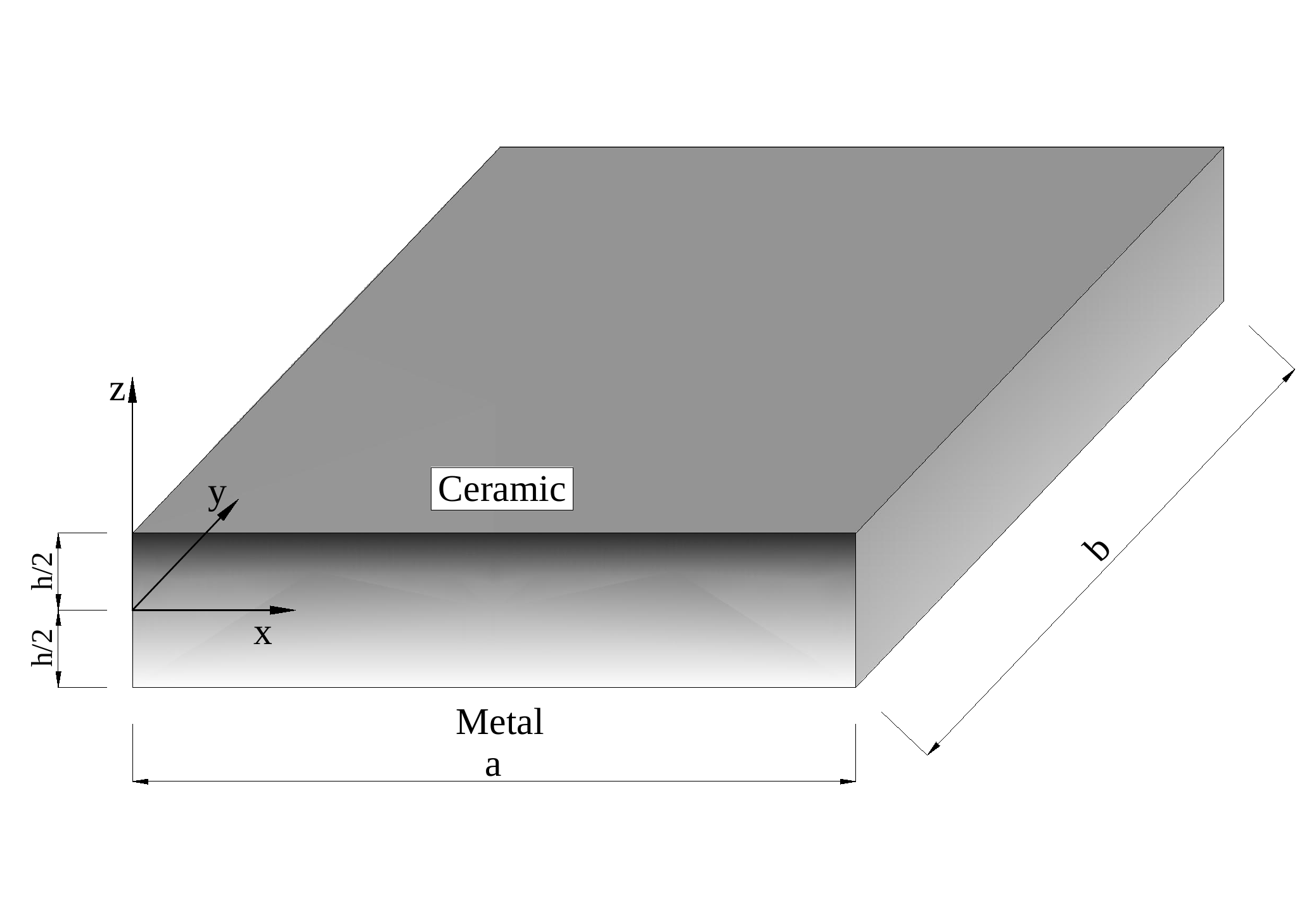}
	\caption{The functionally graded microplate model. \label{fig:fgm_model}}
\end{figure} 

The model configuration of the functional graded material which is made of metal and ceramic is illustrated in Figure \ref{fig:fgm_model}. There are several homogeneous models that are employed to estimate the effective properties of the FGMs. According to the rule of mixtures, the corresponding effective properties of the FGMs could be expressed as follow
\begin{equation}
P_e  =  P_m V_m + P_c V_c,
\label{eqn:fgm_effective}
\end{equation}
where $P_m$ and $P_c$ are the material properties of metal and ceramic phases, respectively, including Young's modulus $E$, density $\rho$ and Poisson's ratio $\nu$. Meanwhile, $V_m$ and $V_c$ represent the volume fraction of metal and ceramic phases, respectively, which could be defined as follow \cite{reddy_analysis_2000}
\begin{equation}
V_c \left( z \right) = \left( {\frac{1}{2} + \frac{z}{h}} \right)^n , \quad V_m  = 1 - V_c , \quad -\dfrac{h}{2} \leq z \leq \dfrac{h}{2},
\label{eqn:volume_distribution}
\end{equation}
where $n$ is the material index. This equation implies smooth variation in material properties of FGM is guaranteed and governed by the material index $n$. As can be inferred from Eq. (\ref{eqn:volume_distribution}), $n=0$ leads to a FGM with fully homogeneous ceramic while the FGM becomes fully metal as $n$ goes toward $\infty$. Nevertheless, the rule of mixtures model fails to describe the interaction between the material phases \cite{vel_exact_2002,qian_static_2004}. Therefore, Mori-Tanaka scheme \cite{mori_average_1973,kiani_thermal_2014} is developed to fill this gap in estimation of FGM's properties by integrating effective bulk modulus $K_e$ and shear modulus $G_e$ which are given by
\begin{equation}
\frac{{K_e  - K_m }}{{K_c  - K_m }} = \frac{{V_c }}{{1 + V_m \frac{{K_c  - K_m }}{{K_m  + \frac{4}{3}G_m }}}}, \quad
\frac{{G_e  - G_m }}{{G_c  - G_m }} = \frac{{V_c }}{{1 + V_m \frac{{G_c  - G_m }}{{G_m  + f_1 }}}},
\end{equation}
where
\begin{equation}
f_1  = \frac{{G_m \left( {9K_m  + 8G_m } \right)}}{{6\left( {K_m  + 2G_m } \right)}}.
\end{equation}
The effective Young's modulus $E_e$ and Poisson's ratio $\nu_e$ are then defined as
\begin{equation}
E_e  = \frac{{9K_e G_e }}{{3K_e  + G_e }}, \quad \nu _e  = \frac{{3K_e  - 2G_e }}{{2\left( {3K_e  + G_e } \right)}}.
\end{equation}

\begin{figure}
	\centering
	\includegraphics[width=1.0\textwidth]{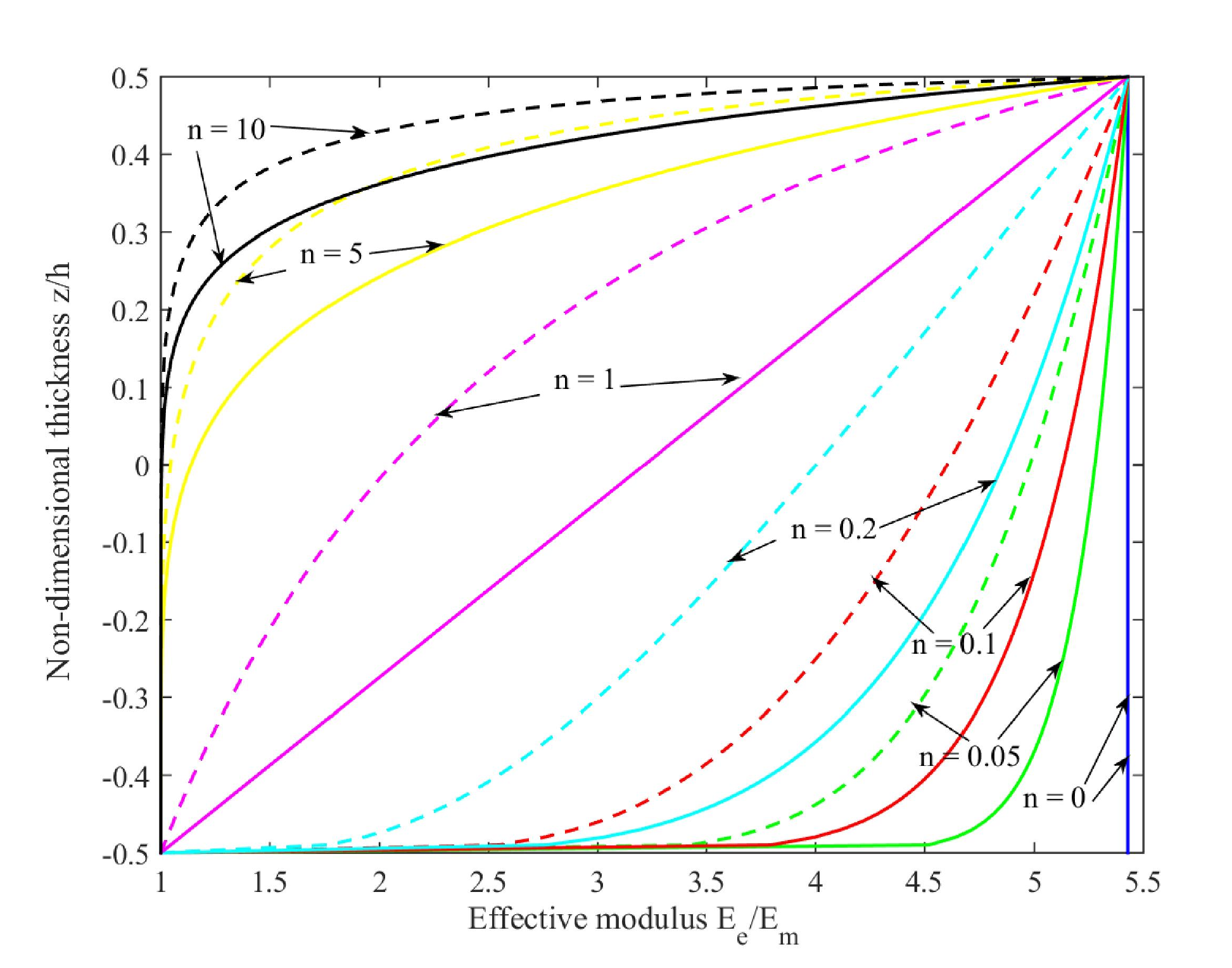}
	\caption{The effective modulus of Al/Al$_2$O$_3$ plates according to the rule of mixtures scheme (in solid lines) and Mori-Tanaka scheme (in dash lines). \label{fig:fgm_variation}}
\end{figure}

The variations in effective Young's modulus of Al/Al$_2$O$_3$ estimated by the rule of mixtures and Mori-Tanaka scheme are depicted in Figure \ref{fig:fgm_variation}. As can be seen, corresponding to the material index $n$, the effective property of FGM is varied continuously from the metal-rich surface at the bottom to the ceramic-rich one at the top of the plate. 

\subsection{A novel seventh-order shear deformation plate theory}
\label{sec:Theory:HSDT}
With regard to the plate theories, the third-order shear deformation model initially proposed by Reddy \cite{reddy_simple_1984} is widely considered as a reliable theory in which no shear correction factor is required. In Reddy's theory, the displacement field, for $z \in \left[ { - h/2; h/2} \right]$, is defined as
\begin{subequations}
\begin{align}
u\left( {x,y,z} \right) &= u_0\left( {x,y} \right)  + z\beta _x\left( {x,y} \right)  + g\left( z \right)\left( {\beta _x\left( {x,y} \right)  + w_{,x} \left( {x,y} \right)} \right), \\
v\left( {x,y,z} \right) &= v_0\left( {x,y} \right)  + z\beta _y\left( {x,y} \right)  + g\left( z \right)\left( {\beta _y\left( {x,y} \right)  + w_{,y}\left( {x,y} \right) } \right),\\
w\left( {x,y,z} \right) &= w_0\left( {x,y} \right) ,
\end{align}
\label{eqn:displacement_TSDT}
\end{subequations}
where comma notation represents derivative, $g(z) = -4z^3/(3h^2)$ and the variables $ {\bf{u}}_0  = \left[ {u_0 \quad v_0 } \right]^T$, $w_0$, and ${\boldsymbol{\beta }} = \left[ {\beta _x \quad \beta _y } \right]^T $ are the membrane displacements, the transverse deflection of the mid-plane surface, and the rotations, respectively. By making further assumptions, $w_0 = w_b + w_s, \beta _x  =  - w_{b,x}, \beta _y  =  - w_{b,y}$,  to the Reddy's theory which contains five unknowns, Senthilnathan \cite{senthilnathan_buckling_1987} initially proposes the four-variable refined plate theory which could be expressed in the generalized form as follows
\begin{subequations}
\begin{align}
u\left( {x,y,z} \right) &= u_0 \left( {x,y} \right) - zw_{b,x} \left( {x,y} \right) + g\left( z \right)w_{s,x} \left( {x,y} \right), \\
 v\left( {x,y,z} \right) &= v_0 \left( {x,y} \right) - zw_{b,y} \left( {x,y} \right) + g\left( z \right)w_{s,y} \left( {x,y} \right), \\
 w\left( {x,y,z} \right) &= w_b \left( {x,y} \right) + w_s \left( {x,y} \right), \label{eqn:deflection_w}
\end{align}
\label{eqn:displacement_RPT4}
\end{subequations}
where $w_b$ and $w_s$ represent bending and shear components of transverse displacement, respectively. The function $g(z) = f(z) - z$ is employed to describe the distribution of transverse shear strains and stresses through the plates' thickness. It is necessary to have the first derivative of $f(z)$ satisfies the tangential zero value at $z=\pm h/2$ such that the traction-free condition at top and bottom surfaces is meet. Consequently, the shear correction factor is no longer required for higher-order shear deformation theory and refined plate theory.

It should be noted that the higher-order shear deformation theory and refined plate theory both fail to capture the thickness stretching effect of normal deformation $(\varepsilon_z \neq 0)$ due to the constant deflection through the plate thickness which can be inferred from Eq. (\ref{eqn:deflection_w}). In order to bypass this shortcoming, a number of theories which consider the thickness stretching effect have been developed \cite{Carrera2011123, mantari_novel_2013, thai_simple_2016}. Zenkour \cite{Zenkour20139041, zenkour_bending_2013} initially proposes the four-variable quasi-3D plate theory accounting for both transverse shear and normal deformations which could be alternatively expressed as follows
\begin{subequations}
\begin{align}
u\left( {x,y,z} \right) &= u_0 \left( {x,y} \right) - zw_{b,x} \left( {x,y} \right) + f\left( z \right)w_{s,x} \left( {x,y} \right), \\
 v\left( {x,y,z} \right) &= v_0 \left( {x,y} \right) - zw_{b,y} \left( {x,y} \right) + f\left( z \right)w_{s,y} \left( {x,y} \right), \\
 w\left( {x,y,z} \right) &= w_b \left( {x,y} \right) + \phi\left( z \right) w_s \left( {x,y} \right).
\end{align}
\label{eqn:displacement_Q3D4}
\end{subequations}

As can be observed, this quasi-3D model has a similar form to that of four-variable refined plate theory shown in Eq. (\ref{eqn:displacement_RPT4}). Indeed, the displacement field based on the refined plate theory could be readily obtained by degrading those of quasi-3D theory in which $f(z)$ and $\phi(z)$ are  replaced by $g(z)$ and $1$, respectively.

A number of distribution functions, $f(z)$ and $\phi(z)$, are available to analyse the FGM plates based on higher-order shear deformation theory, refined plate theory, and quasi-3D plate theory. One may find the general framework to construct such polynomial functions in the recent work of Nguyen et al. \cite{nguyen_general_2016}. In this study, a novel seventh-order function of $f(z)$ and corresponding function $\phi(z)$ are proposed for the four-variable refined plate theory and quasi-3D theory. Those functions are presented in Table \ref{tab:distribution_function} along with other's existing in the literature.
\begin{table}[htbp]
  \centering
  \caption{Various forms of distribution function used for HSDT, RPT, and quasi-3D theories \label{tab:distribution_function}}
    \begin{tabular}{llll}
    \hline
    Theory & $\varepsilon_z$ & $f(z)$     & $\phi_z$ \\
    \hline
    HSDT \cite{reddy_simple_1984} &$= 0$   &  $
z - \frac{4}{3}\frac{{z^3 }}{{h^2 }}$         & -	\\
    RPT \cite{NguyenXuan2014222} & $=0$ & $\arctan \left( {\sin \left( {\frac{\pi }{h}z} \right)} \right)$     & -	\\
    Quasi-3D \cite{thai_simple_2013} & $\neq 0$     & $\frac{h}{\pi }\sin \left( {\frac{{\pi z}}{h}} \right) - z$
     & $f'\left( z \right) + 1$
	\\
    Quasi-3D \cite{Zenkour20139041} & $\neq 0$     & $h\sinh \left( {\frac{z}{h}} \right) - \frac{{4z^3 }}{{3h^2 }}\cosh \left( {\frac{1}{2}} \right)$
    & $\frac{1}{{12}}f'\left( z \right)$
	\\
    Quasi-3D \cite{Nguyen2015191} & $\neq 0$
     & $\frac{\pi }{h}z - \frac{{9\pi }}{{5h^3 }}z^3  + \frac{{28\pi }}{{25h^5 }}z^5$     & $\frac{1}{8}f'\left( z \right)$
	\\
    Present RPT     & $=0$     & $ - 8z + \frac{{10z^3 }}{{h^2 }} + \frac{{6z^5 }}{{5h^4 }} + \frac{{8z^7 }}{{7h^6 }}$
     & -	\\
     Present quasi-3D     & $\neq 0$     & $ - 8z + \frac{{10z^3 }}{{h^2 }} + \frac{{6z^5 }}{{5h^4 }} + \frac{{8z^7 }}{{7h^6 }}$ & $\frac{3}{{20}}f'\left( z \right)$\\
	\hline
    \end{tabular}
\end{table}

According to the displacement field and the strain-displacement relation, which are presented in Eqs. (\ref{eqn:displacement_Q3D4}) and (\ref{eqn:disp_strain_def}), respectively, the following strain expressions could be obtained as

\begin{subequations}
\label{eqn:disp_strain}
\begin{align}
\pmb{{\bf{\varepsilon}}} &= \pmb{{\bf{\varepsilon}}}_0  + z {\boldsymbol{\kappa}}_b  + f\left( z \right){\boldsymbol{\kappa}}_s,
\\
\pmb{\bf{\gamma}} &= \left[ {f'\left( z \right) + \phi \left( z \right)} \right]\pmb{{\bf{\varepsilon}}}_s,
\end{align}
\end{subequations}
where
\begin{equation}
\label{eqn:disp_strain_denotes}
\begin{split}
\pmb{{\bf{\varepsilon}}} &= \left[ \begin{array}{l}
 \varepsilon_x  \\ 
 \varepsilon_y \\ 
 \gamma_{xy}  \\ 
 \end{array} \right], \quad
\pmb{{\bf{\varepsilon}}}_0 = \left[ \begin{array}{l}
 u_{0,x}  \\ 
 v_{0,y}  \\ 
 u_{0,y}  + v_{0,x}  \\ 
 \end{array} \right], \quad
{\boldsymbol{\kappa}}_b  =  - \left[ \begin{array}{l}
 w_{b,xx}  \\ 
 w_{b,yy}  \\ 
 2w_{b,xy}  \\ 
 \end{array} \right], \quad
{\boldsymbol{\kappa}}_s = \left[ \begin{array}{l}
 w_{s,xx}  \\ 
 w_{s,yy}  \\ 
 2w_{s,xy}  \\ 
 \end{array} \right],
 \\
\pmb{\bf{\gamma}} &= \left[ \begin{array}{l}
 \gamma_{xz}  \\ 
 \gamma_{yz} \\ 
 \end{array} \right], \quad \pmb{\bf{\varepsilon}}_s = \left[ \begin{array}{l}
 w_{s,x}  \\ 
 w_{s,y}  \\ 
 \end{array} \right], \quad \varepsilon _z  = \phi '\left( z \right)w_s.
\end{split}
\end{equation}
Using Eqs. (\ref{eqn:displacement_Q3D4}), (\ref{eqn:rotation_displ}), and (\ref{eqn:curvature_rotation}),  rotation vector and curvature vector are respectively expressed by 
\begin{subequations}
\label{eqn:theta_chi}
\begin{align}
\boldsymbol{\theta}&= \left[ \begin{array}{l}
 \theta _x  \\ 
 \theta _y  \\ 
 \theta _z  \\ 
 \end{array} \right] = \frac{1}{2}\left[ \begin{array}{l}
 2w_{b,y}  - \left( {f' - \phi} \right)w_{s,y}  \\ 
  - 2w_{b,x}  + \left( {f' - \phi} \right)w_{s,x}  \\ 
 v_{0,x}  - u_{0,y}  \\ 
 \end{array} \right], 
 \\
 {\boldsymbol{\chi }} &= \left[ \begin{array}{l}
 {\boldsymbol{\chi }}_b  \\ 
 {\boldsymbol{\chi }}_s  \\ 
 \chi _{zz}  \\ 
 \end{array} \right] = \left[ \begin{array}{l}
 {\boldsymbol{\chi }}_{b0}  \\ 
 {\boldsymbol{\chi }}_{s0}  \\ 
 0 \\ 
 \end{array} \right] + \left[ \begin{array}{l}
 f'{\boldsymbol{\chi }}_{b1}  \\ 
 f''{\boldsymbol{\chi }}_{s2}  \\ 
 0 \\ 
 \end{array} \right] + \left[ \begin{array}{l}
 \phi {\boldsymbol{\chi }}_{b3}  \\ 
 \phi '{\boldsymbol{\chi }}_{s4}  \\ 
 0 \\ 
 \end{array} \right],
\end{align}
\end{subequations}
where
\begin{equation}
\label{eqn:curvatures}
\begin{split}
{\boldsymbol{\chi }}_b  &= \left[ \begin{array}{l}
 \chi _{xx}  \\ 
 \chi _{yy}  \\ 
 \chi _{xy}  \\ 
 \end{array} \right],
{\boldsymbol{\chi }}_{b0}  = \frac{1}{2}\left[ \begin{array}{l}
 w_{b,xy}  \\ 
  - w_{b,xy}  \\ 
 w_{b,yy}  - w_{b,xx}  \\ 
 \end{array} \right], {\boldsymbol{\chi }}_{b1}  = \frac{1}{4}\left[ \begin{array}{l}
  - 2w_{s,xy}  \\ 
 2w_{s,xy}  \\ 
  - w_{s,yy}  + w_{s,xx}  \\ 
 \end{array} \right], {\boldsymbol{\chi }}_{b3} = \frac{1}{4}\left[ \begin{array}{l}
 2w_{s,xy}  \\ 
  - 2w_{s,xy}  \\ 
 w_{s,yy}  - w_{s,xx}  \\ 
 \end{array} \right], 
 \\
{\boldsymbol{\chi }}_s  &= \left[ \begin{array}{l}
 \chi _{xz}  \\ 
 \chi _{yz}  \\ 
 \end{array} \right],
{\boldsymbol{\chi }}_{s0}  = \frac{1}{4}\left[ \begin{array}{l}
 v_{0,xx}  - u_{0,xy}  \\ 
 v_{0,xy}  - u_{0,yy}  \\ 
 \end{array} \right], {\boldsymbol{\chi }}_{s2}  = \frac{1}{4}\left[ \begin{array}{l}
  - w_{s,y}  \\ 
 w_{s,x}  \\ 
 \end{array} \right], {\boldsymbol{\chi }}_{s4}  = \frac{1}{4}\left[ \begin{array}{l}
 w_{s,y}  \\ 
  - w_{s,x}  \\ 
 \end{array} \right].
\end{split}
\end{equation}
According to Eq. (\ref{eqn:mcs:couple_stress}), the constitutive relations for classical and modified couple stress theories could be presented in explicit form as follows
\begin{subequations}
\begin{align}
\left\{ \begin{array}{l}
 \sigma _x  \\ 
 \sigma _y  \\ 
 \sigma _z  \\ 
 \sigma _{xy}  \\ 
 \tau _{xz}  \\ 
 \tau _{yz}  \\ 
 \end{array} \right\} &= \left[ {\begin{array}{*{20}c}
   {Q_{11} } & {Q_{12} } & {Q_{13} } & 0 & 0 & 0  \\
   {} & {Q_{22} } & {Q_{23} } & 0 & 0 & 0  \\
   {} & {} & {Q_{33} } & 0 & 0 & 0  \\
   {} & {} & {} & {Q_{66} } & 0 & 0  \\
   {} & {} & {} & {} & {Q_{55} } & 0  \\
   {\text{sym.}} & {} & {} & {} & {} & {Q_{44} }  \\
\end{array}} \right]\left\{ \begin{array}{l}
 \varepsilon _x  \\ 
 \varepsilon _y  \\ 
 \varepsilon _z  \\ 
 \varepsilon _{xy}  \\ 
 \gamma _{xz}  \\ 
 \gamma _{yz}  \\ 
 \end{array} \right\}, 
 \\
m_{ij} &= 2G_el^2\chi_{ij},
\end{align}
\end{subequations}
where, for the proposed quasi-3D theory $(\varepsilon_z \neq 0)$, $Q_{ij}$ are the three-dimensional elastic constant which could be calculated as
\begin{equation}
\begin{split}
Q_{11}  &= Q_{22}  = Q_{33}  = \frac{{\left( {1 - \nu _e } \right)E_e }}{{\left( {1 - 2\nu _e } \right)\left( {1 + \nu _e } \right)}},
\\
Q_{12}  &= Q_{13}  = Q_{23}  = \frac{{\nu _e E_e }}{{\left( {1 - 2\nu _e } \right)\left( {1 + \nu _e } \right)}},
\\
Q_{44}  &= Q_{55}  = Q_{66}  = \frac{{E_e }}{{2\left( {1 + \nu _e } \right)}},
\end{split}
\end{equation}
meanwhile, for the proposed refined plate theory $(\varepsilon_z = 0)$, $Q_{ij}$ are reduced plane-stress elastic constant and are expressed as
\begin{equation}
\begin{split}
Q_{11}  &= Q_{22}  = \frac{{E_e }}{{1 - \nu _e^2 }},
\\
Q_{12}  &= Q_{21}  = \frac{{E_e \nu _e }}{{1 - \nu _e^2 }},
\\
Q_{44}  &= Q_{55}  = Q_{66}  = \frac{{E_e }}{{2\left( {1 + \nu _e } \right)}},
\end{split}
\end{equation}
and the shear modulus $G_e = \dfrac{E_e}{2(1+\nu_e)}$.

In this study, the weak form of the static bending, vibration, and buckling problems are derived using the Hamilton's principle and weak formulation. One could find those straightforward procedures in the literature \cite{reddy_theory_2006, NguyenXuan2014222, Nguyen2015191}. Firstly, the weak form of the static bending of the couple-stress-based microplates subjected to transverse load loading $q_0$ could be expressed in the following compact form 
\begin{equation}
\label{eqn:weakform_bending}
\begin{split}
\int\limits_\Omega  {\delta {\pmb{\bf{\varepsilon }}}_b ^T {\bf{D}}^b {\pmb{\bf{\varepsilon }}}_b d\Omega }  + \int\limits_\Omega  {\delta {\pmb{\bf{\varepsilon }}}_s ^T {\bf{D}}^s {\pmb{\bf{\varepsilon }}}_s d\Omega }  + \int\limits_\Omega  {\delta \left( {{\boldsymbol{\chi }}_b^c } \right)^T {\boldsymbol{D}}_c^b {\boldsymbol{\chi }}_b^c d\Omega }  + \int\limits_\Omega  {\delta \left( {{\boldsymbol{\chi }}_s^c } \right)^T {\bf{D}}_c^s {\boldsymbol{\chi }}_s^c d\Omega }  \\ = \int\limits_\Omega  {\left[ {\delta w_b  + \phi \left( {\frac{h}{2}} \right)\delta w_s } \right]q_0 d\Omega } 
\end{split},
\end{equation}
where the strain tensors and matrices in the first two terms in Eq. (\ref{eqn:weakform_bending}) related to classical elastic theory are
\begin{equation}
{\pmb{\boldsymbol{\varepsilon }}}_b  = \left[ \begin{array}{l}
 {\bf{\varepsilon }}_0  \\ 
 {\bf{\kappa }}_b  \\ 
 {\bf{\kappa }}_s  \\ 
 w_s  \\ 
 \end{array} \right], \quad {\pmb{\boldsymbol{\varepsilon }}}_s  = \left[ \begin{array}{l}
 w_{s,x}  \\ 
 w_{s,y}  \\ 
 \end{array} \right], \quad {\bf{D}}^b  = \left[ {\begin{array}{*{20}c}
   {\bf{A}} & {\bf{B}} & {\bf{E}} & {\bf{X}}  \\
   {\bf{B}} & {\bf{D}} & {\bf{F}} & {{\bf{Y}}^b }  \\
   {\bf{E}} & {\bf{F}} & {\bf{H}} & {{\bf{Y}}^s }  \\
   {\bf{X}} & {{\bf{Y}}^b } & {{\bf{Y}}^s } & {Z_{33} }  \\
\end{array}} \right],
\end{equation}
in which the material matrices are calculated by
\begin{subequations}
\begin{align}
&\left( {{\bf{A}},{\bf{B}},{\bf{D}},{\bf{E}},{\bf{F}},{\bf{H}}} \right) = \int\limits_{ - h/2}^{h/2} {\left[ {1,z,z^2 ,f\left( z \right),zf\left( z \right),f^2 \left( z \right)} \right]\,{\bar{\bf Q}}dz},
\\
&\left( {{\bf{X}},{\bf{Y}}^b ,{\bf{Y}}^s } \right) = \int\limits_{ - h/2}^{h/2} {\left[ {\phi '\left( z \right),z\phi '\left( z \right),f\left( z \right) \phi '\left( z \right)} \right]\,{\tilde{\bf Q}}dz},
\\
&Z_{33}  = \int\limits_{ - h/2}^{h/2} {\left[ {\phi '\left( z \right)} \right]^2 \,Q_{33} dz},
\\
&{\bf{D}}^s  = \int\limits_{ - h/2}^{h/2} {\left[ {f'\left( z \right) + \phi \left( z \right)} \right]^2 {\hat{\bf Q}}dz},
\\
&{\bf{\bar Q}} = \left[ {\begin{array}{*{20}c}
   {Q_{11} } & {Q_{12} } & 0  \\
   {Q_{21} } & {Q_{22} } & 0  \\
   0 & 0 & {Q_{66} }  \\
\end{array}} \right], \quad {\tilde{\bf Q}} = \left[ {\begin{array}{*{20}c}
   {Q_{13} }  \\
   {Q_{23} }  \\
   0  \\
\end{array}} \right], \quad {\hat{\bf Q}} = \left[ {\begin{array}{*{20}c}
   {Q_{44} } & 0  \\
   0 & {Q_{55} }  \\
\end{array}} \right],
\end{align}
\end{subequations}
and where the curvature tensors and matrices in the third and fourth terms in Eq. (\ref{eqn:weakform_bending}) representing the couple stress theory are
\begin{equation}
{\boldsymbol{\chi }}_b^c  = \left[ \begin{array}{l}
 {\boldsymbol{\chi }}_{b0}  \\ 
 {\boldsymbol{\chi }}_{b1}  \\ 
 {\boldsymbol{\chi }}_{b3}  \\ 
 \end{array} \right], \quad {\boldsymbol{\chi }}_s^c  = \left[ \begin{array}{l}
 {\boldsymbol{\chi }}_{s0}  \\ 
 {\boldsymbol{\chi }}_{s2}  \\ 
 {\boldsymbol{\chi }}_{s4}  \\ 
 \end{array} \right], \quad {\bf{D}}_c^b  = \left[ {\begin{array}{*{20}c}
   {{\bf{A}}^c } & {{\bf{B}}^c } & {{\bf{E}}^c }  \\
   {{\bf{B}}^c } & {{\bf{D}}^c } & {{\bf{F}}^c }  \\
   {{\bf{E}}^c } & {{\bf{F}}^c } & {{\bf{H}}^c }  \\
\end{array}} \right], \quad {\bf{D}}_c^s  = \left[ {\begin{array}{*{20}c}
   {{\bf{X}}^c } & {{\bf{Y}}^c } & {{\bf{T}}^c }  \\
   {{\bf{Y}}^c } & {{\bf{Z}}^c } & {{\bf{V}}^c }  \\
   {{\bf{T}}^c } & {{\bf{V}}^c } & {{\bf{W}}^c }  \\
\end{array}} \right],
\end{equation}
in which the material matrices could be defined as
\begin{subequations}
\begin{align}
\left( {{\bf{A}}^c ,{\bf{B}}^c ,{\bf{D}}^c ,{\bf{E}}^c ,{\bf{F}}^c ,{\bf{H}}^c } \right) &= \int\limits_{ - h/2}^{h/2} {\left( {1,f'\left( z \right),\left[ {f'\left( z \right)} \right]^2 ,\phi \left( z \right),f'\left( z \right)\phi \left( z \right),\left[ {\phi \left( z \right)} \right]^2 } \right)\,{\bar{\bf G}}dz}, 
\\
\left( {{\bf{X}}^c ,{\bf{Y}}^c ,{\bf{Z}}^c ,{\bf{T}}^c ,{\bf{V}}^c ,{\bf{W}}^c } \right)\, &= \int\limits_{ - h/2}^{h/2} {\left( {1,f''\left( z \right),\left[ {f''\left( z \right)} \right]^2 ,\phi '\left( z \right),f''\left( z \right)\phi '\left( z \right),\left[ {\phi '\left( z \right)} \right]^2 } \right)\,{\hat{\bf G}}dz},
\end{align}
\end{subequations}
where
\begin{equation}
{\bar{\bf G}} = 2G_e l^2 \left[ {\begin{array}{*{20}c}
   1 & 0 & 0  \\
   0 & 1 & 0  \\
   0 & 0 & 1  \\
\end{array}} \right], \quad {\hat{\bf G}} = 2G_e l^2 \left[ {\begin{array}{*{20}c}
   1 & 0  \\
   0 & 1  \\
\end{array}} \right].
\end{equation}

The weak form of the free vibration of the couple-stress-based microplates is briefly expressed as
\begin{equation}
\label{eqn:weakform_vibration}
\int\limits_\Omega  {\pmb{\delta {\bf{\varepsilon }}}_b ^T {\bf{D}}^b {\pmb{\bf{\varepsilon }}}_b d\Omega }  + \int\limits_\Omega  {\delta {\pmb{\bf{\varepsilon }}}_s ^T {\bf{D}}^s {\pmb{\bf{\varepsilon }}}_s d\Omega }  + \int\limits_\Omega  {\delta \left( {{\boldsymbol{\chi }}_b^c } \right)^T {\bf{D}}_c^b {\boldsymbol{\chi }}_b^c d\Omega }  + \int\limits_\Omega  {\delta \left( {{\boldsymbol{\chi }}_s^c } \right)^T {\bf{D}}_c^s {\boldsymbol{\chi }}_s^c d\Omega }  = \int\limits_\Omega  {\delta {\tilde{\bf u}}^T {\tilde{{\bf{m}}}} \ddot{\tilde{{\bf{u}}}} d\Omega },
\end{equation}
where $\tilde{{\bf{u}}} = \left[ {u_0 \quad - w_{b,x}  \quad w_{s,x}  \quad v_0  \quad  - w_{b,y}  \quad w_{s,y}  \quad w_b  \quad w_s  \quad 0} \right]^T$, and the mass matrix $\tilde{{\bf{m}}}$ is defined by
\begin{subequations}
\begin{align}
{\tilde{\bf m}} = \left[ {\begin{array}{*{20}c}
   {{\bf{I}}_0 } & 0 & 0  \\
   0 & {{\bf{I}}_0 } & 0  \\
   0 & 0 & {{\bf{I}}_1 }  \\
\end{array}} \right]
\quad \text{in which} \quad {\bf{I}}_0  = \left[ {\begin{array}{*{20}c}
   {I_1 } & {I_2 } & {I_4 }  \\
   {I_2 } & {I_3 } & {I_5 }  \\
   {I_4 } & {I_5 } & {I_6 }  \\
\end{array}} \right],\quad {\bf{I}}_1  = \left[ {\begin{array}{*{20}c}
   {I_1 } & {I_7 } & 0  \\
   {I_7 } & {I_8 } & 0  \\
   0 & 0 & 0  \\
\end{array}} \right],
\\
\left( {I_1 ,I_2 ,I_3 ,I_4 ,I_5 ,I_6 ,I_7 ,I_8 } \right) = \int\limits_{ - h/2}^{h/2} {\rho \left[ {1,z,z^2 ,f\left( z \right),zf\left( z \right),f^2 \left( z \right),\phi \left( z \right),\phi ^2 \left( z \right)} \right]dz}.
\end{align}
\end{subequations}

For buckling analysis, the weak form of the couple-stress-based microplates subjected to in-plane loading is of the form
\begin{equation}
\label{eqn:weakform_buckling}
\begin{split}
\int\limits_\Omega  {\delta {\pmb{\bf{\varepsilon }}}_b ^T {\bf{D}}^b {\pmb{\bf{\varepsilon }}}_b d\Omega }  + \int\limits_\Omega  {\delta {\pmb{\bf{\varepsilon }}}_s ^T {\bf{D}}^s {\pmb{\bf{\varepsilon }}}_s d\Omega }  + \int\limits_\Omega  {\delta \left( {{\boldsymbol{\chi }}_b^c } \right)^T {\bf{D}}_c^b {\boldsymbol{\chi }}_b^c d\Omega }  + \int\limits_\Omega  {\delta \left( {{\boldsymbol{\chi }}_s^c } \right)^T {\bf{D}}_c^s {\boldsymbol{\chi }}_s^c d\Omega } \\ + \int\limits_\Omega  {\nabla ^T \delta \left[ {w_b  + \phi \left( 0 \right)w_s } \right]{\bf{N}}_0 \nabla \left[ {w_b  + \phi \left( 0 \right)w_s } \right]d\Omega }  = 0
\end{split},
\end{equation}
where $\nabla ^T  = \left[ {\partial /\partial x\quad \partial /\partial y} \right]^T 
$ and ${\bf{N}}_0  = \left[ {\begin{array}{*{20}c}
   {N_x^0 } & {N_{xy}^0 }  \\
   {N_{xy}^0 } & {N_y^0 }  \\
\end{array}} \right]
$ are the transpose of gradient operator and matrix of pre-buckling loads, respectively.

\section{FG microplate formulation based on NURBS basis functions}
\label{sec:IGA}

In this section, briefing review on NURBS which are served as the basis functions of isogeometric analysis will be presented. It is followed by a novel NURBS-based formulation for couple-stress microplate bending, free vibration and buckling that are relied on refined plate theory and quasi-3D theory.

\subsection{B-splines and NURBS basis functions}
\label{sec:IGA:NURBS}
\begin{figure}
	\centering
	\begin{subfigure}{1.0\textwidth}
		\includegraphics[width=\linewidth]{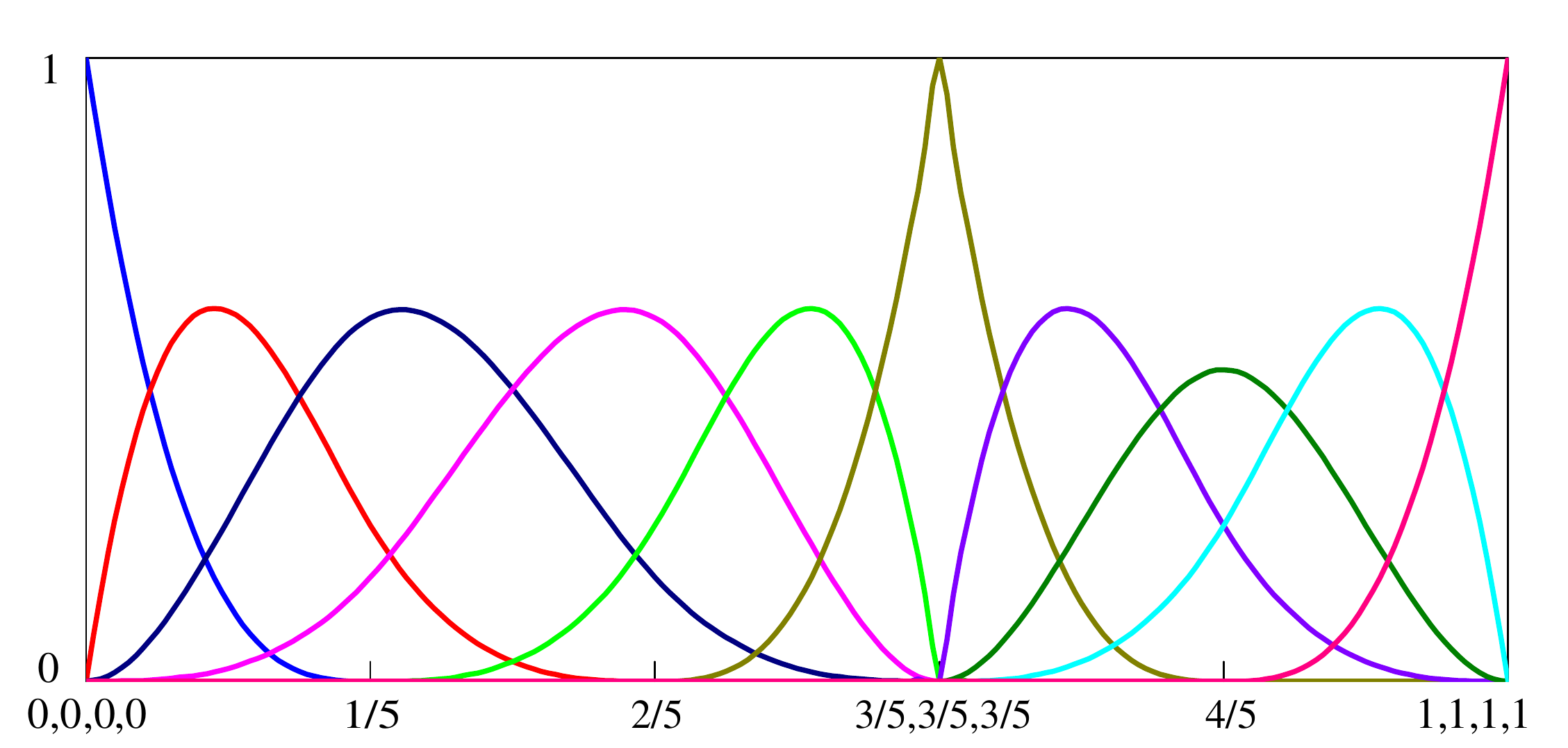}
		\caption{Cubic basis functions corresponding to ${\boldsymbol \Xi} = \{0,0,0,0,\dfrac{1}{5},\dfrac{2}{5},\dfrac{3}{5},\dfrac{3}{5},\dfrac{3}{5},\dfrac{4}{5},1,1,1,1\}$.\label{fig:bsplines_1D}}
	\end{subfigure}
	%\hspace*{\fill}
	\begin{subfigure}{1.0\textwidth}
		\includegraphics[width=\linewidth]{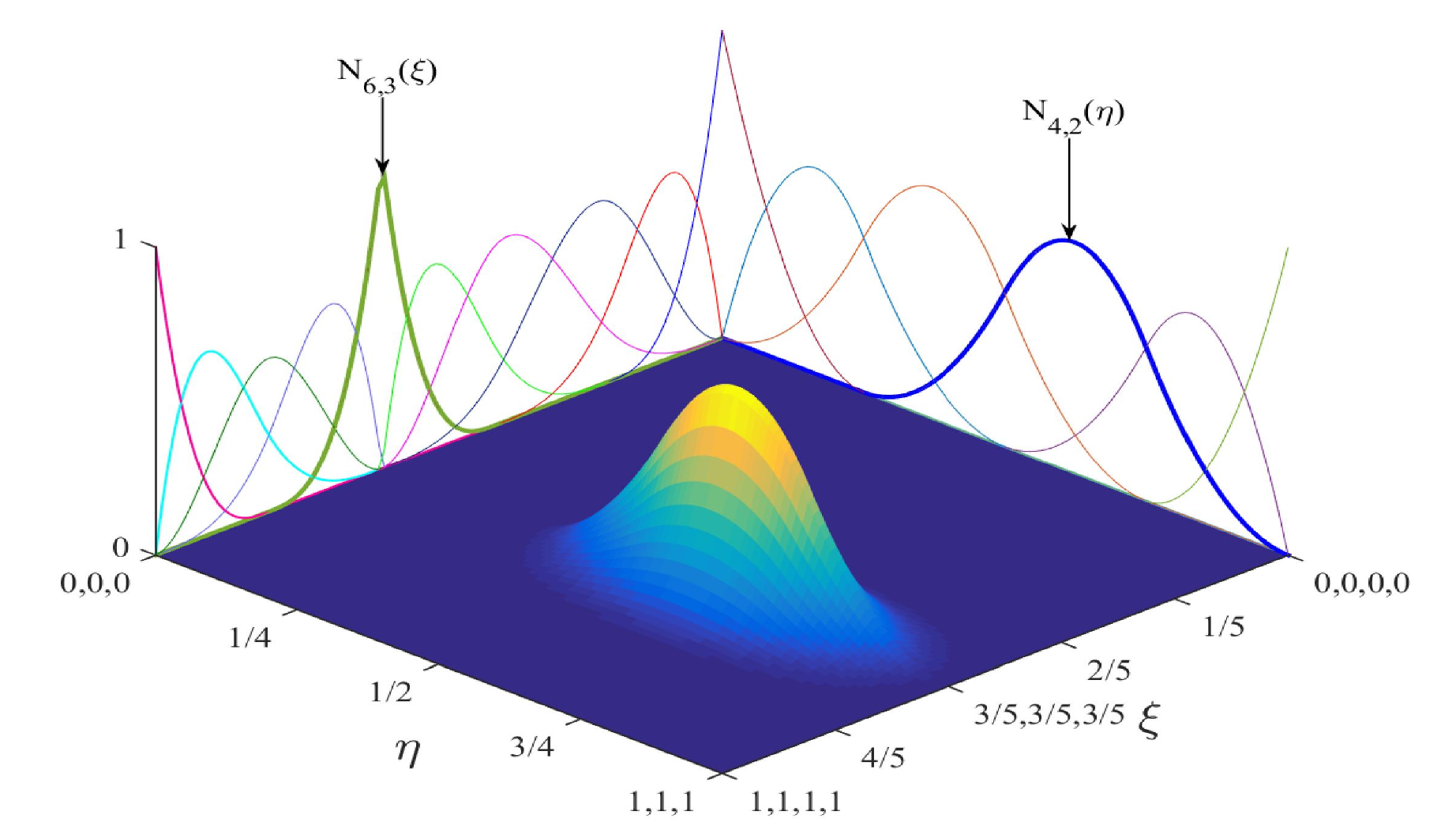}
		\caption{Bivariate B-spline basis functions. \label{fig:bsplines_2D}}
	\end{subfigure}
			\caption{One- and two-dimensional B-spline basis functions. \label{fig:bsplines}}
\end{figure}

The starting point of the NURBS basis functions is non-decreasing knot vector ${\bf{\Xi}}  = \left\{ {\xi _1 ,\xi _2 ,...,\xi _{n + p + 1} } \right\}$ where the $i^{th}$ knot $\xi _i  \in \mathbb{R}$, $n$ represents the number of basis functions, and $p$ denotes the polynomial order. The knot vectors could be either uniform if the knots are equally spaced in the parameter space or open if its first and last knot's values are repeated $p+1$ times. The knot spans which are bounded by knots defines element domain. The B-spline basis functions that are constructed by the Cox-de Boor recursion formula, starting with the zeroth order ($p=0$) are defined by \cite{Hughes20054135,cottrell_isogeometric_2009}
\begin{subequations}
\begin{align}
N_{i,0} \left( \xi  \right) &= \left\{ \begin{array}{l}
 1 \quad \text{if} \quad \xi_i  \le \xi  < \xi _{i + 1} , \\ 
 0 \quad \text{otherwise,} \\ 
 \end{array} \right.
\\
N_{i,p} \left( \xi  \right) &= \frac{{\xi  - \xi _i }}{{\xi _{i + p}  - \xi _i }}N_{i,p - 1} \left( \xi  \right) + \frac{{\xi _{i + p + 1}  - \xi }}{{\xi _{i + p + 1}  - \xi _{i + 1} }}N_{i + 1,p - 1} \left( \xi  \right), \text{for $p\geq 1$},
\end{align}
\end{subequations}
in which the fraction $0/0$ is defined as zero. While the basis functions are smooth, e.g. $C^\infty$ continuity, within this domain, they are $C^{p-m}$ continuity across the knots, where $m$ is the multiplicity of the knot. Therefore, for $p \geq 2$, the basis functions would be of C$^1$ continuity at single knot and knot span as well. Two-dimensional B-splines are obtained by introducing the second knot vector 	${{\bf{H}}}  = \left\{ {\eta _1 ,\eta _2 ,...,\eta _{m + q + 1} } \right\}$ and using tensor product of $\bf{\Xi}$ and ${{\bf{H}}}$ in the parametric dimensions yielding 
\begin{equation}
N_A \left( {\xi ,\eta } \right) = N_{i,p} \left( \xi  \right)M_{j,q} \left( \eta  \right).
\end{equation}
For the illustration purposes, Fig. \ref{fig:bsplines} depicts the one- and two-dimensional B-spline basis functions which are generated from the knot vector ${\bf{\Xi}} = \{0,0,0,0,1/5,2/5,3/5,3/5,3/5,4/5,1,1,1,1\}$ and its combination with the knot vector ${\bf{H}} = \{ 0,0,0,1/4,1/2,3/4,1,1,1\}$, respectively. The non-uniform rational B-splines (NURBS) basis functions are then further defined by providing additional weight $\zeta_A$ to each control point given by \cite{cottrell_isogeometric_2009}
\begin{equation}
R_A \left( {\xi ,\eta } \right) = \frac{{N_A \zeta _A }}{{\sum\limits_{\hat A}^{m \times n} {N_{\hat A} \left( {\xi ,\eta } \right)\zeta _{\hat A} } }}.
\end{equation}
It is noted that the B-spline basis function is served as a special case of the NURBS function. Indeed, if all the individual weights corresponding the control points are assigned an equal constant, the NURBS function yields B-spline function.

\subsection{A novel NURBS-based formulation of modified couple stress theory}
\label{sec:IGA:MCS}
By using the NURBS basis functions, the displacement variables $\bf{u}$ of a microplate could be approximately calculated as follows
\begin{equation}
\label{eqn:u_approx}
{\bf{u}}^h \left( {\xi ,\eta } \right) = \sum\limits_{A=1}^{n \times m} {R_A \left( {\xi ,\eta } \right)} {\bf{q}}_A,
\end{equation}
where $n \times m$ is the number of basis functions and ${\bf{q}}_A = [u_{0A} \quad v_{0A} \quad w_{bA} \quad w_{sA}]^T$ denotes the vector of degrees of freedom associated with the control point A. By substituting the variable approximations Eq. (\ref{eqn:u_approx}) into the strain-displacement relations Eq. (\ref{eqn:disp_strain_denotes}), the in-plane and shear strains could be obtained as follows
\begin{equation}
\label{eqn:strain_approx}
\left[ {{\boldsymbol{\varepsilon }}_0^T \quad {\boldsymbol{\kappa }}_b^T \quad {\boldsymbol{\kappa }}_s^T \quad {\boldsymbol{\varepsilon }}_s^T } \right]^T  = \sum\limits_{A = 1}^{n \times m} {\left[ {\left( {{\bf{B}}_A^m } \right)^T \quad \left( {{\bf{B}}_A^{b1} } \right)^T \quad \left( {{\bf{B}}_A^{b2} } \right)^T  \quad \left( {{\bf{B}}_A^s } \right)^T } \right]^T {\bf{q}}_A },
\end{equation}
where	
\begin{equation}
\label{eqn:Bmatrix_strain}
\begin{split}
{\bf{B}}_A^m  &= \left[ {\begin{array}{*{20}c}
   {R_{A,x} } & 0 & 0 & 0  \\
   0 & {R_{A,y} } & 0 & 0  \\
   {R_{A,y} } & {R_{A,x} } & 0 & 0  \\
\end{array}} \right], \quad {\bf{B}}_A^{b1}  =  - \left[ {\begin{array}{*{20}c}
   0 & 0 & {R_{A,xx} } & 0  \\
   0 & 0 & {R_{A,yy} } & 0  \\
   0 & 0 & {2R_{A,xy} } & 0  \\
\end{array}} \right],
\\
{\bf{B}}_A^{b2}  &= \left[ {\begin{array}{*{20}c}
   0 & 0 & 0 & {R_{A,xx} }  \\
   0 & 0 & 0 & {R_{A,yy} }  \\
   0 & 0 & 0 & {2R_{A,xy} }  \\
\end{array}} \right], \quad {\bf{B}}_A^s  = \left[ {\begin{array}{*{20}c}
   0 & 0 & 0 & {R_{A,x} }  \\
   0 & 0 & 0 & {R_{A,y} }  \\
\end{array}} \right],
\end{split}
\end{equation}
and the curvatures could be obtained by substituting Eq. (\ref{eqn:u_approx}) into  Eq. (\ref{eqn:curvatures}) :
\begin{equation}
\label{eqn:curvature_approx}
\begin{split}
&\left[ {{\boldsymbol{\chi }}_{b0}^T \quad {\boldsymbol{\chi }}_{b1}^T \quad {\boldsymbol{\chi }}_{b3}^T \quad {\boldsymbol{\chi }}_{s0}^T \quad {\boldsymbol{\chi }}_{s2}^T \quad {\boldsymbol{\chi }}_{s4}^T } \right]^T  
\\&= \sum\limits_{A = 1}^{n \times m} {\left[ {\left( {{\tilde{\bf B}}_A^{b0} } \right)^T \quad \left( {{\tilde{\bf B}}_A^{b1} } \right)^T \quad \left( {{\tilde{\bf B}}_A^{b3} } \right)^T \quad \left( {{\tilde{\bf B}}_A^{s0} } \right)^T \quad \left( {{\tilde{\bf B}}_A^{s2} } \right)^T \quad \left( {{\tilde{\bf B}}_A^{s4} } \right)^T } \right]^T {\bf{q}}_A },
\end{split}
\end{equation}
where
\begin{equation}
\label{eqn:Bmatrix_curvature}
\begin{split}
{\tilde{\bf B}}_{A}^{b0}  &= \frac{1}{2}\left[ {\begin{array}{*{20}c}
   0 & 0 & {2R_{A,xy} } & 0  \\
   0 & 0 & { - 2R_{A,xy} } & 0  \\
   0 & 0 & {\left( { - R_{A,xx}  + R_{A,yy} } \right)} & 0  \\
\end{array}} \right], \quad {\tilde{\bf B}}_{A}^{b1}  = \frac{1}{4}\left[ {\begin{array}{*{20}c}
   0 & 0 & 0 & { - 2R_{A,xy} }  \\
   0 & 0 & 0 & {2R_{A,xy} }  \\
   0 & 0 & 0 & {\left( {R_{A,xx}  - R_{A,yy} } \right)}  \\
\end{array}} \right],
\\
{\tilde{\bf B}}_{A}^{b3}  &= \frac{1}{4}\left[ {\begin{array}{*{20}c}
   0 & 0 & 0 & {2R_{A,xy} }  \\
   0 & 0 & 0 & { - 2R_{A,xy} }  \\
   0 & 0 & 0 & {\left( { - R_{A,xx}  + R_{A,yy} } \right)}  \\
\end{array}} \right], \quad  {\tilde{\bf B}}_{A}^{s0}  = \frac{1}{4}\left[ {\begin{array}{*{20}c}
   { - R_{A,xy} } & {R_{A,xx} } & 0 & 0  \\
   { - R_{A,yy} } & {R_{A,xy} } & 0 & 0  \\
\end{array}} \right],
\\
{\tilde{\bf B}}_{A}^{s2}  &= \frac{1}{4}\left[ {\begin{array}{*{20}c}
   0 & 0 & 0 & { - R_{A,y} }  \\
   0 & 0 & 0 & {R_{A,x} }  \\
\end{array}} \right], \quad {\tilde{\bf B}}_{A}^{s4}  = \frac{1}{4}\left[ {\begin{array}{*{20}c}
   0 & 0 & 0 & {R_{A,y} }  \\
   0 & 0 & 0 & { - R_{A,x} }  \\
\end{array}} \right].
\end{split}
\end{equation}

Substituting Eqs. (\ref{eqn:strain_approx}) and (\ref{eqn:curvature_approx}) into Eqs. (\ref{eqn:weakform_bending}), (\ref{eqn:weakform_vibration}), and (\ref{eqn:weakform_buckling}), the matrix form of the global equilibrium equations for static bending, free vibration, and buckling could be respectively presented by 
\begin{subequations}
\begin{align}
{\bf{Kq}} &= {\bf{F}},
\\
\left( {{\bf{K}} - {\bf{\omega }}^2 {\bf{M}}} \right){\bf{q}} &= {\bf{0}},
\\
\left( {{\bf{K}} - {\bf{\lambda }}_{cr} {\bf{K}}_g } \right){\bf{q}} &= {\bf{0}}, 
\end{align}
\end{subequations}
where the global stiffness matrix $\bf{K}$ is the summation of the stiffness matrices corresponding to the classical theory $ {\bf{K}}_s$ and the couple stress theory ${\bf{K}}_c$, i.e. ${\bf{K}} = {\bf{K}}_s+ {\bf{K}}_c$. These matrices are calculated as follow
\begin{subequations}
\label{eqn:global_stiffness_matrix}
\begin{align}
{\bf{K}}_s  &= \int\limits_\Omega  {\left( {\left\{ \begin{array}{l}
 {\bf{B}}^m  \\ 
 {\bf{B}}^{b1}  \\ 
 {\bf{B}}^{b2}  \\ 
 {\bf{B}}^z  \\ 
 \end{array} \right\}^T \left[ {\begin{array}{*{20}c}
   {\bf{A}} & {\bf{B}} & {\bf{E}} & {\bf{X}}  \\
   {\bf{B}} & {\bf{D}} & {\bf{F}} & {{\bf{Y}}^b }  \\
   {\bf{E}} & {\bf{F}} & {\bf{H}} & {{\bf{Y}}^s }  \\
   {\bf{X}} & {{\bf{Y}}^b } & {{\bf{Y}}^s } & {Z_{33} }  \\
\end{array}} \right]\left\{ \begin{array}{l}
 {\bf{B}}^m  \\ 
 {\bf{B}}^{b1}  \\ 
 {\bf{B}}^{b2}  \\ 
 {\bf{B}}^z  \\ 
 \end{array} \right\} + \left( {{\bf{B}}^s } \right)^T {\bf{D}}^s {\bf{B}}^s } \right)d\Omega },
\\
{\bf{K}}_c  &= \int\limits_\Omega  {\left( {\left\{ \begin{array}{l}
 {\tilde{\bf B}}^{b0}  \\ 
 {\tilde{\bf B}}^{b1}  \\ 
 {\tilde{\bf B}}^{b3}  \\ 
 \end{array} \right\}^T \left[ {\begin{array}{*{20}c}
   {{\bf{A}}^c } & {{\bf{B}}^c } & {{\bf{E}}^c }  \\
   {{\bf{B}}^c } & {{\bf{D}}^c } & {{\bf{F}}^c }  \\
   {{\bf{E}}^c } & {{\bf{F}}^c } & {{\bf{H}}^c }  \\
\end{array}} \right]\left\{ \begin{array}{l}
 {\tilde{\bf B}}^{b0}  \\ 
 {\tilde{\bf B}}^{b1}  \\ 
 {\tilde{\bf B}}^{b3}  \\ 
 \end{array} \right\} + \left\{ \begin{array}{l}
 {\tilde{\bf B}}^{s0}  \\ 
 {\tilde{\bf B}}^{s2}  \\ 
 {\tilde{\bf B}}^{s4}  \\ 
 \end{array} \right\}^T \left[ {\begin{array}{*{20}c}
   {{\bf{X}}^c } & {{\bf{Y}}^c } & {{\bf{T}}^c }  \\
   {{\bf{Y}}^c } & {{\bf{Z}}^c } & {{\bf{V}}^c }  \\
   {{\bf{T}}^c } & {{\bf{V}}^c } & {{\bf{W}}^c }  \\
\end{array}} \right]\left\{ \begin{array}{l}
 {\tilde{\bf B}}^{s0}  \\ 
 {\tilde{\bf B}}^{s2}  \\ 
 {\tilde{\bf B}}^{s4}  \\ 
 \end{array} \right\}} \right)d\Omega },
\end{align}
\end{subequations}
in which ${\bf{B}}_A^z  = \left[ {\begin{array}{*{20}c}
   0 & 0 & 0 & {R_A }  \\
\end{array}} \right]$. The load vector $\bf{F}$ is given by 
\begin{equation}
{\bf{F}} = \int\limits_\Omega  {q_0 {\bf{R}}d\Omega }, 
\end{equation}
where ${\bf{R}} = \left[ {\begin{array}{*{20}c}
   0 & 0 & {R_A } & {\phi \left( {\frac{h}{2}} \right)R_A }  \\
\end{array}} \right]^T$, the global mass matrix is computed by
\begin{equation}
{\bf{M}} = \int\limits_\Omega  {{\tilde{\bf R}}^T {\tilde{\bf m}} {\tilde{\bf R}}d\Omega }, 
\end{equation}
in which
\begin{equation}
\begin{split}
\tilde{\bf{ R}} &= \left\{ \begin{array}{l}
 {\bf{R}}_1  \\ 
 {\bf{R}}_2  \\ 
 {\bf{R}}_3  \\ 
 \end{array} \right\}, \quad 
{\bf{R}}_1  = \left[ {\begin{array}{*{20}c}
   {R_A } & 0 & 0 & 0  \\
   0 & 0 & { - R_{A,x} } & 0  \\
   0 & 0 & 0 & {R_{A,x} }  \\
\end{array}} \right],
\\
{\bf{R}}_2  &= \left[ {\begin{array}{*{20}c}
   0 & {R_A } & 0 & 0  \\
   0 & 0 & { - R_{A,y} } & 0  \\
   0 & 0 & 0 & {R_{A,y} }  \\
\end{array}} \right], \quad
{\bf{R}}_3  = \left[ {\begin{array}{*{20}c}
   0 & 0 & {R_A } & 0  \\
   0 & 0 & 0 & {R_A }  \\
   0 & 0 & 0 & 0  \\
\end{array}} \right],
\end{split}
\end{equation}
the geometric stiffness matrix is given as
\begin{equation}
{\bf{K}}_g  = \int\limits_\Omega  {\left( {{\bf{B}}^g } \right)} ^T {\bf{N}}_0 {\bf{B}}^g d\Omega ,	
\end{equation}
where
\begin{equation}
{\bf{B}}^g  = \left[ {\begin{array}{*{20}c}
   0 & 0 & {R_{A,x} } & {\phi \left( 0 \right)R_{A,x} }  \\
   0 & 0 & {R_{A,y} } & {\phi \left( 0 \right)R_{A,y} }  \\
\end{array}} \right],
\end{equation}
and $\omega$ and $\lambda_{cr}$ represent the natural frequency and the critical buckling value, respectively.

As can be observed from Eq. (\ref{eqn:global_stiffness_matrix}), by introducing the distribution function $f(z)$, the four-variable refined plate theory and quasi-3D theory do not require any shear correction factor, which is usually needed if the first-order shear deformation theory is applied, to describe the transverse shear stresses satisfying traction-free conditions. In addition, the expressions of $\bf{B}$ and ${\tilde{\bf{B}}}$ matrices in Eqs. (\ref{eqn:Bmatrix_strain}) and (\ref{eqn:Bmatrix_curvature}), respectively, show the employment of the second-order derivatives of the approximation functions $R_A$. Consequently, C$^1$-continuity approximations are required. This requirement may cause severe issues in finite element analysis which could only be solved by increasing the degrees of freedom resulting in higher number of variables and computational cost. However, within the platform of isogeometric analysis in which NURBS basis functions are employed, the C$^1$-continuity requirement could be naturally satisfied for $p \geq 2$ since the basis functions are C$^{p-1}$ continuous across the knot spans, i.e. elements.

\section{Numerical examples and discussion}
\label{sec:Exp}
In this section, the convergence and verification studies are firstly conducted to demonstrate the validity and accuracy of the novel approaches presented in Section \ref{sec:Theory} and \ref{sec:IGA}. In order to illustrate the efficiency  of the IGA approach in dealing with modified couple stress theory, this section is then continued by the computational analysis of FG rectangular and circular microplates with various types of boundary condition for static bending, free vibration and buckling problems. In these investigations, the FG microplates made of mixtures of metal and ceramic whose material properties presented in Table \ref{tab:exp:material} are used. Throughout the numerical examples, unless otherwise specified, the material length scale $l$ is chosen as $17.6 \times 10^{-6}$m which was suggested by Lam et al. \cite{Lam20031477}. There are two types of boundary condition considered as follows

Simply supported (S)

\quad $v_0 = w_b = w_s = 0$ \quad at $x = 0,a$

\quad $u_0 = w_b = w_s = 0$ \quad at $y = 0,b$

Clamped (C)

\quad $u_0 = v_0 = w_b = w_s = 0$ and $w_{b,x} = w_{b,y} = w_{b,z} = w_{s,x} = w_{s,y} = w_{s,z} = 0$

It should be noted that, within the IGA platform, while the homogeneous boundary conditions corresponded to displacement itself, e.g. $u_0, v_0, w_b, w_s$, are easily treated in a similar way to the traditional finite element method, those require first derivative of displacement, e.g. $w_{b,x}, w_{b,y}, w_{b,z}, w_{s,x}, w_{s,y}, w_{s,z}$, could be done by  assigning zero values for all displacements of control points which are directly related to clamped edges and their adjacent points \cite{Auricchio2007160, Nguyen2015191}.

% Table generated by Excel2LaTeX from sheet 'Q3D_Bending'
\begin{table}[htbp]
  \centering
  \caption{Material properties of FG plates}
    \begin{tabular}{lllll}
    \hline
    Property & \multicolumn{4}{l}{Material} \\
    \cline{2-5}
          & Al    & Al$_2$O$_3$ & ZrO$_2$-1	& ZrO$_2$-2 \\
    \hline
    E (GPa) & 70    & 380   & 200 & 151\\
    $\nu$     & 0.3   & 0.3   & 0.3 & 0.3\\
    $\rho$ (kg/m$^3$) & 2707  & 3800  & 3000 & 3000\\
    \hline
    \end{tabular}
  \label{tab:exp:material}%
\end{table}

\subsection{Convergence and verification studies}
\label{sec:Exp:Verification}

\begin{figure}
	\centering
	\includegraphics[width=1.0\textwidth]{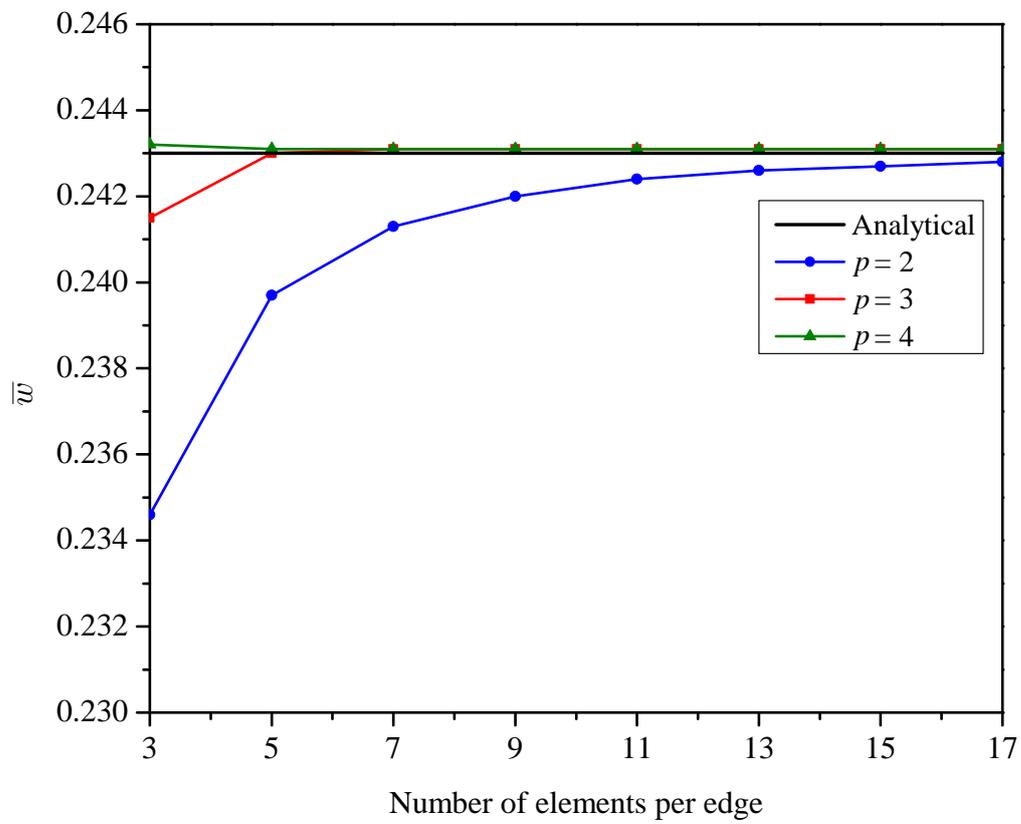}
	\caption{Convergence study of non-dimensional central deflection of homogeneous square microplates. \label{fig:exp:convergence_study}}
\end{figure}

\begin{figure}
	\centering
	\begin{subfigure}{0.8\textwidth}
		\includegraphics[width=\linewidth]{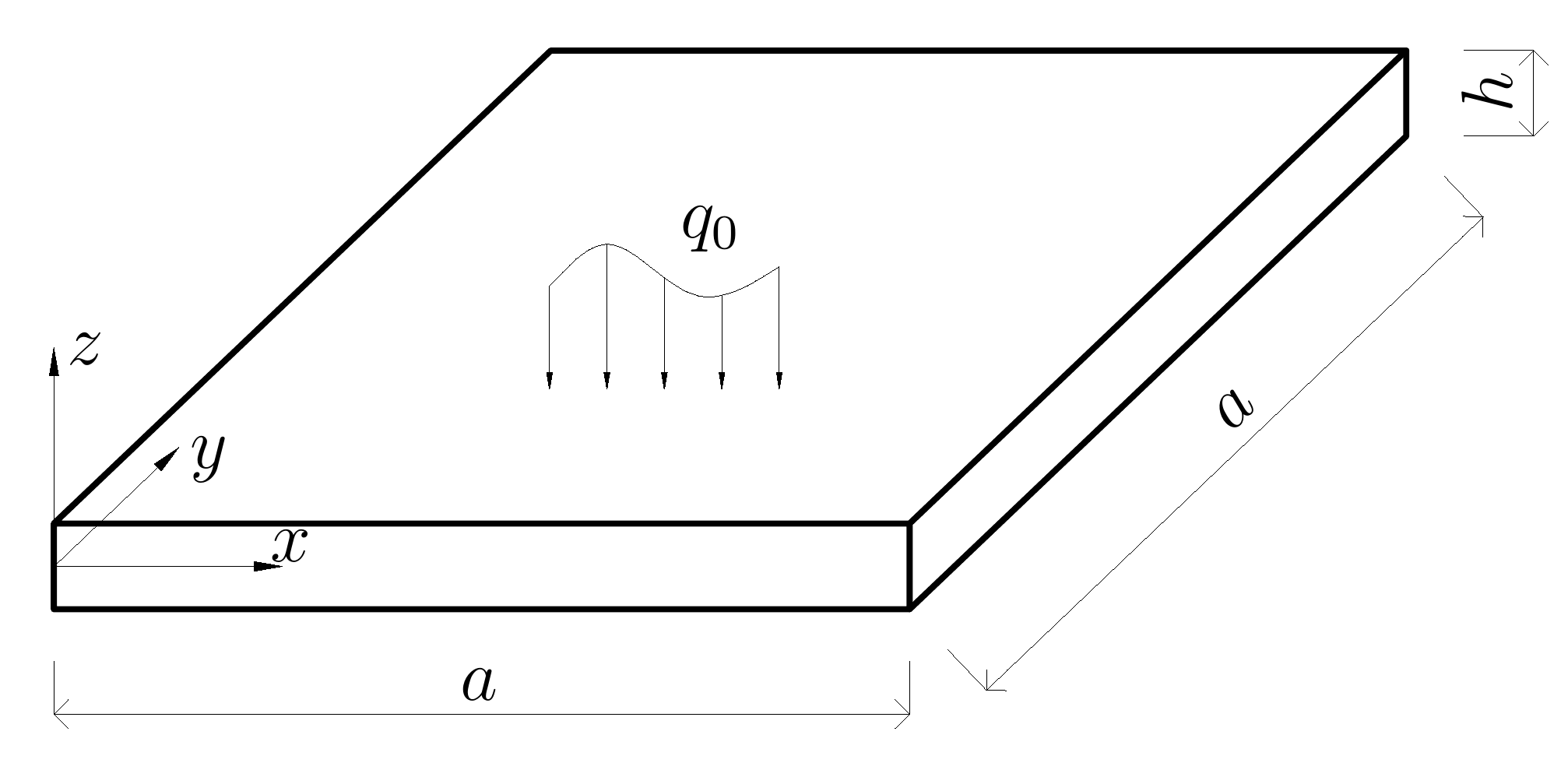}
		\caption{Geometric configuration.}
	\end{subfigure}
	%\hspace*{\fill}
	\begin{subfigure}{0.6\textwidth}
		\includegraphics[width=\linewidth]{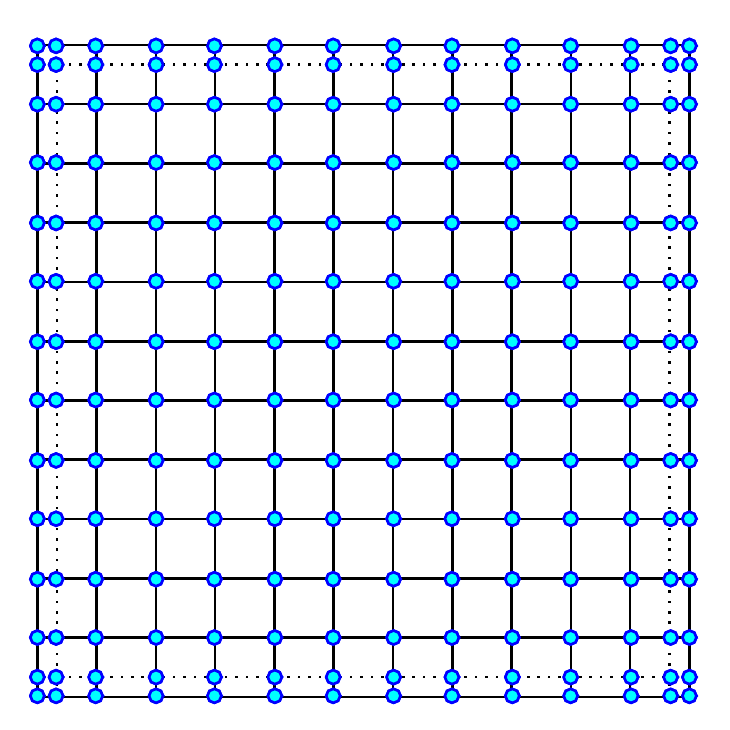}
		\caption{Control point net and 11$\times$11 cubic elements.}
	\end{subfigure}
	\caption{Geometry and element mesh of a square microplate. \label{fig:exp:sq:geometry_mesh}}
\end{figure}

In order to evaluate the convergence and reliability of the novel shape functions and IGA approach proposed in Section \ref{sec:Theory} and \ref{sec:IGA}, the MCST-based size-dependent analysis of homogeneous fully simply-supported (SSSS) square microplate which is shown in Fig. \ref{fig:exp:sq:geometry_mesh} is conducted using RPT model. Moderately thick plates $(a/h = 20)$ with four different values of material length scale ratio $(l/h = 0, 0.2, 0.6, 1)$ are investigated. For each case, eight different finite element meshes are analysed to study the convergence rate of the proposed IGA approach. As can be seen from Table \ref{tab:exp:convergence}, while fast convergence of the analysis for polynomial order $p=3$ and $p=4$ is archived, solutions using $p=2$ experience relatively slower convergence rate toward analytical solutions reported by Thai and Kim \cite{Thai20131636}. Fig. \ref{fig:exp:convergence_study} presents the convergence study of non-dimensional central deflection of homogeneous square microplates with respect to the analytical solutions \cite{Thai20131636} for the case of $l/h = 0.2$. Based on the convergence study, the cubic ($p=3$) NURBS element mesh of $11\times11$  is relatively sufficient for all analysis cases. Therefore, this mesh whose geometry shown in Fig. \ref{fig:exp:sq:geometry_mesh} will be used throughout the next examples unless otherwise specified.

% Table generated by Excel2LaTeX from sheet 'Q3D_Bending'
\begin{table}[htbp]
  \centering
  \caption{Convergence of non-dimensional central deflection $\bar{w}=\dfrac{10Eh^3}{q_0L^4}w\left(a/2,a/2,0\right)$  of SSSS homogeneous square plate subjected to sinusoidally distributed load, $a/h=20$}
    \begin{tabular}{lllllllllll}
    \hline
    $l/h$ & $p$ & \multicolumn{8}{l}{Element Mesh} & Analytical \cite{Thai20131636}\\
\cline{3-10}          &       & 3$\times$3   & 5$\times$5   & 7$\times$7   & 9$\times$9   & 11$\times$11 & 13$\times$13 & 15$\times$15 & 17$\times$17 &  \\
    \hline
    0     &       &       &       &       &       &       &       &       &       & 0.2842 \\
          & 2   & 0.2734 & 0.2799 & 0.2819 & 0.2828 & 0.2833 & 0.2835 & 0.2837 & 0.2838 &  \\
          & 3   & 0.2823 & 0.2841 & 0.2842 & 0.2842 & 0.2842 & 0.2842 & 0.2842 & 0.2842 &  \\
          & 4   & 0.2843 & 0.2842 & 0.2842 & 0.2842 & 0.2842 & 0.2842 & 0.2842 & 0.2842 &  \\
    0.2   &       &       &       &       &       &       &       &       &       & 0.2430 \\
          & 2   & 0.2346 & 0.2397 & 0.2413 & 0.2420 & 0.2424 & 0.2426 & 0.2427 & 0.2428 &  \\
          & 3   & 0.2415 & 0.2430 & 0.2431 & 0.2431 & 0.2431 & 0.2431 & 0.2431 & 0.2431 &  \\
          & 4   & 0.2432 & 0.2431 & 0.2431 & 0.2431 & 0.2431 & 0.2431 & 0.2431 & 0.2431 &  \\
    0.6   &       &       &       &       &       &       &       &       &       & 0.1124 \\
          & 2   & 0.1098 & 0.1115 & 0.1121 & 0.1123 & 0.1125 & 0.1125 & 0.1126 & 0.1126 &  \\
          & 3   & 0.1120 & 0.1127 & 0.1127 & 0.1127 & 0.1127 & 0.1127 & 0.1127 & 0.1127 &  \\
          & 4   & 0.1127 & 0.1127 & 0.1127 & 0.1127 & 0.1127 & 0.1127 & 0.1127 & 0.1127 &  \\
    1     &       &       &       &       &       &       &       &       &       & 0.0542 \\
          & 2   & 0.0532 & 0.0539 & 0.0541 & 0.0542 & 0.0543 & 0.0543 & 0.0543 & 0.0543 &  \\
          & 3   & 0.0541 & 0.0544 & 0.0544 & 0.0544 & 0.0544 & 0.0544 & 0.0544 & 0.0544 &  \\
          & 4   & 0.0544 & 0.0544 & 0.0544 & 0.0544 & 0.0544 & 0.1127 & 0.1127 & 0.1127 &  \\
    \hline
    \end{tabular}
  \label{tab:exp:convergence}
\end{table}

Further investigation on the accuracy of the proposed method is conducted using FG plates made of alumina and aluminum (Al/Al$_2$O$_3$). In this case,  without considering couple stress effects, proposed RPT $(\varepsilon_z = 0)$ and quasi-3D $(\varepsilon_z \neq 0)$ theories  are applied in analysis of SSSS square plates using the rule of mixtures scheme. The plates are subjected to uniformly and sinusoidally distributed loads which are defined as $q_0$ and $q_0sin\left( \dfrac{\pi x}{a} \right)  sin\left( \dfrac{\pi y}{a}\right)  $, respectively. As can be observed in Table \ref{tab:exp:verification}, the present results are in good agreement with those available in published works using various 2-D and quasi-3D theories. Apparently, these investigations have confirmed the validity and reliability of the present study.

% Table generated by Excel2LaTeX from sheet 'Q3D_Bending'
\begin{table}[htbp]
  \centering
  \caption{Comparison of non-dimensional deflection $\bar{w}=\dfrac{10E_ch^3}{q_0L^4}w\left(a/2,b/2,0\right)$  of SSSS Al/Al$_2$O$_3$ square plates (rule of mixtures scheme)}
    \begin{tabular}{lllllllll}
    \hline
    $n$     & Theory & $\varepsilon_z = 0$ &       &       &       & $\varepsilon_z \neq 0$ &       &  \\
   \cline{3-5} \cline{7-9}
          &       & $a/h=4$ & $a/h=10$    & $a/h=100$   &       & $a/h=4$ & $a/h=10$    & $a/h=100$ \\
          \hline
    \multicolumn{9}{l}{Sinusoidally distributed load} \\
    1     & Zenkour \cite{Zenkour20139041} & -     & -     & -     &       & 0.6828 & 0.5592 & 0.5624 \\
          & Neves et al. \cite{Neves2012711} & -     & -     & -     &       & 0.6997 & 0.5845 & 0.5624 \\
          & Mantari and Soares \cite{mantari_quasi-3d_2014} & -     & -     & -     &       & 0.693 & 0.569 & 0.545 \\
          & Carrera et al. \cite{Carrera2011123} & 0.7289 & 0.5890 & 0.5625 &       & 0.7171 & 0.5875 & 0.5625 \\
          & Akavci and Tanrikulu \cite{Akavci2015203} & 0.7282 & 0.5889 & 0.5625 &       & 0.6908 & 0.5691 & 0.5457 \\
          & Present$^\dagger$ & 0.7284 & 0.5889 & 0.5625 &       & 0.6935 & 0.5691 & 0.5460 \\
    4     & Zenkour \cite{Zenkour20139041} & -     & -     & -     &       & 1.1001 & 0.8404 & 0.7933 \\
          & Neves et al. \cite{Neves2012711} & -     & -     & -     &       & 1.1178 & 0.8750 & 0.8286 \\
          & Mantari and Soares \cite{mantari_quasi-3d_2014} & -     & -     & -     &       & 1.085 & 0.838 & 0.793 \\
          & Carrera et al. \cite{Carrera2011123} & 1.1673 & 0.8828 & 0.8286 &       & 1.1585 & 0.8821 & 0.8286 \\
          & Akavci and Tanrikulu \cite{Akavci2015203} & 1.1613 & 0.8818 & 0.8287 &       & 1.0983 & 0.8417 & 0.7925 \\
          & Present$^\dagger$ & 1.1590 & 0.8813 & 0.8287 &       & 1.0868 & 0.8392 & 0.7933 \\
    10    & Zenkour \cite{Zenkour20139041} & -     & -     & -     &       & 1.3391 & 0.9806 & 0.9140 \\
          & Neves et al. \cite{Neves2012711} & -     & -     & -     &       & 1.3490 & 0.8750 & 0.8286 \\
          & Mantari and Soares \cite{mantari_quasi-3d_2014} & -     & -     & -     &       & 1.308 & 0.972 & 0.911 \\
          & Carrera et al. \cite{Carrera2011123} & 1.3925 & 1.0090 & 0.9361 &       & 1.3745 & 1.0072 & 0.9361 \\
          & Akavci and Tanrikulu \cite{Akavci2015203} & 1.3917 & 1.0089 & 0.9362 &       & 1.3352 & 0.9818 & 0.9141 \\
          & Present$^\dagger$& 1.3902 & 1.0086 & 0.9362 &       & 1.3116 & 0.9748 & 0.9132 \\
          \\
              \multicolumn{9}{l}{Uniformly distributed load} \\
    1     & Zenkour \cite{Zenkour200667} & -     & 0.9287 & -     &       & -     & -     & - \\
          & Akavci and Tanrikulu \cite{Akavci2015203} & -     & 0.9288 & -     &       & -     & 0.8977 & - \\
          & Present$^\dagger$& 1.1319 & 0.9288 & 0.8904 &       & 1.0788 & 0.8978 & 0.8642 \\
    4     & Zenkour \cite{Zenkour200667} & -     & 1.3890 & -     &       & -     & -     & - \\
          & Akavci and Tanrikulu \cite{Akavci2015203} & -     & 1.3888 & -     &       & -     & 1.3259 & - \\
          & Present$^\dagger$& 1.7928 & 1.3882 & 1.3116 &       & 1.6827 & 1.3223 & 1.2556 \\
    10    & Zenkour \cite{Zenkour200667} & -     & 1.5876 & -     &       & -     & -     & - \\
          & Akavci and Tanrikulu \cite{Akavci2015203} & -     & 1.5875 & -     &       & -     & 1.5453 & - \\
          & Present$^\dagger$& 2.1432 & 1.5870 & 1.4818 &       & 2.0244 & 1.5347 & 1.4454 \\
    \hline
    \multicolumn{9}{l}{$^\dagger$Proposed RPT and quasi-3D models are used for the case of $\varepsilon_z = 0$ and $\varepsilon_z \neq 0$, respectively.}
    \end{tabular}
  \label{tab:exp:verification}
\end{table}

\subsection{Static bending analysis}
\label{sec:Exp:Static}

In this section, the static bending analysis of FG microplates based on modified couple stress theory will be investigated. The SSSS square microplates whose materials follow the rule of mixtures. The aspect ratio $a/h$, material length scale ratio $l/h$, and material index $n$ are taken into account. Table \ref{tab:exp:bending_SSSS} presents the comparison of non-dimensional central deflection of SSSS square plate with those of Thai and Kim \cite{Thai20131636}. While the results generated from the proposed RPT theory are in very good agreement, quasi-3D theory yields slightly different responses in term of displacement. This is due to the consideration of the thickness stretching effect in the quasi-3D theory.

\begin{figure}
	\centering
	\includegraphics[width=1.0\textwidth]{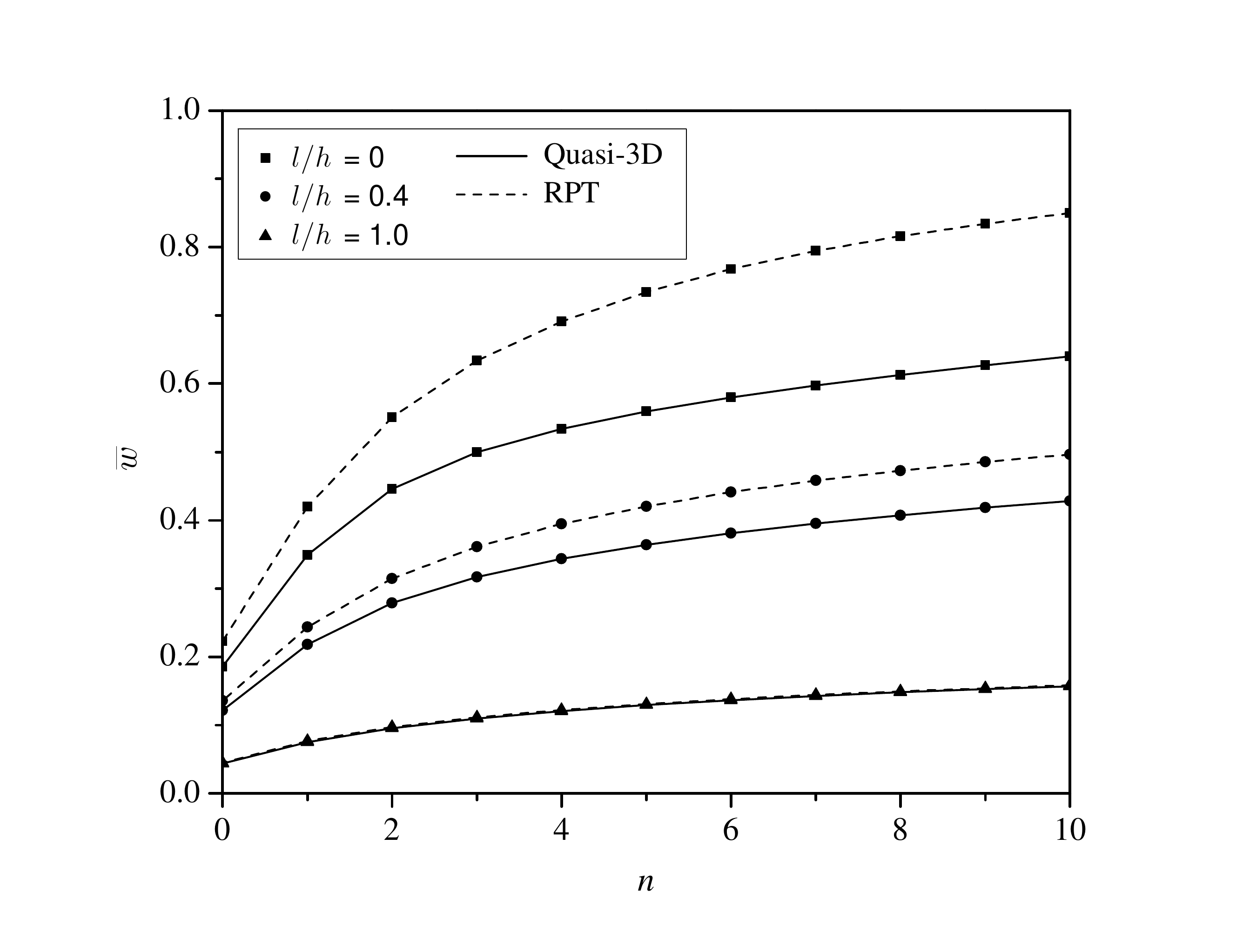}
	\caption{Effects of material index $n$ and material length scale ratio $l/h$ on the central deflection of CCCC Al/Al$_2$O$_3$ square microplates subjected to uniformly distributed load, $a/h=5$ (rule of mixtures scheme). \label{fig:exp:bending_effects}}
\end{figure}

\begin{figure}
	\centering
	\begin{subfigure}{0.49\textwidth}
		\includegraphics[width=\linewidth]{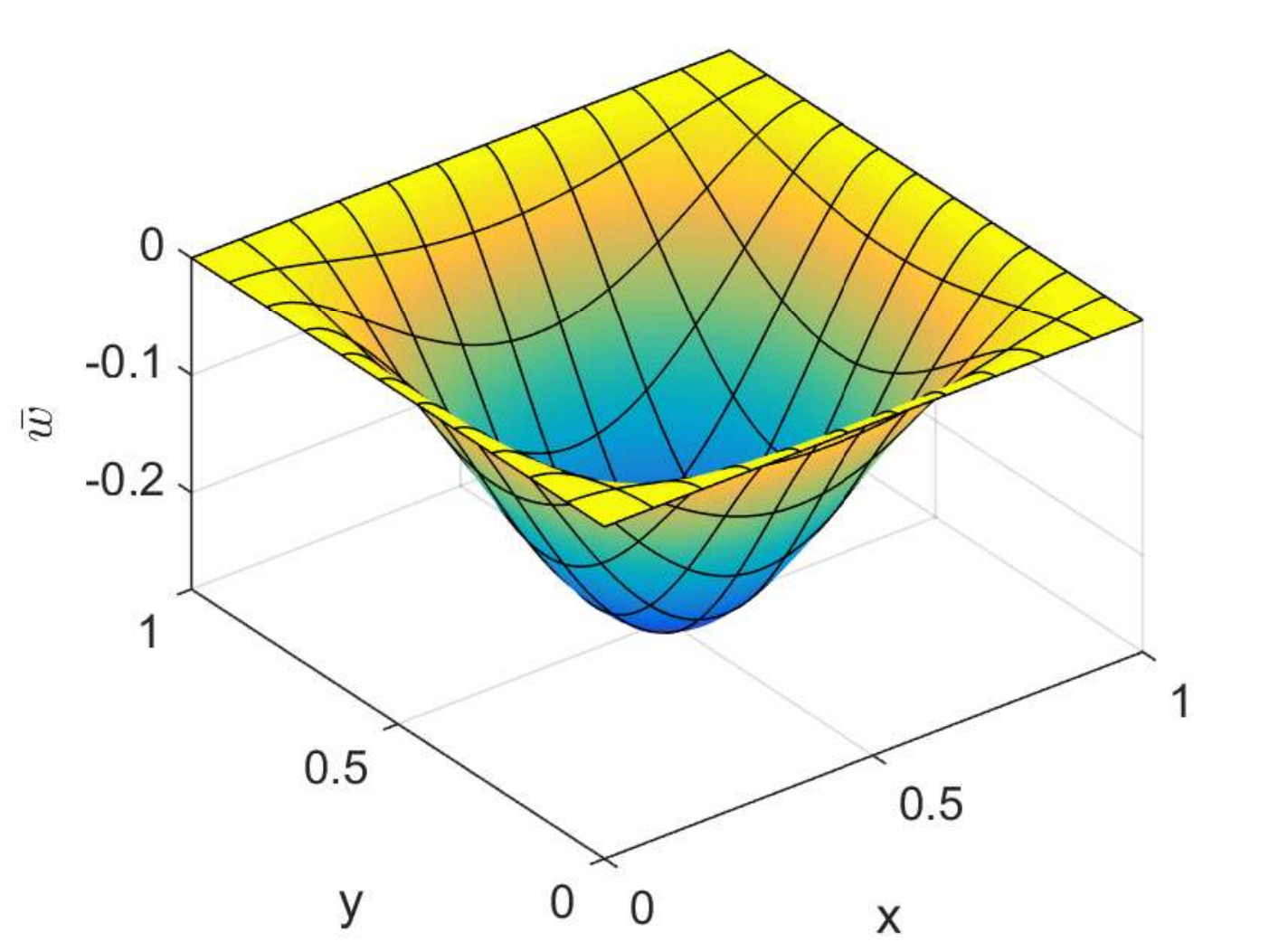}
		\caption{CCCC.}
	\end{subfigure}
	%\hspace*{\fill}
	\begin{subfigure}{0.49\textwidth}
		\includegraphics[width=\linewidth]{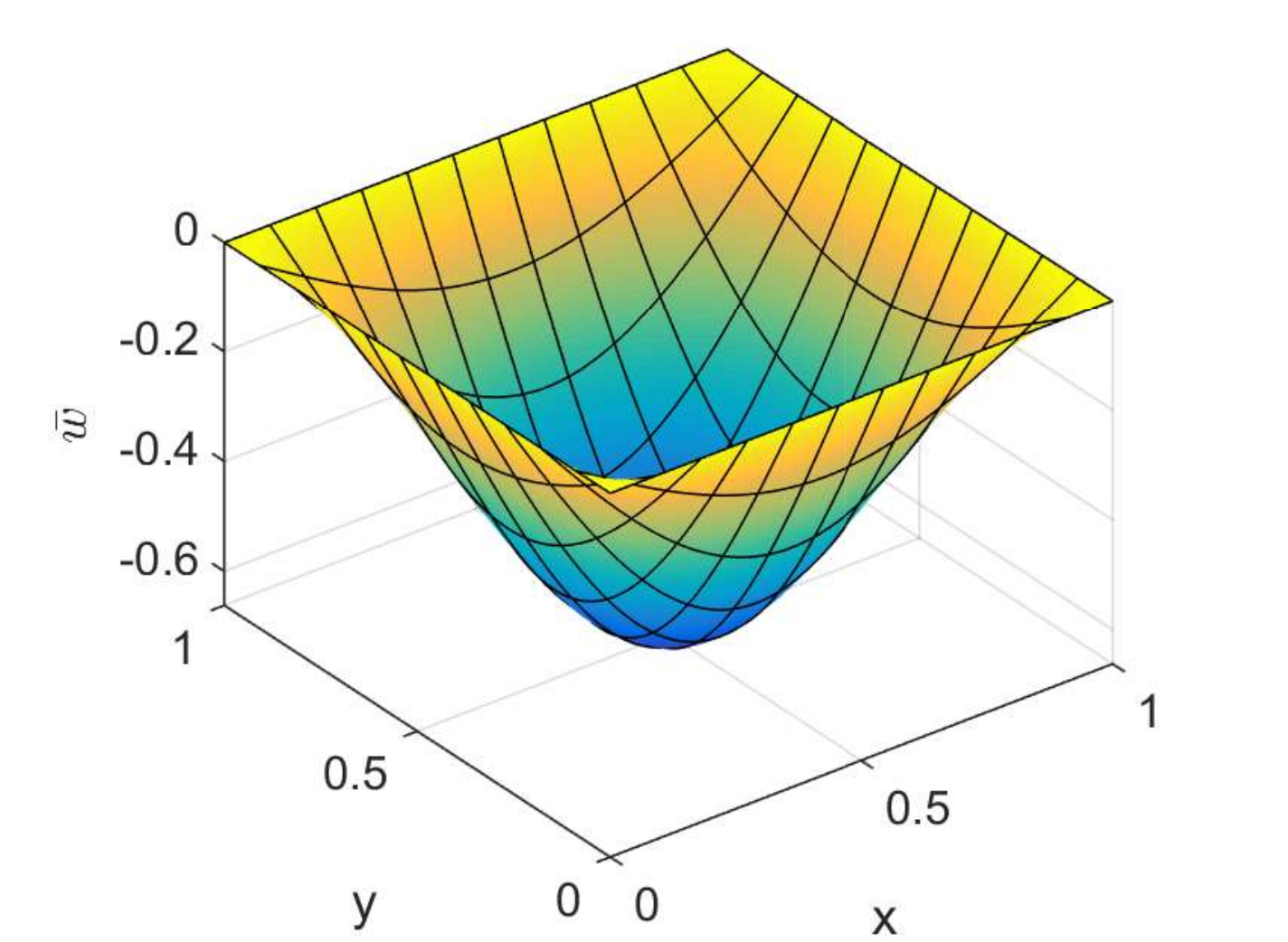}
		\caption{SSSS.}
	\end{subfigure}
	\begin{subfigure}{0.49\textwidth}
		\includegraphics[width=\linewidth]{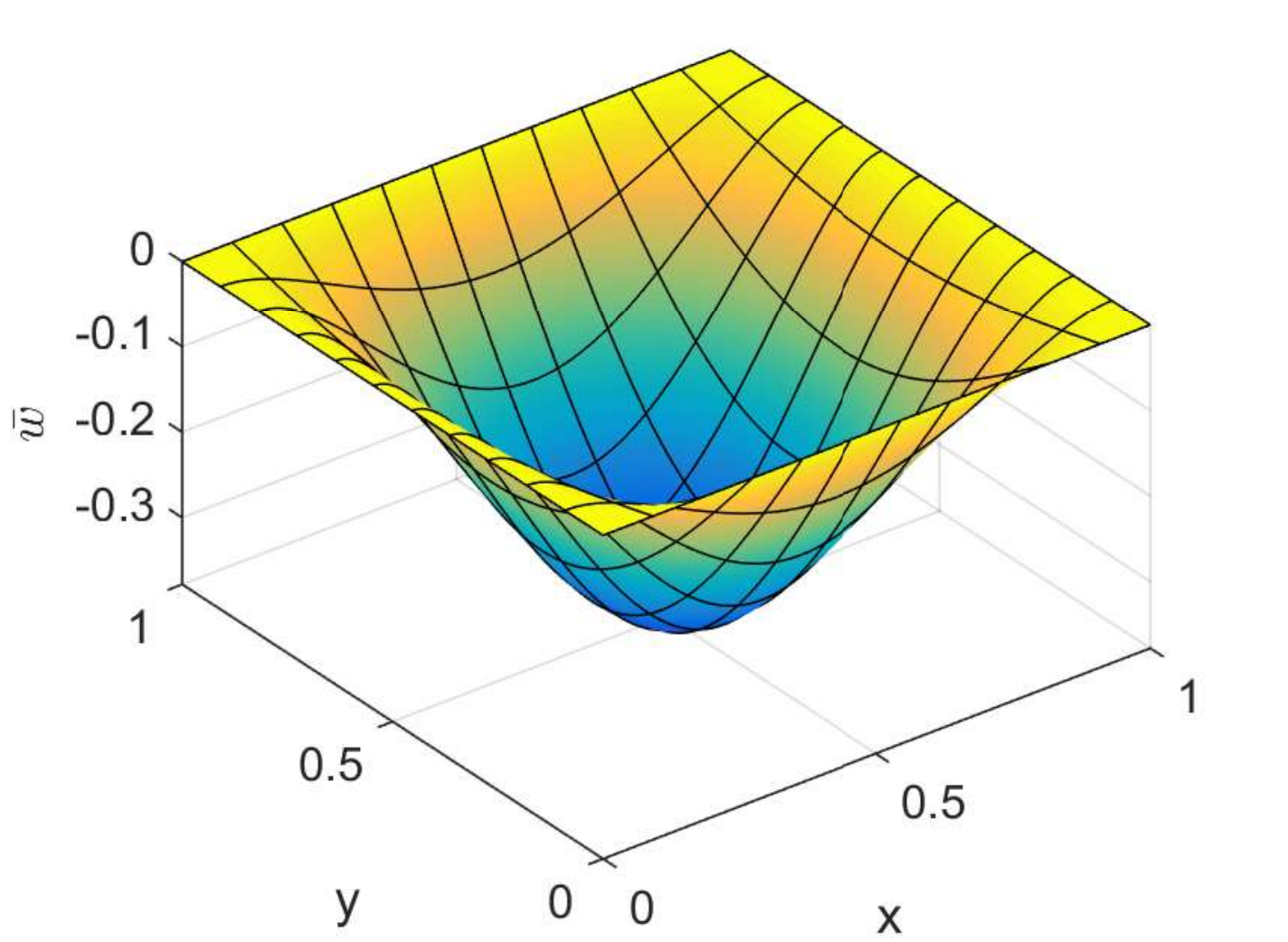}
		\caption{SCSC.}
	\end{subfigure}
	\begin{subfigure}{0.49\textwidth}
		\includegraphics[width=\linewidth]{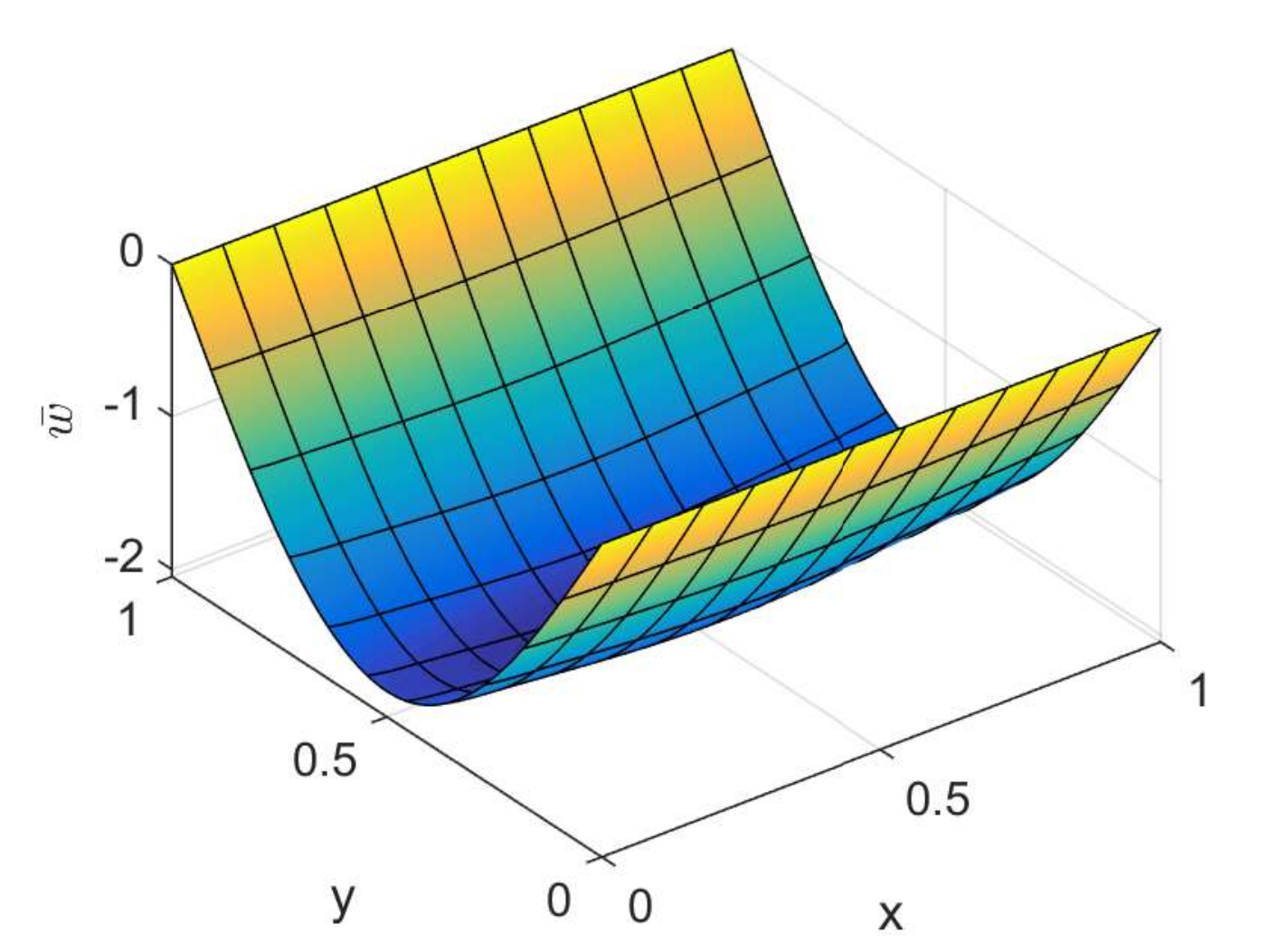}
		\caption{SFSF.}
	\end{subfigure}
	\caption{Deformed configuration of Al/Al$_2$O$_3$ square microplates. \label{fig:exp:bending:deformed}}
\end{figure}

The bending responses of CCCC square Al/Al$_2$O$_3$ microplates under sinusoidally and uniformly distributed loads are further studied and presented in Table \ref{tab:exp:bending_CCCC}. It is noted that since there is no study on the static behaviours of CCCC microplates using MCST are reported in the literature, the results are compared with those generated from Reddy's HSDT model \cite{reddy_simple_1984} using the proposed IGA approach. As can be seen, the results based on Reddy's model yields excellent agreement with proposed RPT-based solutions. The effects of material index $n$ and material length scale $l$ on the central displacement of CCCC square Al/Al$_2$O$_3$ plate are depicted in Fig. \ref{fig:exp:bending_effects} in which data are generated by the proposed RPT and quasi-3D theories for three different ratios $l/h$ of 0, 0.4 and 1.0. As can be observed, the increase of material index $n$ leads to a rise of plates' central deflection due to the decrease in plates' stiffness. On the contrary, the growth of material length scale ratio $l/h$ is followed by a decline of displacement due to the increase in magnitude of stiffness matrix. In other word, for specific material length scale $l$, the thinner the microplates are, the higher plates' stiffness would be. It can be further observed that the discrepancy in term of central deflection by the proposed RPT and quasi-3D is significantly decreased as $l/h$ increase and vanishes when $l/h=1.0$. Fig. \ref{fig:exp:bending:deformed} depicts the deformed configurations of the Al/Al$_2$O$_3$ square microplates with various boundary conditions subjected to sinusoidally distributed load in which $a/h, l/h$ and $n$ are equal to $5, 0.4$ and $10$, respectively. It should be noted that the deformed shape of the microplates are scaled up for illustration purpose.

% Table generated by Excel2LaTeX from sheet 'Sheet1'
\begin{table}[htbp]
\begin{adjustwidth}{-1.7cm}{}
  \centering
  \caption{Non-dimensional deflection $\bar{w}=\dfrac{10E_ch^3}{q_0L^4}w\left(a/2,b/2,0\right)$  of SSSS Al/Al$_2$O$_3$ square microplates subjected to sinusoidally distributed load (rule of mixtures scheme)}
    \begin{tabular}{lllllllllllll}
    \hline
    $a/h$   & $l/h$ & $n = 0$ &       &       &       & $n = 1$ &       &       &       & $n = 10$ &       &  \\
    \cline{3-5} \cline{7-9} \cline{11-13}
          &       & RPT & Quasi-3D & TSDT\cite{Thai20131636} &       & RPT & Quasi-3D & TSDT\cite{Thai20131636} &       & RPT & Quasi-3D & TSDT\cite{Thai20131636} \\
          \hline
    5     & 0     & 0.3433 & 0.3360 & 0.3433 &       & 0.6688 & 0.6401 & 0.6688 &       & 1.2271 & 1.1663 & 1.2276 \\
          & 0.2   & 0.2898 & 0.2853 & 0.2875 &       & 0.5505 & 0.5321 & 0.5468 &       & 1.0400 & 1.0019 & 1.0247 \\
          & 0.4   & 0.1975 & 0.1965 & 0.1934 &       & 0.3601 & 0.3537 & 0.3535 &       & 0.7140 & 0.7043 & 0.6908 \\
          & 0.6   & 0.1292 & 0.1296 & 0.1251 &       & 0.2288 & 0.2274 & 0.2224 &       & 0.4694 & 0.4711 & 0.4514 \\
          & 0.8   & 0.0871 & 0.0879 & 0.0838 &       & 0.1517 & 0.1520 & 0.1464 &       & 0.3174 & 0.3220 & 0.3052 \\
          & 1     & 0.0614 & 0.0623 & 0.0588 &       & 0.1060 & 0.1069 & 0.1017 &       & 0.2242 & 0.2289 & 0.2158 \\
    20    & 0     & 0.2842 & 0.2836 & 0.2842 &       & 0.5689 & 0.5516 & 0.5689 &       & 0.9537 & 0.9280 & 0.9538 \\
          & 0.2   & 0.2431 & 0.2427 & 0.2430 &       & 0.4739 & 0.4619 & 0.4737 &       & 0.8313 & 0.8120 & 0.8303 \\
          & 0.4   & 0.1695 & 0.1694 & 0.1693 &       & 0.3157 & 0.3105 & 0.3153 &       & 0.6001 & 0.5906 & 0.5986 \\
          & 0.6   & 0.1127 & 0.1127 & 0.1124 &       & 0.2029 & 0.2008 & 0.2025 &       & 0.4102 & 0.4061 & 0.4090 \\
          & 0.8   & 0.0767 & 0.0768 & 0.0765 &       & 0.1352 & 0.1343 & 0.1349 &       & 0.2842 & 0.2825 & 0.2834 \\
          & 1     & 0.0544 & 0.0544 & 0.0542 &       & 0.0947 & 0.0943 & 0.0944 &       & 0.2038 & 0.2031 & 0.2033 \\
    100   & 0     & 0.2804 & 0.2803 & 0.2804 &       & 0.5625 & 0.5460 & 0.5625 &       & 0.9362 & 0.9132 & 0.9362 \\
          & 0.2   & 0.2401 & 0.2400 & 0.2401 &       & 0.4689 & 0.4574 & 0.4689 &       & 0.8176 & 0.8001 & 0.8176 \\
          & 0.4   & 0.1677 & 0.1677 & 0.1677 &       & 0.3128 & 0.3076 & 0.3128 &       & 0.5925 & 0.5833 & 0.5925 \\
          & 0.6   & 0.1116 & 0.1116 & 0.1116 &       & 0.2012 & 0.1990 & 0.2011 &       & 0.4062 & 0.4018 & 0.4061 \\
          & 0.8   & 0.0760 & 0.0760 & 0.0760 &       & 0.1341 & 0.1332 & 0.1341 &       & 0.2820 & 0.2799 & 0.2820 \\
          & 1     & 0.0539 & 0.0539 & 0.0539 &       & 0.0939 & 0.0934 & 0.0939 &       & 0.2024 & 0.2014 & 0.2024 \\
    \hline
    \end{tabular}
  \label{tab:exp:bending_SSSS}
\end{adjustwidth}
\end{table}

% Table generated by Excel2LaTeX from sheet 'Sheet1'
\begin{table}[htbp]
  \begin{adjustwidth}{-1.9cm}{}
  \centering
  \caption{Non-dimensional deflection $\bar{w}=\dfrac{10E_ch^3}{q_0L^4}w\left(a/2,b/2,0\right)$  of CCCC Al/Al$_2$O$_3$ square microplates (rule of mixtures scheme)}
    \begin{tabular}{lllllllllllll}
    \hline
    $a/h$   & $l/h$ & \multicolumn{3}{l}{$n = 0$} &       &\multicolumn{3}{l}{$n = 1$}      &       & \multicolumn{3}{l}{$n = 10$}  \\
    \cline{3-5} \cline{7-9} \cline{11-13}
          &       & RPT & Quasi-3D & IGA-Reddy &       & RPT & Quasi-3D & IGA-Reddy &       & RPT & Quasi-3D & IGA-Reddy \\
          \hline
     \multicolumn{13}{l}{Sinusoidally distributed load}\\
    5     & 0     & 0.1601 & 0.1359 & 0.1601 &       & 0.3021 & 0.2554 & 0.3020 &       & 0.6111 & 0.4740 & 0.6113 \\
          & 0.2   & 0.1378 & 0.1197 & 0.1377 &       & 0.2555 & 0.2214 & 0.2554 &       & 0.5178 & 0.4199 & 0.5183 \\
          & 0.4   & 0.0974 & 0.0883 & 0.0973 &       & 0.1751 & 0.1586 & 0.1751 &       & 0.3568 & 0.3132 & 0.3575 \\
          & 0.6   & 0.0655 & 0.0616 & 0.0655 &       & 0.1151 & 0.1081 & 0.1151 &       & 0.2358 & 0.2204 & 0.2364 \\
          & 0.8   & 0.0450 & 0.0435 & 0.0449 &       & 0.0779 & 0.0752 & 0.0779 &       & 0.1602 & 0.1559 & 0.1606 \\
          & 1     & 0.0321 & 0.0316 & 0.0320 &       & 0.0551 & 0.0542 & 0.0551 &       & 0.1136 & 0.1134 & 0.1139 \\
    20    & 0     & 0.1035 & 0.0950 & 0.1035 &       & 0.2065 & 0.1863 & 0.2065 &       & 0.3505 & 0.3150 & 0.3505 \\
          & 0.2   & 0.0919 & 0.0849 & 0.0919 &       & 0.1797 & 0.1638 & 0.1797 &       & 0.3150 & 0.2857 & 0.3151 \\
          & 0.4   & 0.0688 & 0.0645 & 0.0688 &       & 0.1294 & 0.1204 & 0.1294 &       & 0.2419 & 0.2237 & 0.2420 \\
          & 0.6   & 0.0485 & 0.0462 & 0.0485 &       & 0.0882 & 0.0837 & 0.0882 &       & 0.1746 & 0.1647 & 0.1747 \\
          & 0.8   & 0.0343 & 0.0331 & 0.0343 &       & 0.0611 & 0.0587 & 0.0611 &       & 0.1258 & 0.1204 & 0.1258 \\
          & 1     & 0.0250 & 0.0243 & 0.0250 &       & 0.0438 & 0.0425 & 0.0438 &       & 0.0926 & 0.0896 & 0.0926 \\
    100   & 0     & 0.0999 & 0.0955 & 0.0999 &       & 0.2003 & 0.1872 & 0.2003 &       & 0.3336 & 0.3131 & 0.3336 \\
          & 0.2   & 0.0889 & 0.0853 & 0.0889 &       & 0.1746 & 0.1643 & 0.1746 &       & 0.3013 & 0.2842 & 0.3013 \\
          & 0.4   & 0.0668 & 0.0646 & 0.0668 &       & 0.1262 & 0.1204 & 0.1262 &       & 0.2336 & 0.2228 & 0.2336 \\
          & 0.6   & 0.0473 & 0.0461 & 0.0473 &       & 0.0863 & 0.0835 & 0.0863 &       & 0.1701 & 0.1640 & 0.1701 \\
          & 0.8   & 0.0336 & 0.0329 & 0.0336 &       & 0.0599 & 0.0584 & 0.0599 &       & 0.1232 & 0.1199 & 0.1233 \\
          & 1     & 0.0244 & 0.0241 & 0.0244 &       & 0.0430 & 0.0422 & 0.0430 &       & 0.0910 & 0.0891 & 0.0910 \\
          \\
    \multicolumn{13}{l}{Uniformly distributed load} \\
    5     & 0     & 0.2239 & 0.1860 & 0.2238 &       & 0.4220 & 0.3500 & 0.4219 &       & 0.8557 & 0.6408 & 0.8559 \\
          & 0.2   & 0.1924 & 0.1641 & 0.1924 &       & 0.3566 & 0.3040 & 0.3565 &       & 0.7233 & 0.5701 & 0.7239 \\
          & 0.4   & 0.1358 & 0.1215 & 0.1358 &       & 0.2443 & 0.2186 & 0.2442 &       & 0.4976 & 0.4286 & 0.4985 \\
          & 0.6   & 0.0914 & 0.0851 & 0.0913 &       & 0.1606 & 0.1495 & 0.1605 &       & 0.3288 & 0.3034 & 0.3295 \\
          & 0.8   & 0.0627 & 0.0602 & 0.0627 &       & 0.1087 & 0.1041 & 0.1087 &       & 0.2234 & 0.2154 & 0.2239 \\
          & 1     & 0.0447 & 0.0439 & 0.0447 &       & 0.0769 & 0.0752 & 0.0769 &       & 0.1584 & 0.1570 & 0.1588 \\
    20    & 0     & 0.1436 & 0.1300 & 0.1436 &       & 0.2864 & 0.2553 & 0.2864 &       & 0.4863 & 0.4307 & 0.4863 \\
          & 0.2   & 0.1275 & 0.1164 & 0.1275 &       & 0.2493 & 0.2248 & 0.2493 &       & 0.4372 & 0.3912 & 0.4373 \\
          & 0.4   & 0.0956 & 0.0887 & 0.0956 &       & 0.1797 & 0.1657 & 0.1797 &       & 0.3360 & 0.3072 & 0.3361 \\
          & 0.6   & 0.0674 & 0.0637 & 0.0674 &       & 0.1227 & 0.1156 & 0.1227 &       & 0.2428 & 0.2269 & 0.2429 \\
          & 0.8   & 0.0478 & 0.0457 & 0.0478 &       & 0.0850 & 0.0813 & 0.0850 &       & 0.1750 & 0.1664 & 0.1751 \\
          & 1     & 0.0348 & 0.0336 & 0.0348 &       & 0.0610 & 0.0589 & 0.0610 &       & 0.1289 & 0.1240 & 0.1289 \\
    100   & 0     & 0.1384 & 0.1314 & 0.1384 &       & 0.2775 & 0.2577 & 0.2775 &       & 0.4622 & 0.4310 & 0.4622 \\
          & 0.2   & 0.1232 & 0.1174 & 0.1232 &       & 0.2421 & 0.2265 & 0.2421 &       & 0.4176 & 0.3915 & 0.4176 \\
          & 0.4   & 0.0927 & 0.0891 & 0.0927 &       & 0.1751 & 0.1664 & 0.1751 &       & 0.3242 & 0.3075 & 0.3242 \\
          & 0.6   & 0.0657 & 0.0637 & 0.0657 &       & 0.1199 & 0.1155 & 0.1199 &       & 0.2363 & 0.2268 & 0.2363 \\
          & 0.8   & 0.0467 & 0.0456 & 0.0467 &       & 0.0833 & 0.0810 & 0.0833 &       & 0.1713 & 0.1661 & 0.1713 \\
          & 1     & 0.0340 & 0.0334 & 0.0340 &       & 0.0598 & 0.0586 & 0.0598 &       & 0.1266 & 0.1236 & 0.1266]\\
    \hline
    \end{tabular}
  \label{tab:exp:bending_CCCC}
  \end{adjustwidth}
\end{table}

\subsection{Free vibration analysis}
\label{sec:Exp:Vibration}

In this part, the free vibration analysis of FG microplates based on modified couple stress theory will be discussed. The proposed Quasi-3D model is initially tested for normal-scale SSSS Al/ZrO$_2$-1 plates with various theories taking into account the normal shear deformation. It can be seen that the comparison results which are given in Table \ref{tab:exp:vib_Q3DTest} agree well with those of other published works. The proposed RPT and quasi-3D models using IGA are further tested for homogeneous square microplates. The results which are shown in Table \ref{tab:exp:vib_isotropic} are compared with analytical solutions generated from CPT by Yin et al.  \cite{Yin2010386} and TSDT by Thai and Kim \cite{Thai20131636} . As can be observed, while the proposed RPT model yields slight discrepancy with respect to Yin et al.'s \cite{Yin2010386} due to their CPT assumption neglecting shear deformations, it shows excellent agreement with Thai and Kim's \cite{Thai20131636}, especially as plates become thinner, i.e. $a/h$ ratio is relatively large. Due to the consideration of thickness stretching effects, the proposed quasi-3D model gives slightly different results in comparison with other results. 

% Table generated by Excel2LaTeX from sheet 'Sheet1'
\begin{table}[htbp]
  \centering
  \caption{Comparison of non-dimensional natural frequencies $\bar{\omega} = \omega \dfrac{a^2}{h} \sqrt{\dfrac{\rho_m}{E_m}} $ of SSSS Al/ZrO$_2$-1 plates (Mori-Tanaka scheme)}
    \begin{tabular}{llllllll}
    \hline
    Theory & \multicolumn{3}{l}{$n=1$} &       & \multicolumn{3}{l}{$a/h=5$} \\
\cline{2-4}\cline{6-8}          & $a/h=5$ & $a/h=10$    & $a/h=20$    &       & $n=2$   & $n=3$     & $n=5$\\
    \hline
    Vel and Batra \cite{Vel2004703} & 5.4806 & 5.9609 & 6.1076 &       & 5.4923 & 5.5285 & 5.5632 \\
    Matsunaga \cite{Matsunaga2008499} & 5.7123 & 6.1932 & 6.3390 &       & 5.6599 & 5.6757 & 5.7020 \\
    Neves et al. \cite{Neves2012711} & 5.4825 & 5.9600 & 6.1200 &       & 5.4950 & 5.5300 & 5.5625 \\
    Belabed et al. \cite{Belabed2014274} & 5.4800 & 5.9700 & 6.1200 &       & 5.5025 & 5.5350 & 5.5625 \\
    Alijani and Amabili \cite{Alijani201489} & 5.4796 & 5.9578 & 6.1040 &       & 5.4919 & 5.5279 & 5.5633 \\
    Akavci and Tanrikulu \cite{Akavci2015203} & 5.4829 & 5.9676 & 6.1160 &       & 5.5064 & 5.5388 & 5.5644 \\
    Present  & 5.5172 & 6.0023 & 6.1505 &       & 5.5324 & 5.5642 & 5.5886\\
    \hline
    \end{tabular}
  \label{tab:exp:vib_Q3DTest}
\end{table}

% Table generated by Excel2LaTeX from sheet 'Vibration'
\begin{table}[htbp]
  \centering
  \caption{Comparison of non-dimensional natural frequencies $\bar{\omega} = \omega \dfrac{a^2}{h} \sqrt{\dfrac{\rho}{E}} $ of SSSS homogeneous microplates}
    \begin{tabular}{llllllll}
    \hline
    $a/h$   & Theory & \multicolumn{6}{l}{$l/h$} \\
\cline{3-8}          &       & 0     & 0.2   & 0.4   & 0.6   & 0.8   & 1\\
    \hline
    5     & CPT \cite{Yin2010386} & 5.9734 & 6.4556 & 7.7239 & 9.4673 & 11.4713 & 13.6213 \\
          & TSDT \cite{Thai20131636} & 5.2813 & 5.7699 & 7.0330 & 8.7389 & 10.6766 & 12.7408 \\
          & RPT (Present) &5.2813&5.7496&6.9667&8.6191&9.8943&9.9791
 \\
          & Quasi-3D (Present)&5.3090 &5.7622&6.9438&8.5509&9.8943&9.9791

 \\
    20    & CPT \cite{Yin2010386} & 5.9734 & 6.4556 & 7.7239 & 9.4673 & 11.4713 & 13.6213 \\
          & TSDT \cite{Thai20131636} & 5.9199 & 6.4027 & 7.6708 & 9.4116 & 11.4108 & 13.5545 \\
          & RPT (Present)&5.9199&6.4009&7.6646&9.4005& 11.3945& 13.5330
 \\
          & Quasi-3D (Present)&5.9235&6.4030&7.6633&9.3952&11.3854&13.5202

 \\
    100   & CPT \cite{Yin2010386} & 5.9734 & 6.4556 & 7.7239 & 9.4673 & 11.4713 & 13.6213 \\
          & TSDT \cite{Thai20131636} & 5.9712 & 6.4535 & 7.7217 & 9.4651 & 11.4689 & 13.6186 \\
          & RPT (Present)&5.9712&6.4534&7.7215&9.4646&11.4682&13.6178
 \\
          & Quasi-3D (Present)&5.9723&6.4544&7.7222&9.4650&11.4683&13.6177

\\
    \hline
    \end{tabular}
  \label{tab:exp:vib_isotropic}
\end{table}

\begin{figure}
	\centering
	\includegraphics[width=1.0\textwidth]{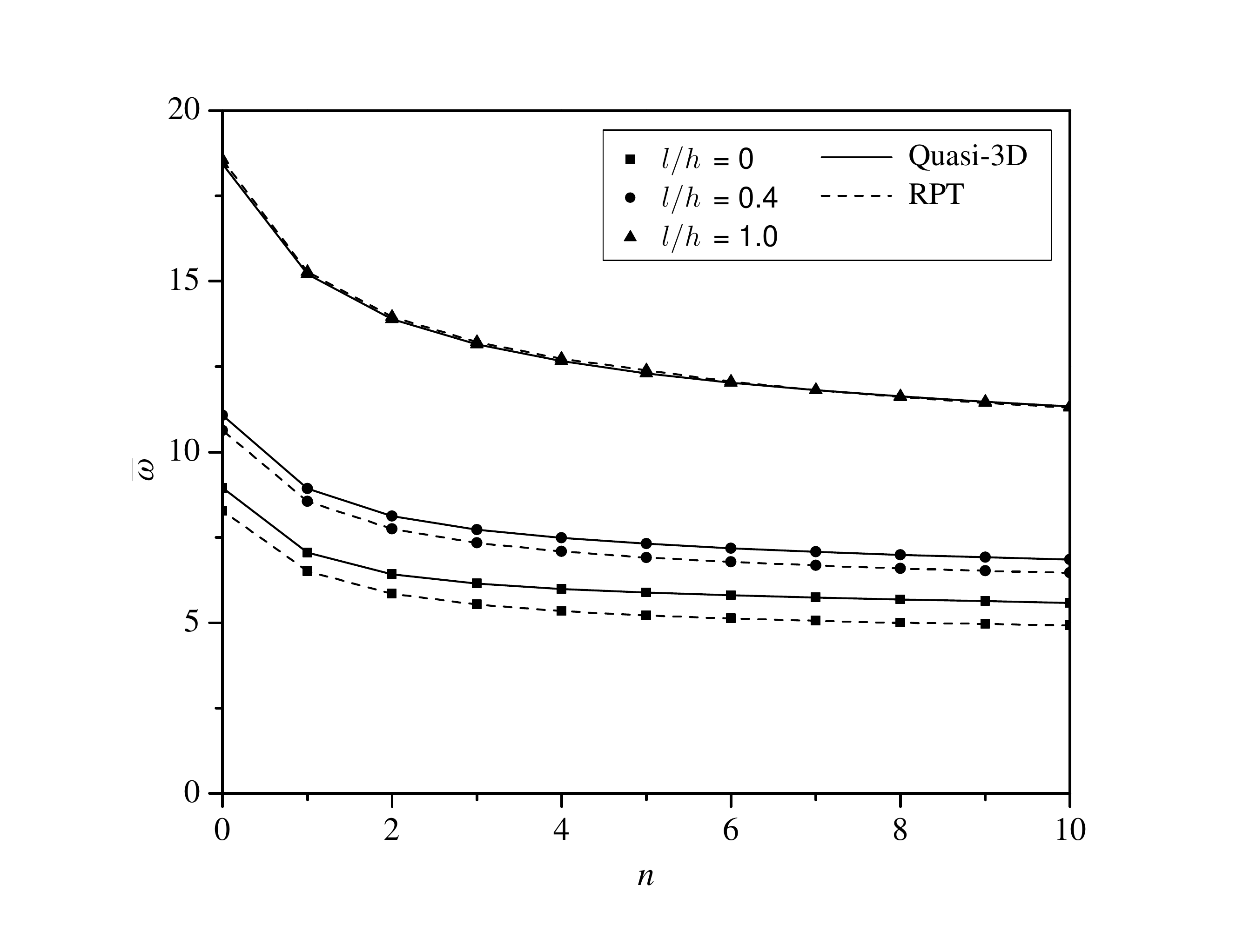}
	\caption{Effects of material index $n$ and material length scale ratio $l/h$ on the natural frequency of CCCC Al/Al$_2$O$_3$ square microplates, $a/h=5$ (rule of mixtures scheme). \label{fig:exp:vib_effects}}
\end{figure}

\begin{figure}
	\centering
	\begin{subfigure}{0.49\textwidth}
		\includegraphics[width=\linewidth]{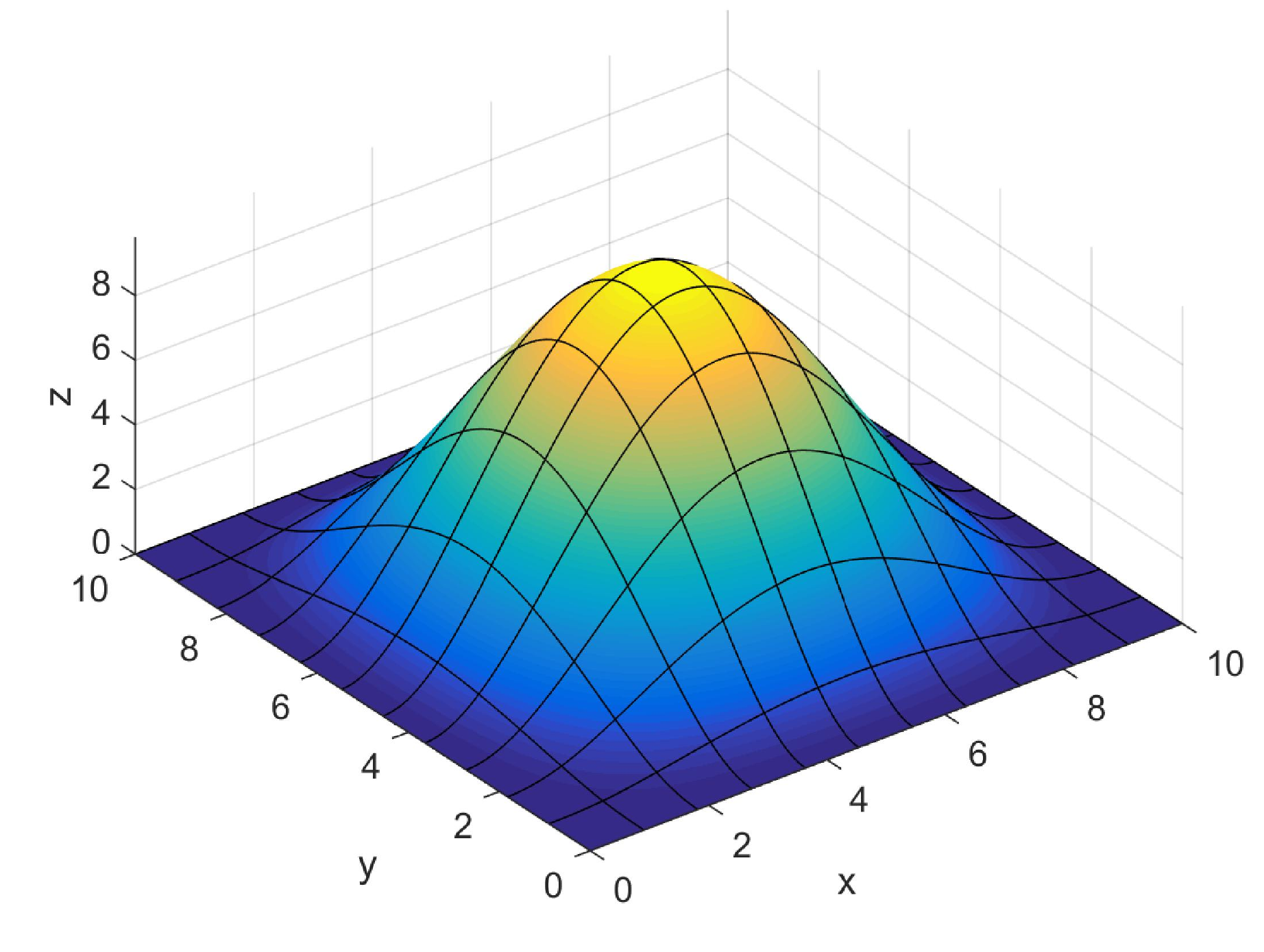}
		\caption{$\bar{\omega}_1 = 15.4413$.}
	\end{subfigure}
	%\hspace*{\fill}
	\begin{subfigure}{0.49\textwidth}
		\includegraphics[width=\linewidth]{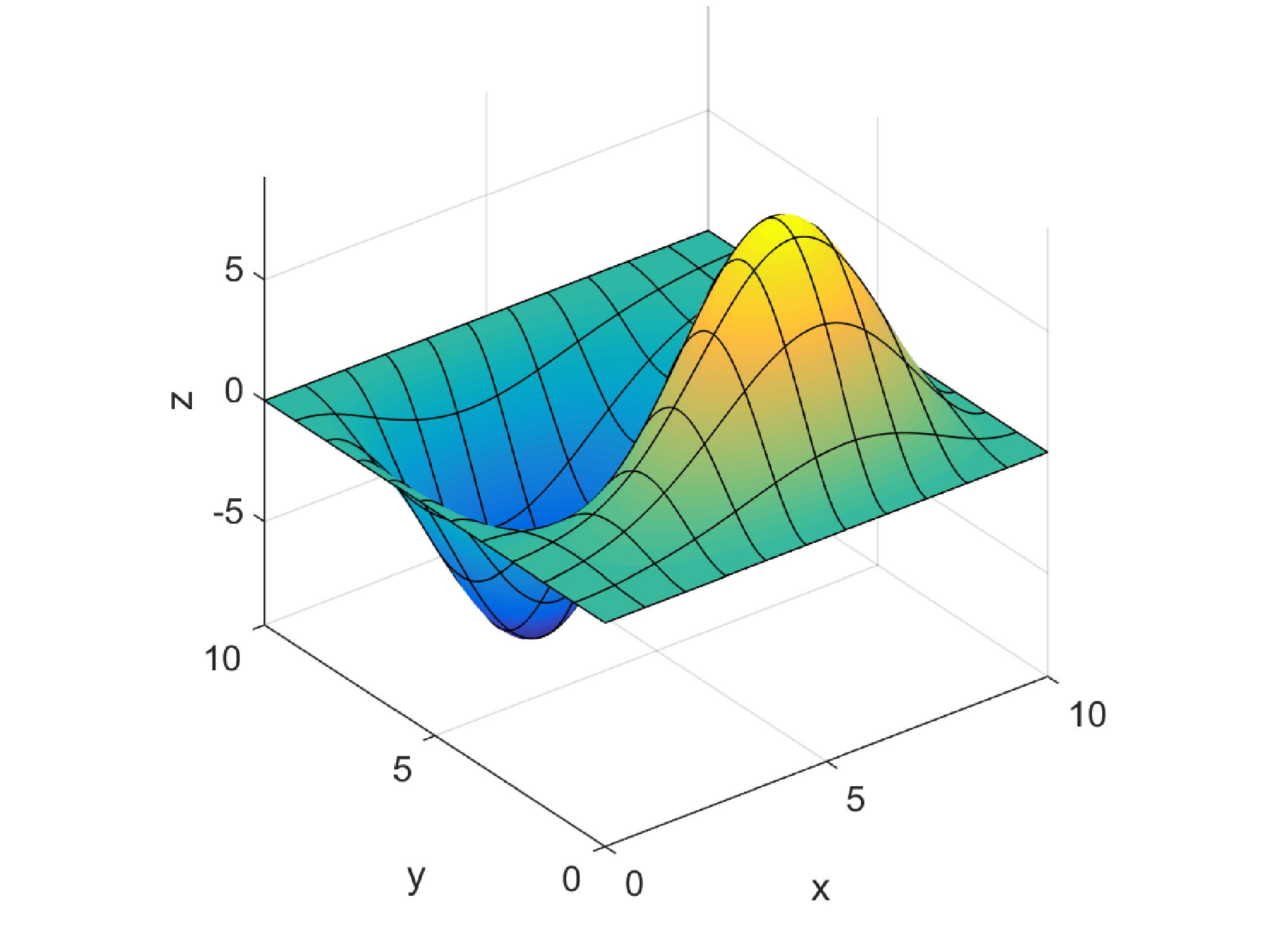}
		\caption{$\bar{\omega}_2 = 29.3267$.}
	\end{subfigure}
	\begin{subfigure}{0.49\textwidth}
		\includegraphics[width=\linewidth]{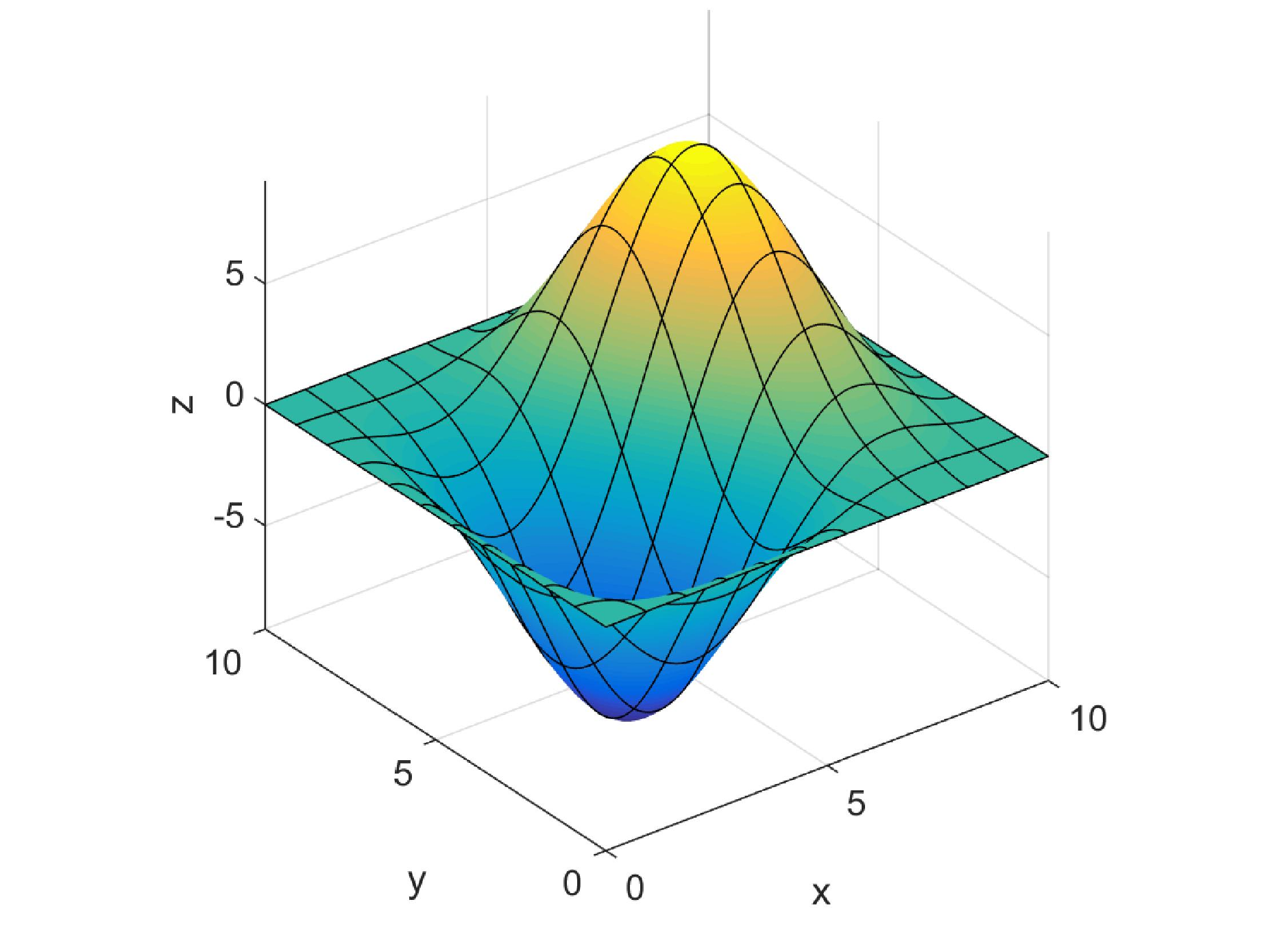}
		\caption{$\bar{\omega}_3 = 29.3267$.}
	\end{subfigure}
	\begin{subfigure}{0.49\textwidth}
		\includegraphics[width=\linewidth]{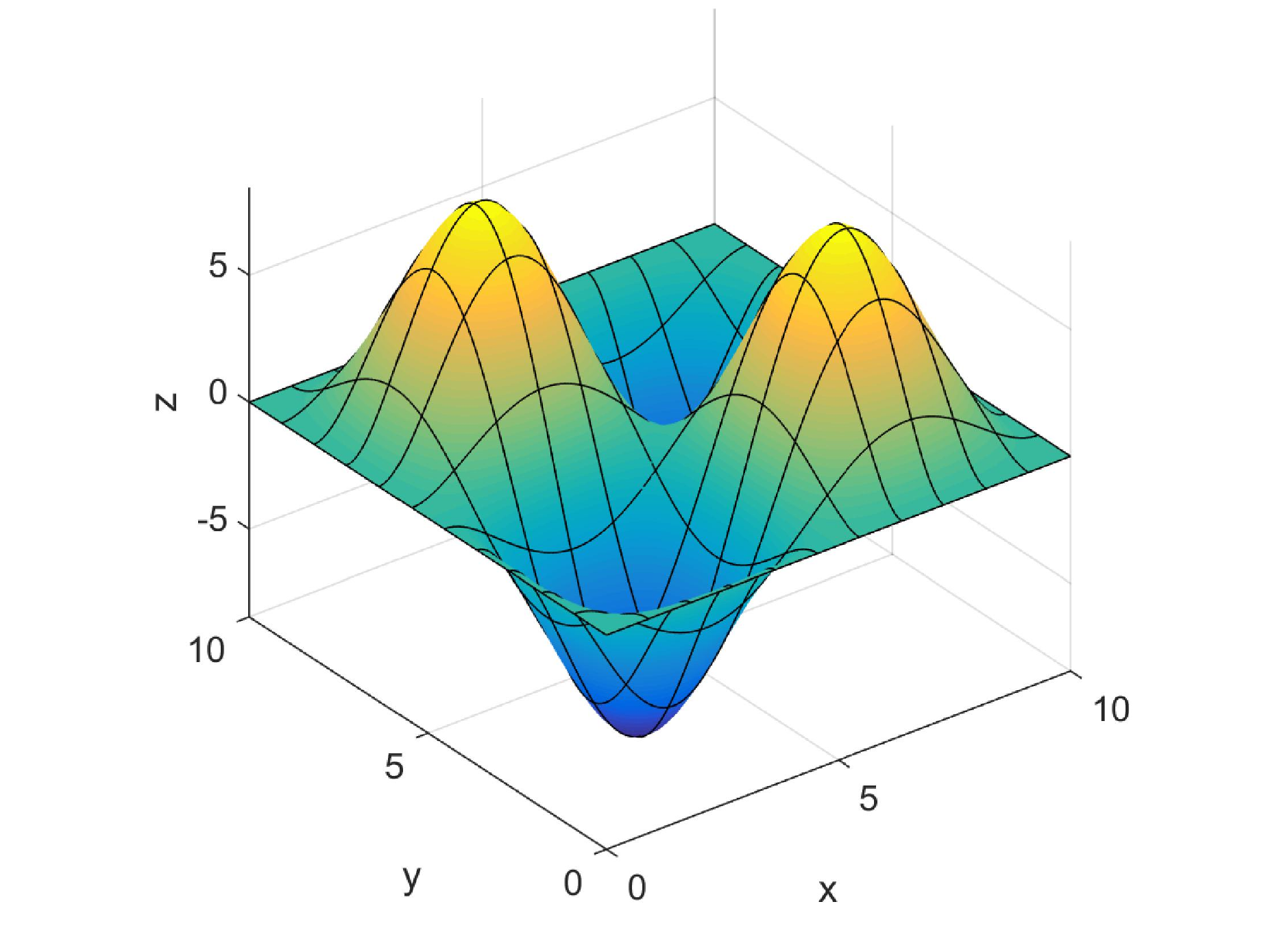}
		\caption{$\bar{\omega}_4 = 41.5955$.}
	\end{subfigure}
	\begin{subfigure}{0.49\textwidth}
		\includegraphics[width=\linewidth]{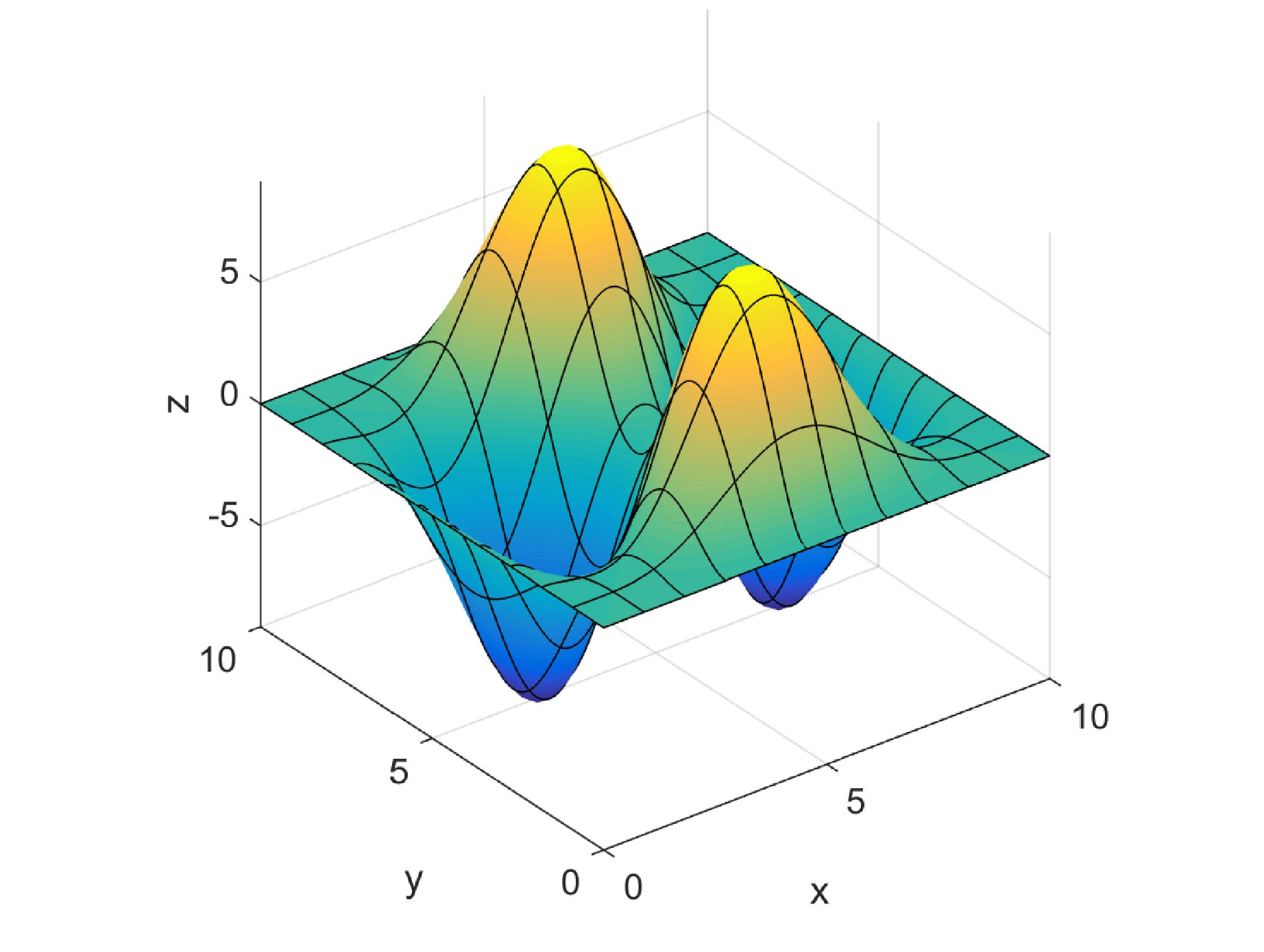}
		\caption{$\bar{\omega}_5 = 48.2657$.}
	\end{subfigure}
	\begin{subfigure}{0.49\textwidth}
		\includegraphics[width=\linewidth]{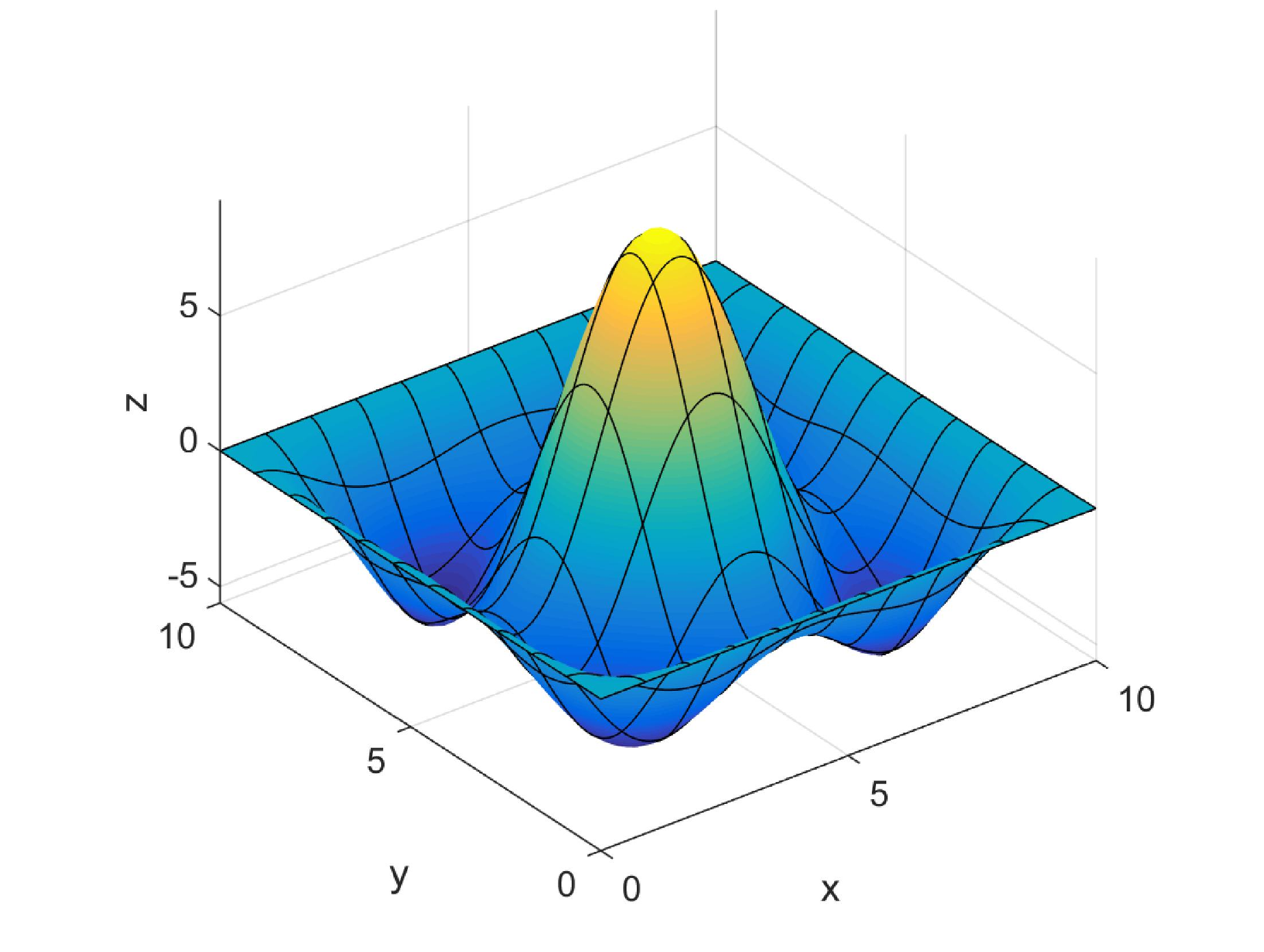}
		\caption{$\bar{\omega}_6 = 48.4504$.}
	\end{subfigure}
	\caption{The first six free vibration mode shapes of Al/Al$_2$O$_3$ CCCC square microplates. \label{fig:exp:vib:sq_6}}
\end{figure}

Table \ref{tab:exp:vib_square_comparison} presents non-dimensional natural frequency of SSSS Al/Al$_2$O$_3$ square microplates. The results are compared with those of Thai and Kim \cite{Thai20131636} in which analytical approach based on TSDT model is employed. Thick $(a/h=5)$, moderately thick $(a/h=20)$ and thin $(a/h=100)$ microplates are considered. The results reveal good agreement between RPT model and TSDT \cite{Thai20131636}, especially when the material length scale ratio $l/h$ is small, e.g. 0 or 0.2. On the contrary, the discrepancy becomes larger as $l/h$ gets closer to 1. However, this phenomenon just happens for thick plates and tends to be vague as the plates become thinner. Meanwhile, the quasi-3D gives slightly different results compared to that of RPT model. General observation from Table \ref{tab:exp:vib_square_comparison} reveals that the higher material length scale ratio is chosen, the bigger natural frequencies of the plates could be obtained as a result of the increase in plate's stiffness. Fig. \ref{fig:exp:vib_effects} provides a closer look at the effects of material index $n$ and material length scale $l$ on the plates' natural frequencies which are computed using proposed RPT and quasi-3D models. Similar to the previous case of bending analysis, the plate's stiffness decreases as a result of rising in material index $n$ and falling in material length scale ratio $l/h$ which leads to the decrease of natural frequency of the plates. In addition, the discrepancy in term of frequency results predicted by the proposed RPT and quasi-3D becomes less significant as $l/h$ gets bigger and is almost zero difference when $l/h =1.0$. The first six natural frequencies of Al/Al$_2$O$_3$ square microplates with different types of boundary conditions are given in Table \ref{tab:exp:vib:sq_6} in which the results are generated for $n=1$ and $l/h = 0.2$. The present quasi-3D results show good agreement with those of Zenkour's quasi-3D model \cite{Zenkour20139041} using the proposed IGA approach. The first six mode shapes corresponding to the quasi-3D vibration analysis of CCCC microplates with $a/h=10$ are presented in Fig. \ref{fig:exp:vib:sq_6}.

% Table generated by Excel2LaTeX from sheet 'Sheet1'
\begin{table}[htbp]
 \begin{adjustwidth}{-1.8cm}{}
  \centering
  \caption{Non-dimensional natural frequency $\bar{\omega} = \omega \dfrac{a^2}{h} \sqrt{\dfrac{\rho_c}{E_c}} $ of SSSS Al/Al$_2$O$_3$ square plates (rule of mixtures scheme)}
    \begin{tabular}{lllllllllllll}
    \hline
	\cline{3-13}
   $a/h$  &   $l/h$    & $n = 0$     &       &       &       & $n = 1$     &       &       &       & $n = 10$    &       &  \\
\cline{3-5}\cline{7-9}\cline{11-13}          &       & RPT & Quasi-3D & TSDT\cite{Thai20131636} &       & RPT & Quasi-3D & TSDT\cite{Thai20131636} &       & RPT & Quasi-3D & TSDT\cite{Thai20131636} \\
    \hline
    5     & 0     & 5.2813 & 5.3090 & 5.2813 &       & 4.0781 & 4.1521 & 4.0781 &       & 3.2519 & 3.3126 & 3.2514 \\
          & 0.2   & 5.7496 & 5.7622 & 5.7699 &       & 4.4959 & 4.5542 & 4.5094 &       & 3.5312 & 3.5740 & 3.5548 \\
          & 0.4   & 6.9667 & 6.9438 & 7.0330 &       & 5.5620 & 5.5865 & 5.6071 &       & 4.2584 & 4.2627 & 4.3200 \\
          & 0.6   & 8.6191 & 8.5509 & 8.7389 &       & 6.9822 & 6.9681 & 7.0662 &       & 5.2471 & 5.2115 & 5.3335 \\
          & 0.8   & 9.8943 & 9.8943 & 10.6766 &       & 8.2313 & 8.2313 & 8.7058 &       & 5.8571 & 5.8571 & 6.4759 \\
          & 1     & 9.9791 & 9.9791 & 12.7408 &       & 8.3019 & 8.3019 & 10.4397 &       & 5.9073 & 5.9073 & 7.6895 \\
    20    & 0     & 5.9199 & 5.9235 & 5.9199 &       & 4.5228 & 4.5919 & 4.5228 &       & 3.7623 & 3.8129 & 3.7622 \\
          & 0.2   & 6.4009 & 6.4030 & 6.4027 &       & 4.9556 & 5.0179 & 4.9568 &       & 4.0299 & 4.0761 & 4.0323 \\
          & 0.4   & 7.6646 & 7.6633 & 7.6708 &       & 6.0714 & 6.1203 & 6.0756 &       & 4.7428 & 4.7794 & 4.7488 \\
          & 0.6   & 9.4005 & 9.3952 & 9.4116 &       & 7.5739 & 7.6107 & 7.5817 &       & 5.7369 & 5.7640 & 5.7453 \\
          & 0.8   & 11.3945 & 11.3854 & 11.4108 &       & 9.2768 & 9.3042 & 9.2887 &       & 6.8914 & 6.9106 & 6.9013 \\
          & 1     & 13.5330 & 13.5202 & 13.5545 &       & 11.0882 & 11.1082 & 11.1042 &       & 8.1384 & 8.1510 & 8.1494 \\
    100   & 0     & 5.9712 & 5.9723 & 5.9712 &       & 4.5579 & 4.6263 & 4.5579 &       & 3.8058 & 3.8533 & 3.8058 \\
          & 0.2   & 6.4534 & 6.4544 & 6.4535 &       & 4.9922 & 5.0546 & 4.9922 &       & 4.0724 & 4.1168 & 4.0725 \\
          & 0.4   & 7.7215 & 7.7222 & 7.7217 &       & 6.1124 & 6.1635 & 6.1126 &       & 4.7837 & 4.8215 & 4.7840 \\
          & 0.6   & 9.4646 & 9.4650 & 9.4651 &       & 7.6220 & 7.6630 & 7.6224 &       & 5.7778 & 5.8090 & 5.7782 \\
          & 0.8   & 11.4682 & 11.4683 & 11.4689 &       & 9.3339 & 9.3673 & 9.3344 &       & 6.9341 & 6.9600 & 6.9345 \\
          & 1     & 13.6178 & 13.6177 & 13.6186 &       & 11.1554 & 11.1832 & 11.1560 &       & 8.1842 & 8.2060 & 8.1846 \\
    \hline
    \end{tabular}
  \label{tab:exp:vib_square_comparison}
  \end{adjustwidth}
\end{table}

% Table generated by Excel2LaTeX from sheet 'Vibration'
\begin{table}[htbp]
  \centering
  \caption{The first six non-dimensional natural frequencies $\bar{\omega} = \omega \dfrac{a^2}{h} \sqrt{\dfrac{\rho_m}{E_m}} $ of Al/Al$_2$O$_3$ square plates (Mori-Tanaka scheme)}
    \begin{tabular}{lllllllll}
    \hline
    BC    & $a/h$   & Theory & \multicolumn{6}{l}{Mode} \\
\cline{4-9}          &       &       & 1     & 2     & 3     & 4     & 5     & 6 \\
    \hline
    SSSS  & 5     & IGA-Zenkour & 7.9366 & 13.8049 & 13.8049 & 17.3259 & 17.3259 & 19.5422 \\
          &       & Quasi-3D (Present) & 7.8883 & 13.8049 & 13.8049 & 17.2045 & 17.2045 & 19.5422 \\
          &       & RPT (Present) & 7.7844 & 13.8049 & 13.8049 & 16.9943 & 16.9943 & 19.5422 \\
          & 10    & IGA-Zenkour & 8.5940 & 20.3230 & 20.3230 & 27.5893 & 27.5893 & 31.7475 \\
          &       & Quasi-3D (Present) & 8.5607 & 20.2031 & 20.2031 & 27.5893 & 27.5893 & 31.5547 \\
          &       & RPT (Present) & 8.4401 & 19.9188 & 19.9188 & 27.5893 & 27.5893 & 31.1386 \\
          & 100   & IGA-Zenkour & 8.8399 & 21.7851 & 21.7851 & 35.3333 & 43.0764 & 43.0764 \\
          &       & Quasi-3D (Present) & 8.8390 & 21.7802 & 21.7802 & 35.3221 & 43.0573 & 43.0573 \\
          &       & RPT (Present) & 8.7127 & 21.4598 & 21.4598 & 34.8171 & 42.4074 & 42.4074 \\
    CCCC  & 5     & IGA-Zenkour & 12.7213 & 22.6661 & 22.6661 & 27.9021 & 27.9021 & 31.1450 \\
          &       & Quasi-3D (Present) & 13.1029 & 22.9300 & 22.9300 & 27.8791 & 27.8791 & 31.3005 \\
          &       & RPT (Present) & 12.1531 & 22.0683 & 22.0683 & 26.3182 & 26.3182 & 30.5398 \\
          & 10    & IGA-Zenkour & 15.2379 & 28.9665 & 28.9665 & 41.2221 & 47.8649 & 47.9221 \\
          &       & Quasi-3D (Present) & 15.4413 & 29.3267 & 29.3267 & 41.5955 & 48.2657 & 48.4504 \\
          &       & RPT (Present) & 14.3638 & 27.8276 & 27.8276 & 39.9305 & 46.0435 & 46.4772 \\
          & 100   & IGA-Zenkour & 16.0795 & 32.6473 & 32.6473 & 48.5912 & 58.0858 & 58.3951 \\
          &       & Quasi-3D (Present) & 16.0540 & 32.6082 & 32.6082 & 48.5477 & 58.0353 & 58.3402 \\
          &       & RPT (Present) & 15.5212 & 31.5708 & 31.5708 & 47.0903 & 56.2566 & 56.5207 \\
    \hline
    \end{tabular}
  \label{tab:exp:vib:sq_6}
\end{table}

\begin{figure}
	\centering
	\begin{subfigure}{0.7\textwidth}
		\includegraphics[width=\linewidth]{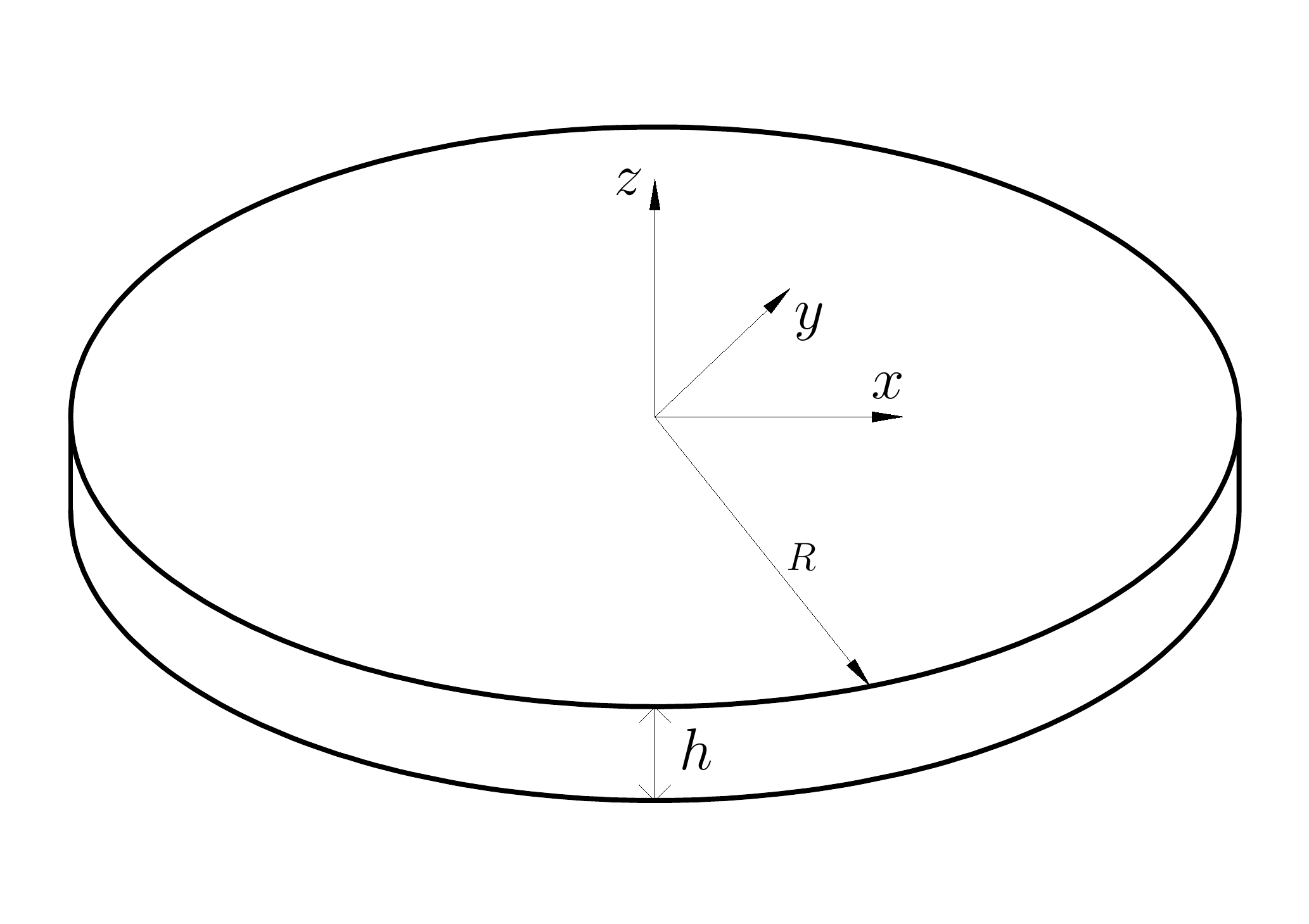}
		\caption{Geometric configuration.}
	\end{subfigure}
	%\hspace*{\fill}
	\begin{subfigure}{0.6\textwidth}
		\includegraphics[width=\linewidth]{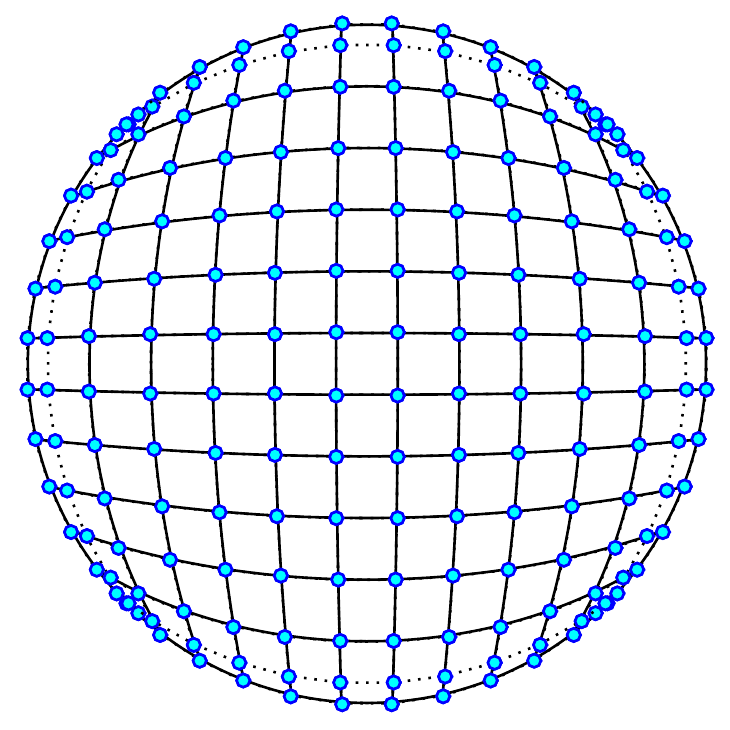}
		\caption{Control point net and 11$\times$11 cubic elements.}
	\end{subfigure}
	\caption{Geometry and element mesh of a circular microplate. \label{fig:exp:cir:geometry_mesh}}
\end{figure}

\begin{figure}
	\centering
	\begin{subfigure}{0.49\textwidth}
		\includegraphics[width=\linewidth]{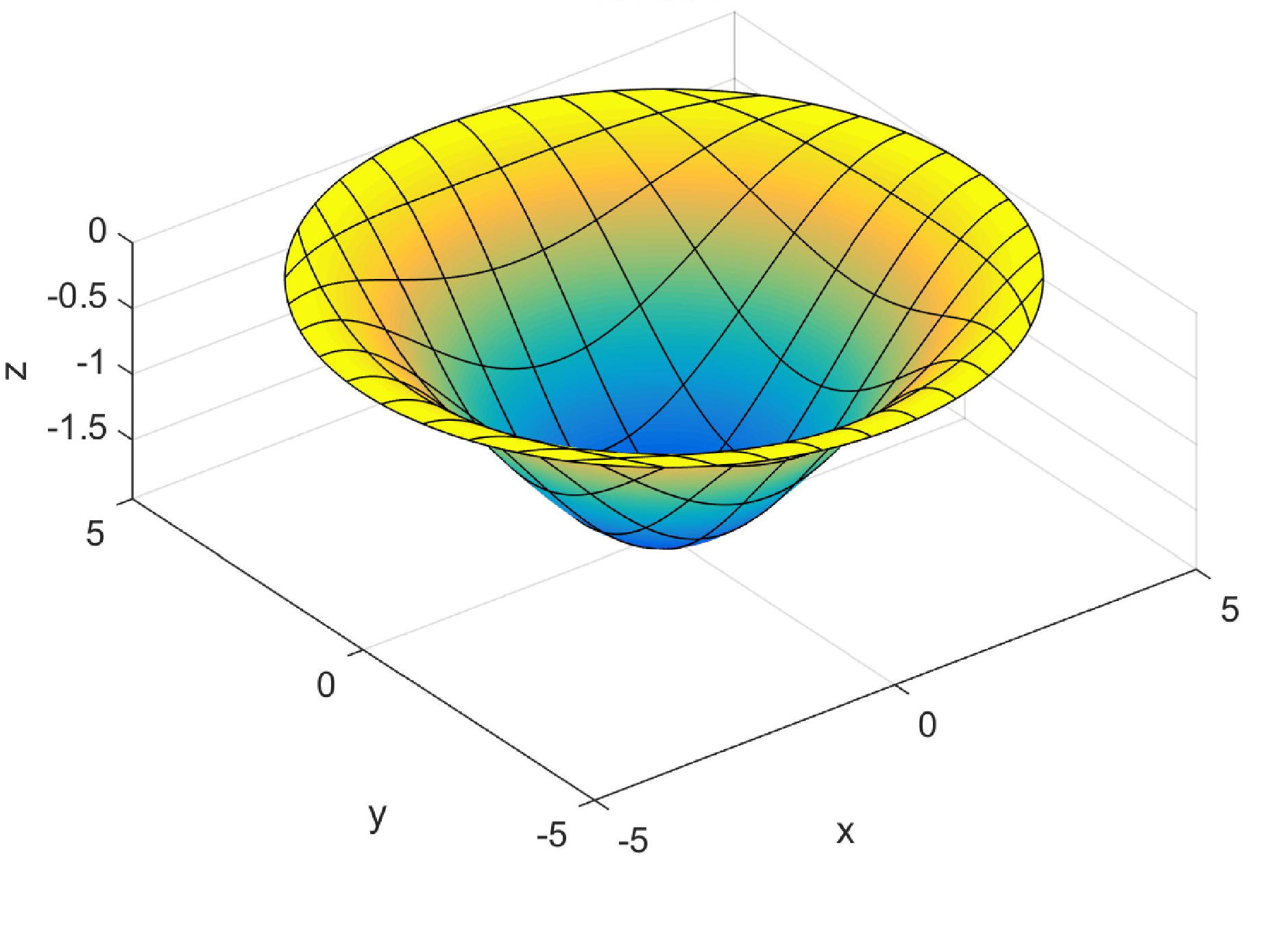}
		\caption{$\bar{\omega}_1 = 7.3195$.}
	\end{subfigure}
	%\hspace*{\fill}
	\begin{subfigure}{0.49\textwidth}
		\includegraphics[width=\linewidth]{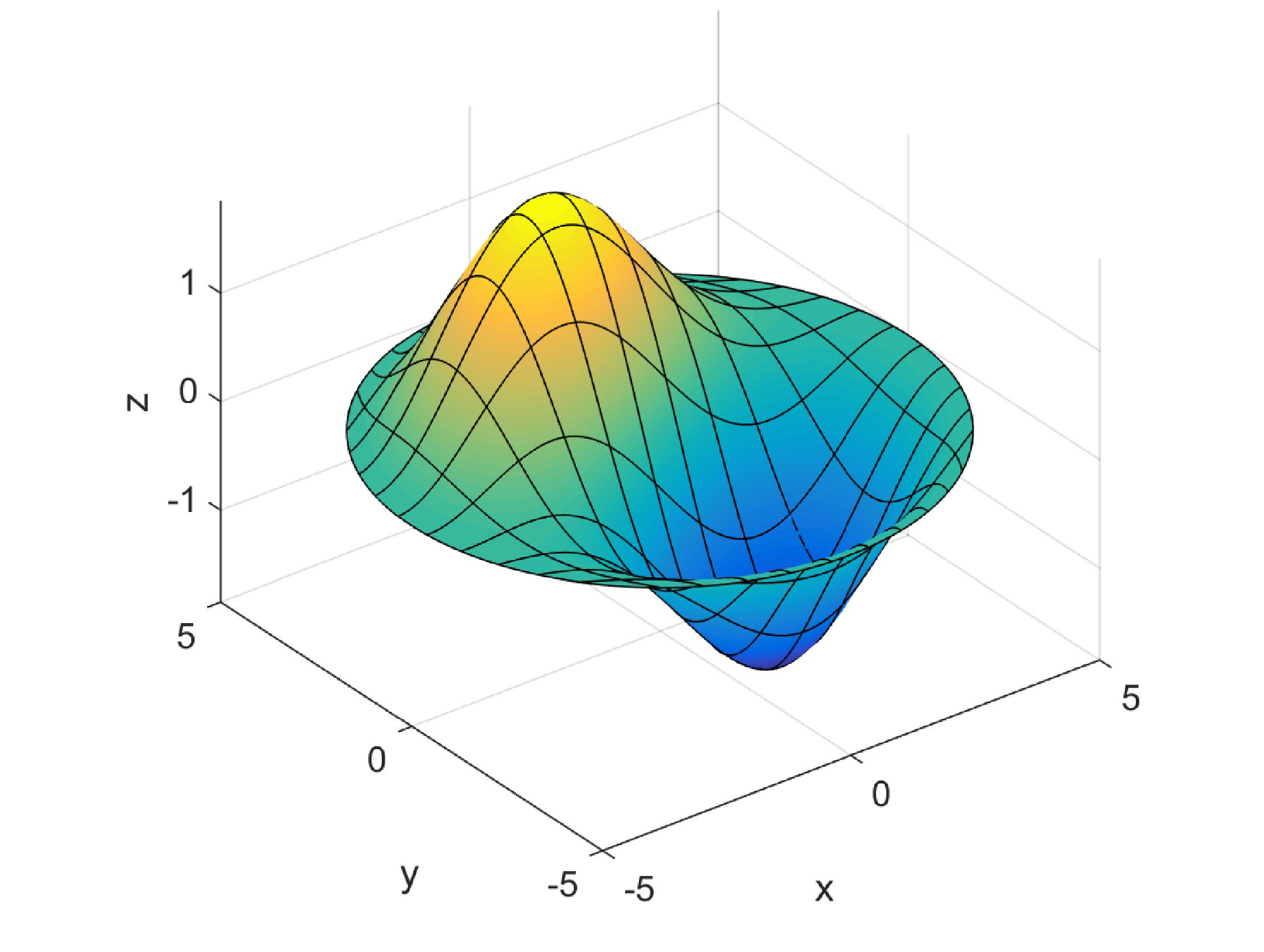}
		\caption{$\bar{\omega}_2 = 14.0950$.}
	\end{subfigure}
	\begin{subfigure}{0.49\textwidth}
		\includegraphics[width=\linewidth]{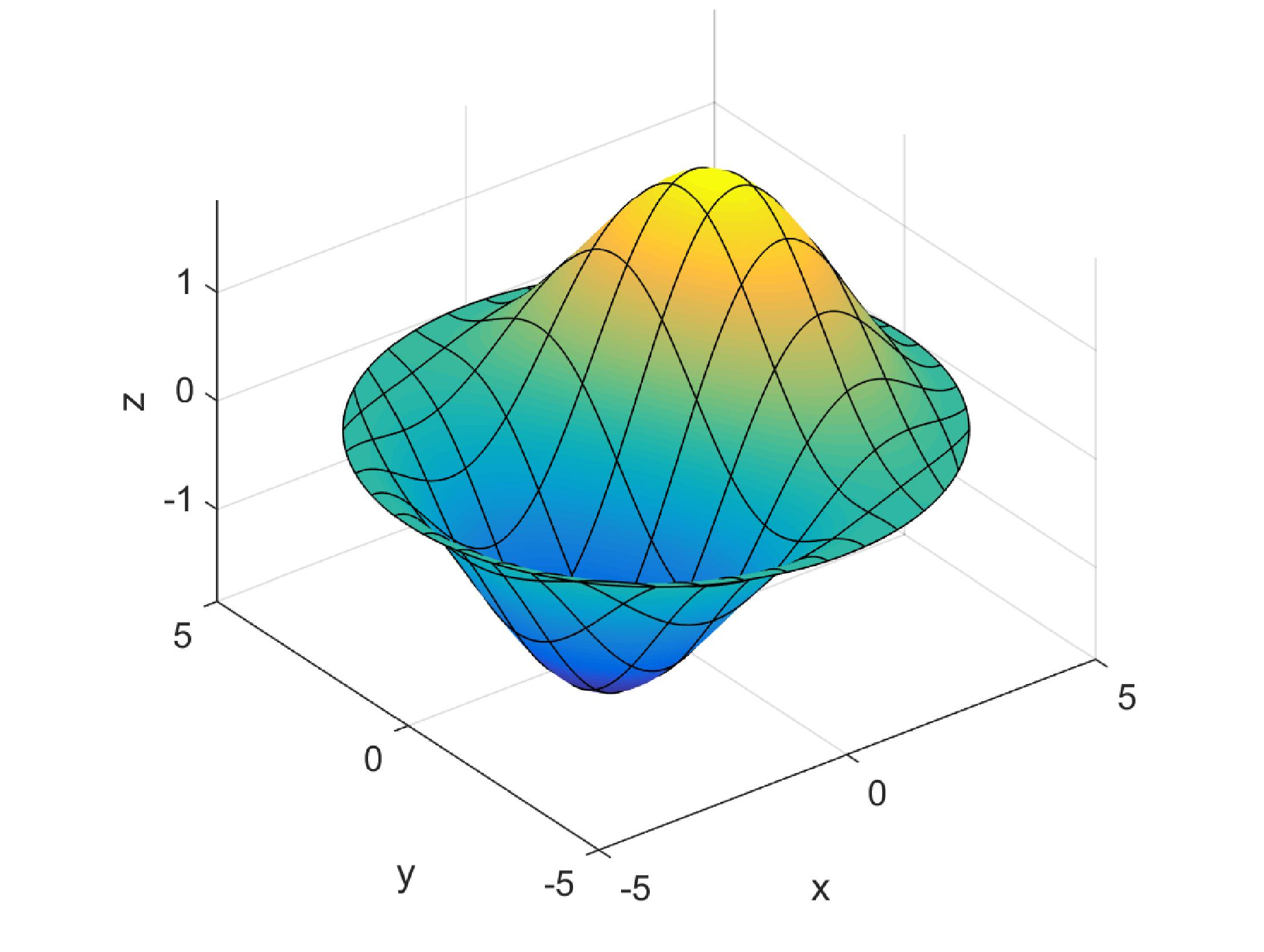}
		\caption{$\bar{\omega}_3 = 14.0950$.}
	\end{subfigure}
	\begin{subfigure}{0.49\textwidth}
		\includegraphics[width=\linewidth]{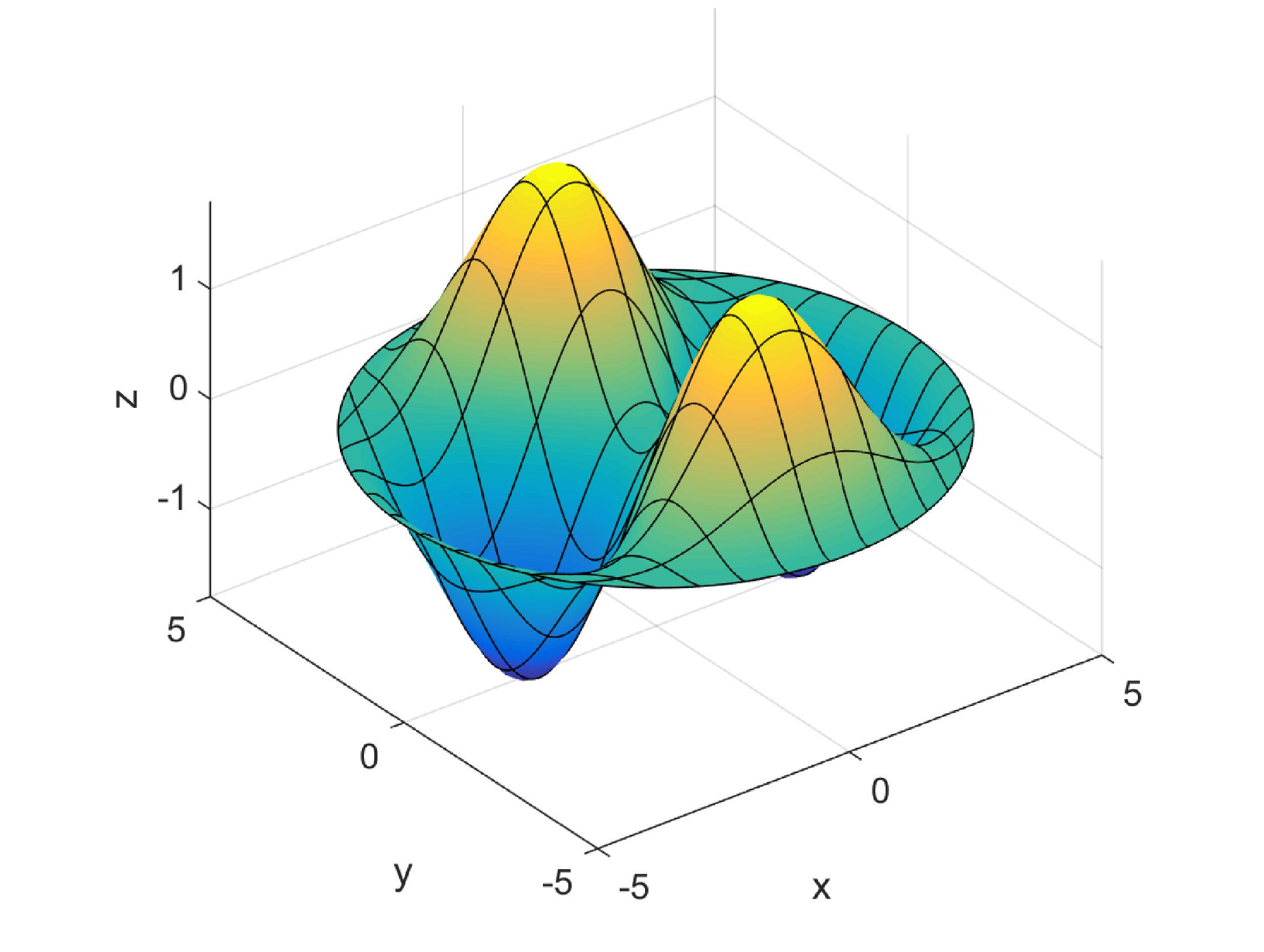}
		\caption{$\bar{\omega}_4 = 21.3095$.}
	\end{subfigure}
	\begin{subfigure}{0.49\textwidth}
		\includegraphics[width=\linewidth]{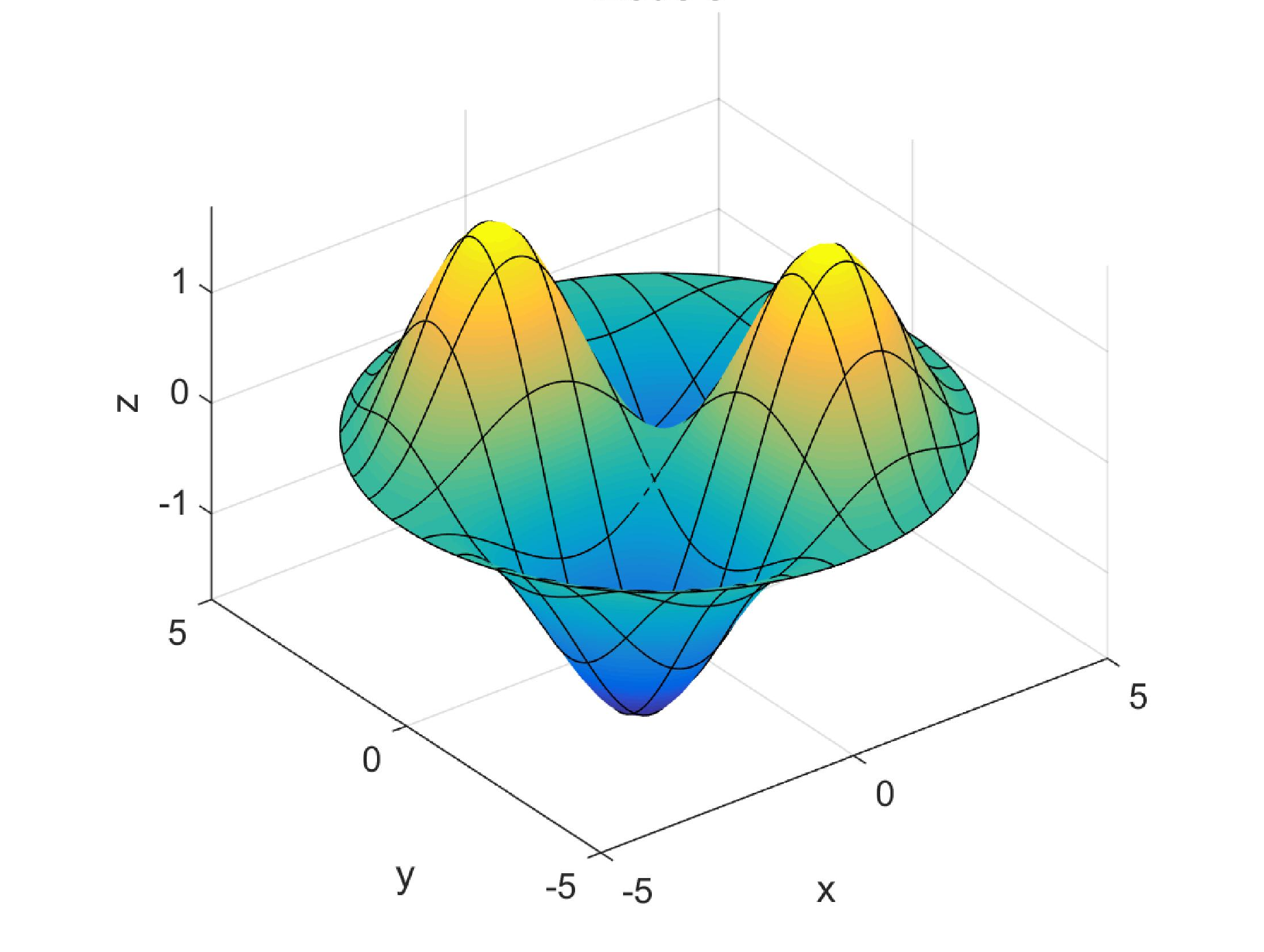}
		\caption{$\bar{\omega}_5 = 21.7503$.}
	\end{subfigure}
	\begin{subfigure}{0.49\textwidth}
		\includegraphics[width=\linewidth]{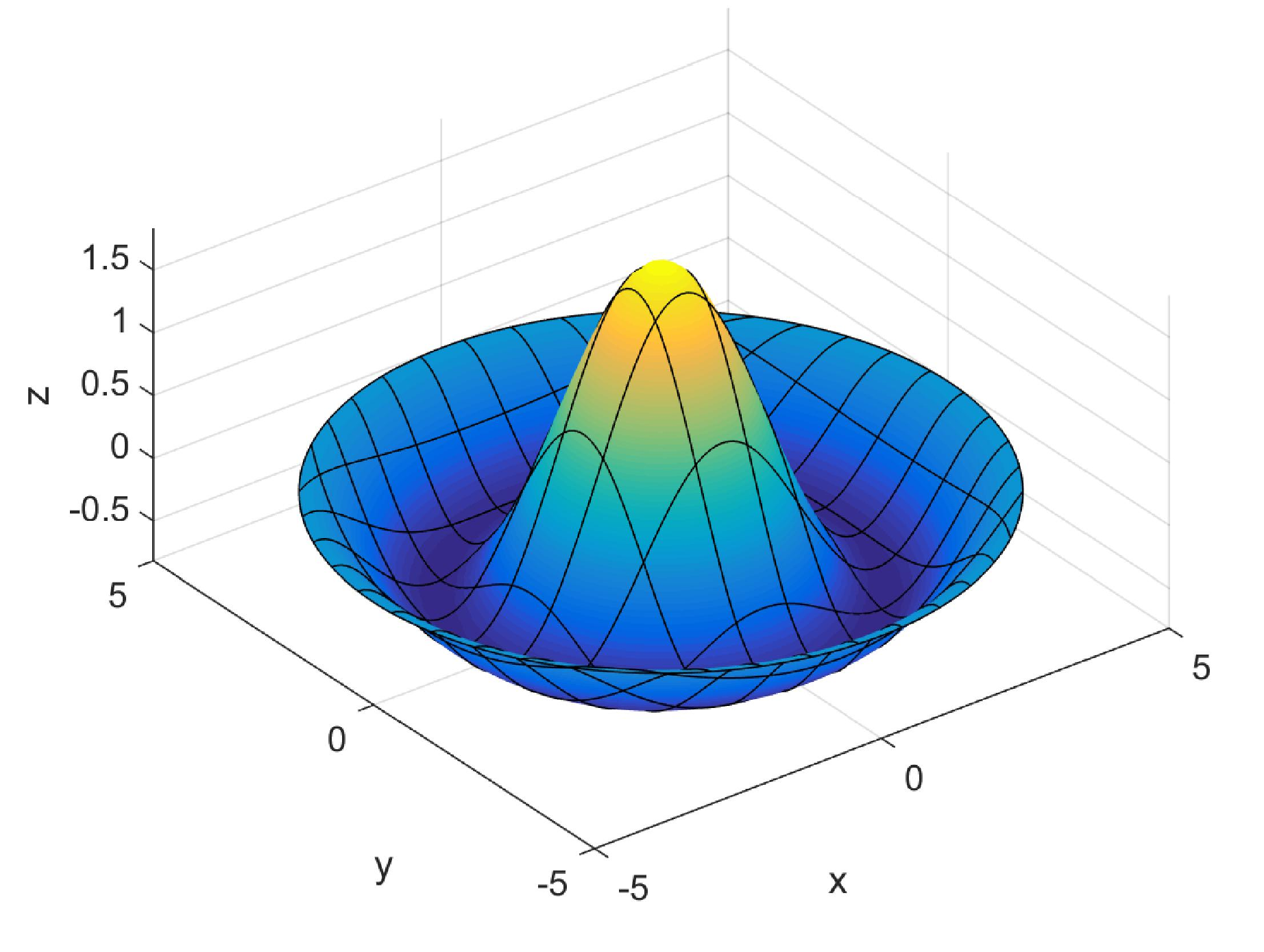}
		\caption{$\bar{\omega}_6 = 23.9885$.}
	\end{subfigure}
	\caption{The first six free vibration mode shapes of clamped Al/Al$_2$O$_3$ circular microplates. \label{fig:exp:vib:cir_6}}
\end{figure}

Next, free vibration of circular plates whose geometry configuration and mesh are shown in Fig. \ref{fig:exp:cir:geometry_mesh} will be investigated. Since there is no publication on the vibration behaviours of FG circular microplates based on the modified couple stress theory is found, the investigation of circulars could be served as benchmark examples. Table \ref{tab:exp:vib:cir:isotropic} presents the fundamental natural frequencies which are firstly tested for homogeneous plates. Natural frequencies of plates without considering size-dependent effects $(l=0)$ are compared with results reported by Mohammandi et al. \cite{Mohammadi201332} and Nguyen et al. \cite{Nguyen2015191}. The results show very good agreement between the theories, especially proposed RPT and Nguyen et al.'s \cite{Nguyen2015191} which also uses another polynomial-based RPT model. Table \ref{tab:exp:vib:cir:FGM} presents the first six natural frequencies of Al/Al$_2$O$_3$ circular plates with simple and clamped supports. The plates' thickness $h$ and material index $n$ are set as $0.2R$ and 1, respectively. As previous case of square microplates, Zenkour's quasi-3D model \cite{Zenkour20139041} using proposed IGA approach is added for reference purpose since there is no study can be found for this analysis problem. The six fundamental mode shapes of clamped plates with $l/h = 0.2$ using quasi-3D model are printed in Fig. \ref{fig:exp:vib:cir_6}.

% Table generated by Excel2LaTeX from sheet 'Vibration'
\begin{table}[htbp]
  \centering
  \caption{The first six non-dimensional natural frequencies $\bar{\omega} = \omega R^2 \sqrt{\dfrac{\rho h}{D^\dagger}}  $ of homogeneous circular microplates}
    \begin{tabular}{llllllll}
    \hline
    $l/h$ & Theory & \multicolumn{6}{l}{Mode} \\
\cline{3-8}          &       & 1     & 2     & 3     & 4     & 5     & 6 \\
    \hline
   \multicolumn{8}{l}{Simple support} \\
    0     & Mohammadi et al. \cite{Mohammadi201332} & 4.9345 & 13.8981 & 25.6132 & 29.7198 & 39.9571 & 48.4788 \\
          & Nguyen et al. \cite{Nguyen2015191} & 4.9304 & 13.8587 & 25.4798 & 29.5390 & 39.6331 & 48.0046 \\
          & RPT (Present) & 4.9304 & 13.8591 & 25.4799 & 29.5456 & 39.6518 & 48.0402 \\
          & Quasi-3D (Present) & 4.9385 & 13.8701 & 25.4983 & 29.5691 & 39.6881 & 48.0906 \\
    0.2   & RPT (Present) & 4.9925 & 14.5095 & 26.5426 & 31.1786 & 41.8855 & 50.8427 \\
          & Quasi-3D (Present) & 5.0024 & 14.5206 & 26.5613 & 31.1981 & 41.9148 & 50.8808 \\
    0.4   & RPT (Present) & 5.1213 & 16.2743 & 29.4369 & 35.5996 & 47.7406 & 58.5547 \\
          & Quasi-3D (Present) & 5.1365 & 16.2857 & 29.4529 & 35.6092 & 47.7542 & 58.5621 \\
    0.6   & RPT (Present) & 5.2422 & 18.8078 & 33.6370 & 41.9096 & 55.8684 & 69.6631 \\
          & Quasi-3D (Present) & 5.2649 & 18.8192 & 33.6497 & 41.9065 & 55.8623 & 69.6331 \\
    0.8   & RPT (Present) & 5.3324 & 21.8342 & 38.6993 & 49.3881 & 65.3972 & 82.8177 \\
          & Quasi-3D (Present) & 5.3642 & 21.8450 & 38.7080 & 49.3713 & 65.3694 & 82.7497 \\
    1     & RPT (Present) & 5.3954 & 25.1769 & 44.3292 & 57.5842 & 75.8223 & 97.1772 \\
          & Quasi-3D (Present) & 5.4379 & 25.1869 & 44.3335 & 57.5538 & 75.7721 & 97.0720 \\ \\
    \multicolumn{8}{l}{Clamped support} \\
    0     & Mohammadi et al. \cite{Mohammadi201332} & 10.2158 & 21.2604 & 34.8772 & 39.7706 & 51.0295 & 60.8290 \\
          & Nguyen et al. \cite{Nguyen2015191} & 10.1839 & 21.1433 & 34.5892 & 39.3624 & 50.4385 & 59.9580 \\
          & RPT (Present) & 10.1842 & 21.1459 & 34.5885 & 39.3832 & 50.4865 & 60.0416 \\
          & Quasi-3D (Present) & 10.4466 & 21.6458 & 35.2774 & 40.2833 & 51.5045 & 61.3186 \\
    0.2   & RPT (Present) & 10.8087 & 22.4449 & 36.3961 & 41.8103 & 53.5802 & 63.7743 \\
          & Quasi-3D (Present) & 11.0612 & 22.9236 & 37.1800 & 42.6664 & 54.5550 & 64.9738 \\
    0.4   & RPT (Present) & 12.4963 & 25.9527 & 41.2953 & 48.3406 & 61.7419 & 73.9729 \\
          & Quasi-3D (Present) & 12.7255 & 26.3811 & 41.9956 & 49.0934 & 62.6149 & 74.9922 \\
    0.6   & RPT (Present) & 14.8897 & 30.9242 & 48.3696 & 57.5674 & 73.1139 & 88.5334 \\
          & Quasi-3D (Present) & 15.0927 & 31.2968 & 48.9732 & 58.2073 & 73.8637 & 89.3634 \\
    0.8   & RPT (Present) & 17.7051 & 36.7699 & 56.8123 & 68.4011 & 86.4224 & 105.6689 \\
          & Quasi-3D (Present) & 17.8850 & 37.0938 & 57.3265 & 68.9439 & 87.0506 & 106.3411 \\
    1     & RPT (Present) & 20.7715 & 43.1360 & 66.1009 & 80.1939 & 100.9243 & 124.3039 \\
          & Quasi-3D (Present) & 20.9330 & 43.4206 & 66.5388 & 80.6591 & 101.4425 & 124.8498 \\
    \hline
    \multicolumn{8}{l}{$^\dagger D = \dfrac{Eh^3}{12(1-\nu^2)}$}
    \end{tabular}
  \label{tab:exp:vib:cir:isotropic}
\end{table}

% Table generated by Excel2LaTeX from sheet 'Vibration'
\begin{table}[htbp]
  \centering
  \caption{The first six non-dimensional natural frequencies $\bar{\omega} = \omega R^2 \sqrt{\dfrac{\rho_c h}{D_c^\dagger}}  $ of Al/Al$_2$O$_3$ circular microplates (Mori-Tanaka scheme)}
    \begin{tabular}{llllllll}
    \hline
    $l/h$ & Theory & \multicolumn{6}{l}{Mode} \\
\cline{3-8}          &       & 1     & 2     & 3     & 4     & 5     & 6\\
    \hline
  \multicolumn{8}{l}{Simple support}\\
    0     & IGA-Zenkour& 3.4629 & 8.9254 & 8.9254 & 12.9890 & 12.9890 & 15.4780 \\
          & Quasi-3D (Present) & 3.4132 & 8.8258 & 8.8258 & 12.9916 & 12.9916 & 15.3331 \\
          & RPT (Present) & 3.3572 & 8.6722 & 8.6722 & 12.9121 & 12.9121 & 15.0490 \\
    0.2   & IGA-Zenkour& 3.5224 & 9.3961 & 9.3961 & 13.1894 & 13.1894 & 16.2612 \\
          & Quasi-3D (Present) & 3.4675 & 9.2928 & 9.2928 & 13.1928 & 13.1928 & 16.1022 \\
          & RPT (Present) & 3.4118 & 9.1595 & 9.1595 & 13.1037 & 13.1037 & 15.8706 \\
    0.4   & IGA-Zenkour& 3.6392 & 10.6411 & 10.6411 & 13.3115 & 13.3115 & 18.2644 \\
          & Quasi-3D (Present) & 3.5698 & 10.5288 & 10.5288 & 13.3143 & 13.3143 & 18.1203 \\
          & RPT (Present) & 3.5114 & 10.4415 & 10.4415 & 13.2326 & 13.2326 & 18.0361 \\
    0.6   & IGA-Zenkour& 3.7577 & 12.2959 & 12.2959 & 13.5601 & 13.5601 & 18.4956 \\
          & Quasi-3D (Present) & 3.6687 & 12.1873 & 12.1873 & 13.5445 & 13.5445 & 18.4939 \\
          & RPT (Present) & 3.6018 & 12.1378 & 12.1378 & 13.4872 & 13.4872 & 18.4475 \\
    0.8   & IGA-Zenkour& 3.8632 & 13.2220 & 13.2220 & 14.8766 & 14.8766 & 18.6662 \\
          & Quasi-3D (Present) & 3.7546 & 13.2030 & 13.2030 & 14.7552 & 14.7552 & 18.6655 \\
          & RPT (Present) & 3.6721 & 13.1245 & 13.1245 & 14.7850 & 14.7850 & 18.6192 \\
    1     & IGA-Zenkour& 3.9548 & 13.4288 & 13.4288 & 17.1268 & 17.1268 & 18.8623 \\
          & Quasi-3D (Present) & 3.8303 & 13.4243 & 13.4243 & 16.9751 & 16.9751 & 18.8618 \\
          & RPT (Present) & 3.9557 & 13.5531 & 13.5531 & 17.4735 & 17.4735 & 18.8633 \\ \\
    \multicolumn{8}{l}{Clamped support} \\
    0     & IGA-Zenkour& 6.7718 & 12.9968 & 12.9968 & 19.8267 & 19.8608 & 21.9673 \\
          & Quasi-3D (Present) & 6.8745 & 13.1770 & 13.1770 & 20.0202 & 20.0388 & 22.2840 \\
          & RPT (Present) & 6.3384 & 12.4133 & 12.4133 & 19.0879 & 19.1363 & 20.9877 \\
    0.2   & IGA-Zenkour& 7.2163 & 13.9289 & 13.9289 & 21.1653 & 21.5889 & 23.7182 \\
          & Quasi-3D (Present) & 7.3195 & 14.0950 & 14.0950 & 21.3095 & 21.7503 & 23.9885 \\
          & RPT (Present) & 6.8208 & 13.4172 & 13.4172 & 20.5401 & 20.9799 & 22.8902 \\
    0.4   & IGA-Zenkour& 8.4091 & 16.4035 & 16.4035 & 24.6620 & 25.9011 & 25.9011 \\
          & Quasi-3D (Present) & 8.5121 & 16.5390 & 16.5390 & 24.7273 & 25.8504 & 25.8504 \\
          & RPT (Present) & 8.0909 & 16.0391 & 16.0391 & 24.2526 & 24.3345 & 24.3345 \\
    0.6   & IGA-Zenkour& 10.0860 & 19.8463 & 19.8463 & 25.9424 & 25.9424 & 28.6525 \\
          & Quasi-3D (Present) & 10.1880 & 19.9431 & 19.9431 & 25.8948 & 25.8948 & 28.6525 \\
          & RPT (Present) & 9.8463 & 19.6321 & 19.6321 & 24.3655 & 24.3655 & 28.6525 \\
    0.8   & IGA-Zenkour& 12.0447 & 23.8379 & 23.8379 & 25.9920 & 25.9920 & 28.9380 \\
          & Quasi-3D (Present) & 12.1441 & 23.8850 & 23.8850 & 25.9507 & 25.9507 & 28.9380 \\
          & RPT (Present) & 11.8741 & 23.7016 & 23.7016 & 24.4627 & 24.4627 & 28.9380 \\
    1     & IGA-Zenkour& 14.1674 & 26.0476 & 26.0476 & 28.1435 & 28.1435 & 29.3010 \\
          & Quasi-3D (Present) & 14.2615 & 25.9953 & 25.9953 & 28.1492 & 28.1492 & 29.3010 \\
          & RPT (Present) & 14.0568 & 24.4150 & 24.4150 & 28.2217 & 28.2217 & 29.3010 \\

    \hline
    \multicolumn{8}{l}{$^\dagger D_c = \dfrac{E_ch^3}{12(1-\nu_c^2)}$}
    \end{tabular}
  \label{tab:exp:vib:cir:FGM}
\end{table}

\subsection{Buckling analysis}
\label{sec:Exp:Buckling} 

In this stage, buckling behaviours of square and circular FG microplates will be discussed. In order to verify the proposed method and models in dealing with buckling analysis, first attempt is to calculate the critical buckling load of SSSS FG microplates bearing biaxial load. The results are compared with analytical solutions based on CPT and FSDT by Thai and Choi \cite{Thai2013142} and RPT by He et al. \cite{He2015107} for which material properties are $E_1 = 14.4 $GPa, $\rho_1 = 12.2 \times 10^3 $kg/m$^3$, $E_2 = 1.44 $GPa, $\rho_2 = 1.22 \times 10^3 $kg/m$^3$, $\nu_1 = \nu_2 = 0.38$. As can be seen in Table \ref{tab:exp:buck:sq:comparison}, although there is discrepancy for the case of thick microplates $(a/h = 5)$, the results predicted by proposed RPT theory are in good agreement with those calculated by FSDT \cite{Thai2013142} and RPT \cite{He2015107} as plates become thinner. On the other hand, while CPT-based solutions are of remarkable difference to that of FDST and RPTs due to the ignorance of shear deformations, especially for thick plates, the proposed quasi-3D theory which takes into account normal deformation yields similar results with respect to those of shear deformable theories of FSDT and RPTs.

% Table generated by Excel2LaTeX from sheet 'Buckling'
\begin{table}[htbp]
%  \centering
  \begin{adjustwidth}{-2cm}{}
  \caption{Comparison of non-dimensional critical buckling load $\bar{P}_{cr} = \dfrac{P_{cr} a^2}{E_2h^3} $ of square FG microplates (rule of mixtures scheme)}
    \begin{tabular}{lllllllllllll}
    \hline
    $l/h$   & Theory & \multicolumn{3}{l}{$a/h=5$} &  & \multicolumn{3}{l}{$a/h=10$} &       & \multicolumn{3}{l}{$a/h=20$}\\
\cline{3-5}\cline{7-9}\cline{11-13}          &        & $n=0$   & $n=1$   & $n=10$  &       & $n=0$   & $n=1$   & $n=10$  &       & $n=0$   & $n=1$   & $n=10$ \\
    \hline
    0     & CPT \cite{Thai2013142}  & 19.2255 & 8.2145 & 3.8359 &       & 19.2255 & 8.2145 & 3.8359 &       & 19.2255 & 8.2145 & 3.8359 \\
          & FSDT \cite{Thai2013142}  & 15.3228 & 6.8576 & 2.9979 &       & 18.0746 & 7.8273 & 3.5853 &       & 18.9243 & 8.1142 & 3.7700 \\
          & RPT \cite{He2015107} & 15.3322 & 6.8611 & 2.7672 &       & 18.0754 & 7.8276 & 3.4969 &       & 18.9243 & 8.1142 & 3.7450 \\
          & RPT$^\dagger$   & 15.3321 & 6.8610 & 2.7702 &       & 18.0756 & 7.8277 & 3.4982 &       & 18.9244 & 8.1143 & 3.7454 \\
          & Quasi-3D$^\dagger$ & 15.3629 & 7.3905 & 3.0118 &       & 18.1561 & 8.5396 & 3.8921 &       & 18.9675 & 8.8639 & 4.1850 \\
    0.2   & CPT \cite{Thai2013142}  & 22.0863 & 9.7879 & 4.3560 &       & 22.0863 & 9.7879 & 4.3560 &       & 22.0863 & 9.7879 & 4.3560 \\
          & FSDT \cite{Thai2013142}  & 17.6150 & 8.1715 & 3.4076 &       & 20.7607 & 9.3241 & 4.0710 &       & 21.7387 & 9.6675 & 4.2809 \\
          & RPT \cite{He2015107} & 18.0422 & 8.3399 & 3.3619 &       & 20.9025 & 9.3767 & 4.0513 &       & 21.7771 & 9.6815 & 4.2752 \\
          & RPT$^\dagger$   & 17.8878 & 8.2820 & 3.2917 &       & 20.8497 & 9.3581 & 4.0246 &       & 21.7628 & 9.6766 & 4.2677 \\
          & Quasi-3D$^\dagger$ & 17.7286 & 8.7153 & 3.4728 &       & 20.8583 & 10.0344 & 4.3958 &       & 21.7852 & 10.4160 & 4.7009 \\
    0.4   & CPT \cite{Thai2013142}  & 30.6685 & 14.5082 & 5.9164 &       & 30.6685 & 14.5082 & 5.9164 &       & 30.6685 & 14.5082 & 5.9164 \\
          & FSDT \cite{Thai2013142}  & 24.2899 & 11.9922 & 4.6013 &       & 28.7478 & 13.7742 & 5.5151 &       & 30.1625 & 14.3167 & 5.8102 \\
          & RPT \cite{He2015107} & 26.1539 & 12.7754 & 5.0407 &       & 29.3735 & 14.0232 & 5.6631 &       & 30.3324 & 14.3832 & 5.8505 \\
          & RPT$^\dagger$   & 25.5457 & 12.5322 & 4.8371 &       & 29.1700 & 13.9459 & 5.5925 &       & 30.2773 & 14.3626 & 5.8312 \\
          & Quasi-3D$^\dagger$ & 24.8060 & 12.6741 & 4.8557 &       & 28.9624 & 14.5168 & 5.9066 &       & 30.2381 & 15.0722 & 6.2486 \\
    0.6   & CPT \cite{Thai2013142}  & 44.9723 & 22.3753 & 8.5171 &       & 44.9723 & 22.3753 & 8.5171 &       & 44.9723 & 22.3753 & 8.5171 \\
          & FSDT \cite{Thai2013142}  & 34.7856 & 17.9838 & 6.4804 &       & 41.8271 & 21.0597 & 7.8802 &       & 44.1369 & 22.0292 & 8.3472 \\
          & RPT \cite{He2015107} & 39.6393 & 20.1658 & 7.7001 &       & 43.4732 & 21.7657 & 8.2906 &       & 44.5855 & 22.2188 & 8.4589 \\
          & RPT$^\dagger$   & 38.2867 & 19.5858 & 7.3772 &       & 43.0329 & 21.5846 & 8.1871 &       & 44.4673 & 22.1708 & 8.4312 \\
          & Quasi-3D$^\dagger$ & 36.5415 & 19.2256 & 7.1597 &       & 42.4620 & 21.9814 & 8.4246 &       & 44.3258 & 22.8320 & 8.8281 \\
    0.8   & CPT \cite{Thai2013142}  & 64.9976 & 33.3892 & 12.1581 &       & 64.9976 & 33.3892 & 12.1581 &       & 64.9976 & 33.3892 & 12.1581 \\
          & FSDT \cite{Thai2013142}  & 48.2915 & 25.6654 & 8.9020 &       & 59.6657 & 30.9928 & 11.1065 &       & 63.5656 & 32.7517 & 11.8745 \\
          & RPT \cite{He2015107} & 58.4862 & 30.5105 & 11.3322 &       & 63.1958 & 32.6036 & 11.9349 &       & 64.5348 & 33.1882 & 12.1011 \\
          & RPT$^\dagger$   & 56.0961 & 29.4240 & 10.9005 &       & 62.4358 & 32.2693 & 11.8036 &       & 64.3321 & 33.0999 & 12.0666 \\
          & Quasi-3D$^\dagger$ & 52.8623 & 28.3151 & 10.3843 &       & 61.3467 & 32.4199 & 11.9498 &       & 64.0474 & 33.6948 & 12.4394 \\
    1     & CPT \cite{Thai2013142}  & 90.7444 & 47.5499 & 16.8393 &       & 90.7444 & 47.5499 & 16.8393 &       & - & - & - \\
          & FSDT \cite{Thai2013142}  & 63.8913 & 34.4981 & 11.7042 &       & 81.8269 & 43.3274 & 15.1152 &       & - & - & - \\
          & RPT \cite{He2015107} & 82.6938 & 43.8094 & 15.9522 &       & 88.5416 & 46.5372 & 16.6033 &       & 90.1804 & 47.2914 & 16.7793 \\
          & RPT$^\dagger$   & 78.9675 & 42.0388 & 15.4071 &       & 87.3775 & 45.9981 & 16.4431 &       & 89.8715 & 47.1494 & 16.7376 \\
          & Quasi-3D$^\dagger$ & 73.6925 & 39.8872 & 14.5287 &       & 85.6043 & 45.8223 & 16.4819 &       & 89.4018 & 47.6596 & 17.0825 \\
    \hline
    \multicolumn{13}{l}{$^\dagger$Present}
    \end{tabular}
  \label{tab:exp:buck:sq:comparison}
  \end{adjustwidth}
\end{table}

\begin{figure}
	\centering
	\begin{subfigure}{0.49\textwidth}
		\includegraphics[width=\linewidth]{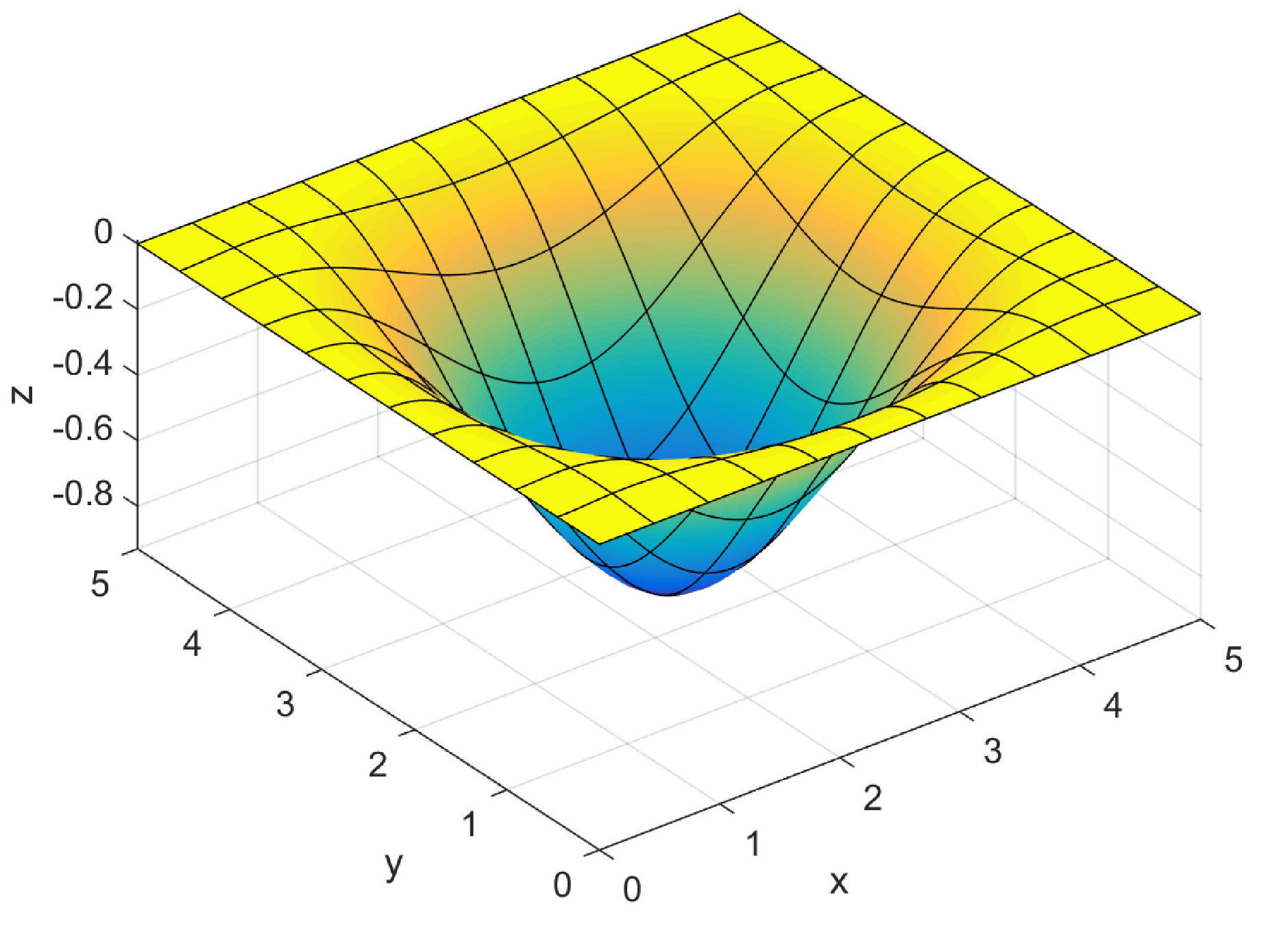}
		\caption{$\bar{P}_1 = 52.1032$.}
	\end{subfigure}
	%\hspace*{\fill}
	\begin{subfigure}{0.49\textwidth}
		\includegraphics[width=\linewidth]{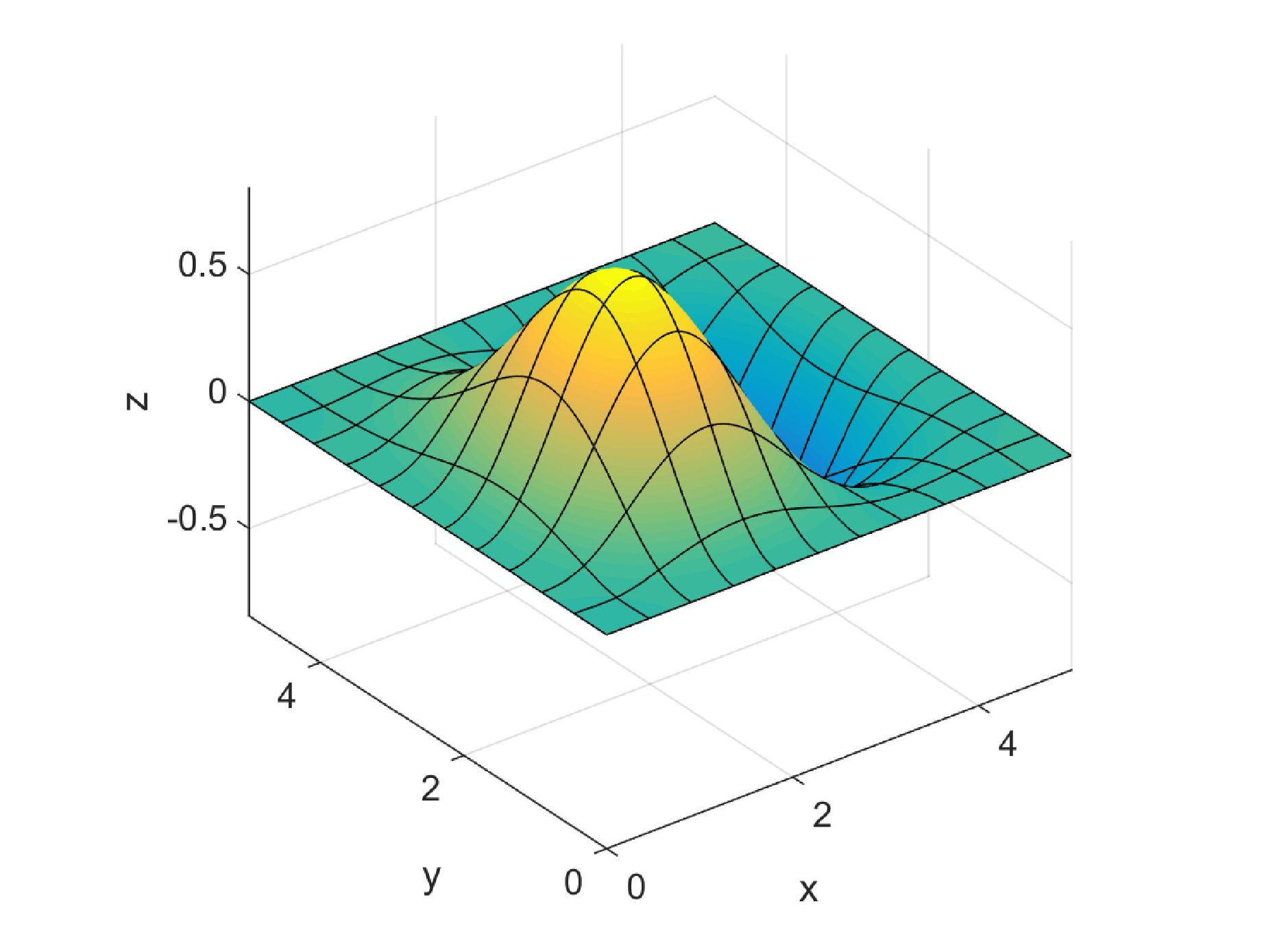}
		\caption{$\bar{P}_2 = 66.3156$.}
	\end{subfigure}
	\begin{subfigure}{0.49\textwidth}
		\includegraphics[width=\linewidth]{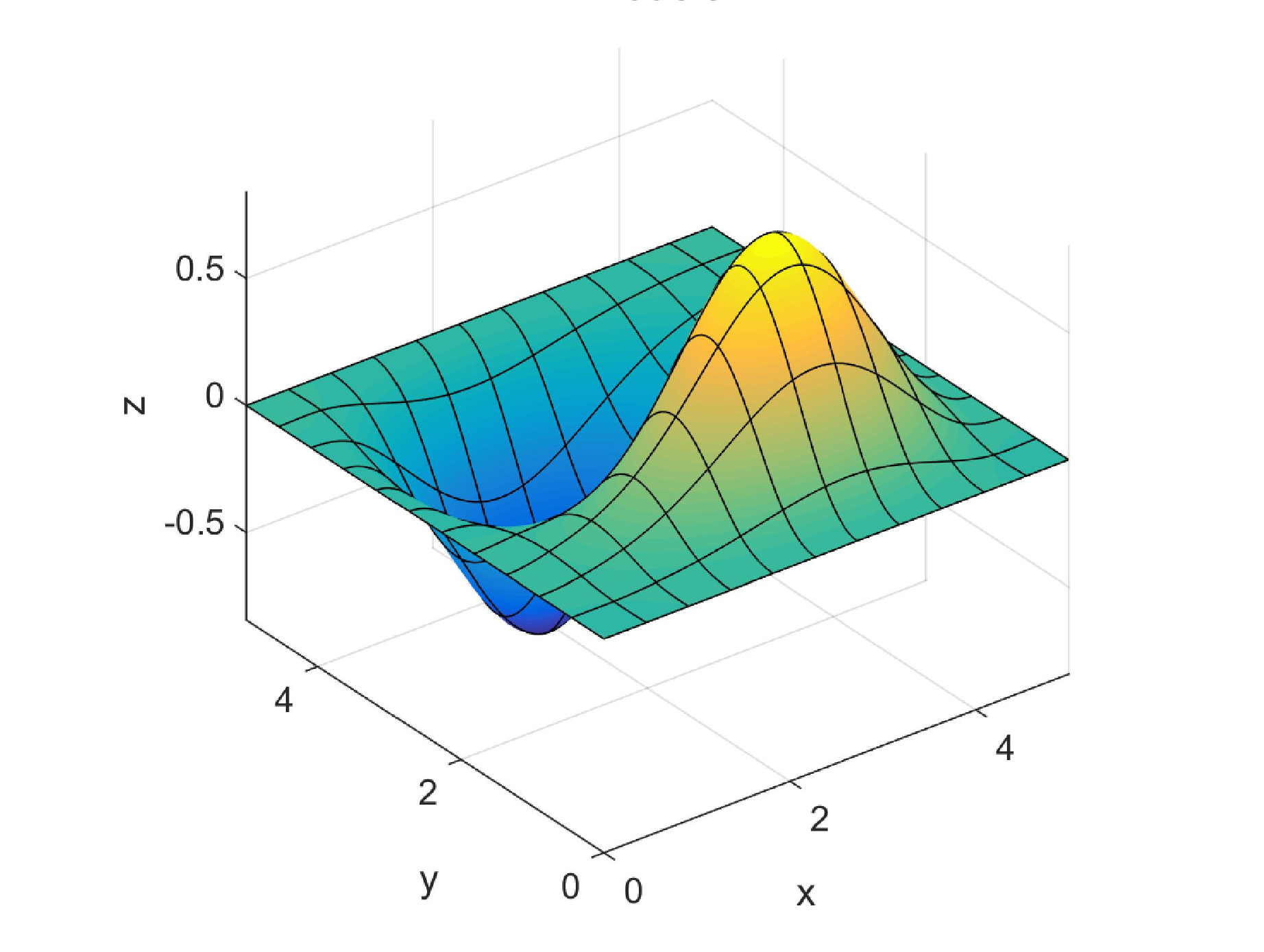}
		\caption{$\bar{P}_3 = 66.3156$.}
	\end{subfigure}
	\begin{subfigure}{0.49\textwidth}
		\includegraphics[width=\linewidth]{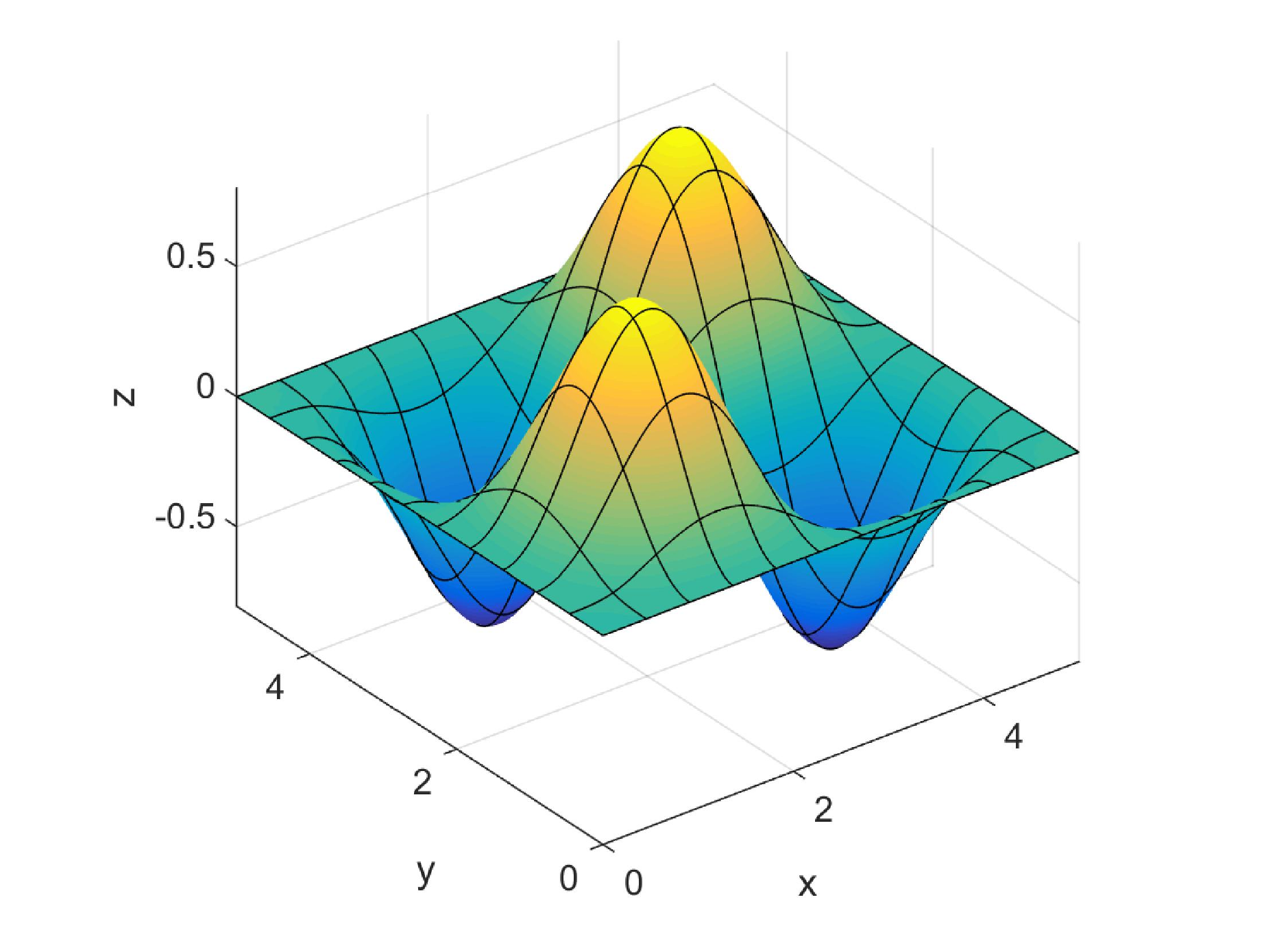}
		\caption{$\bar{P}_4 = 77.2179$.}
	\end{subfigure}
	\begin{subfigure}{0.49\textwidth}
		\includegraphics[width=\linewidth]{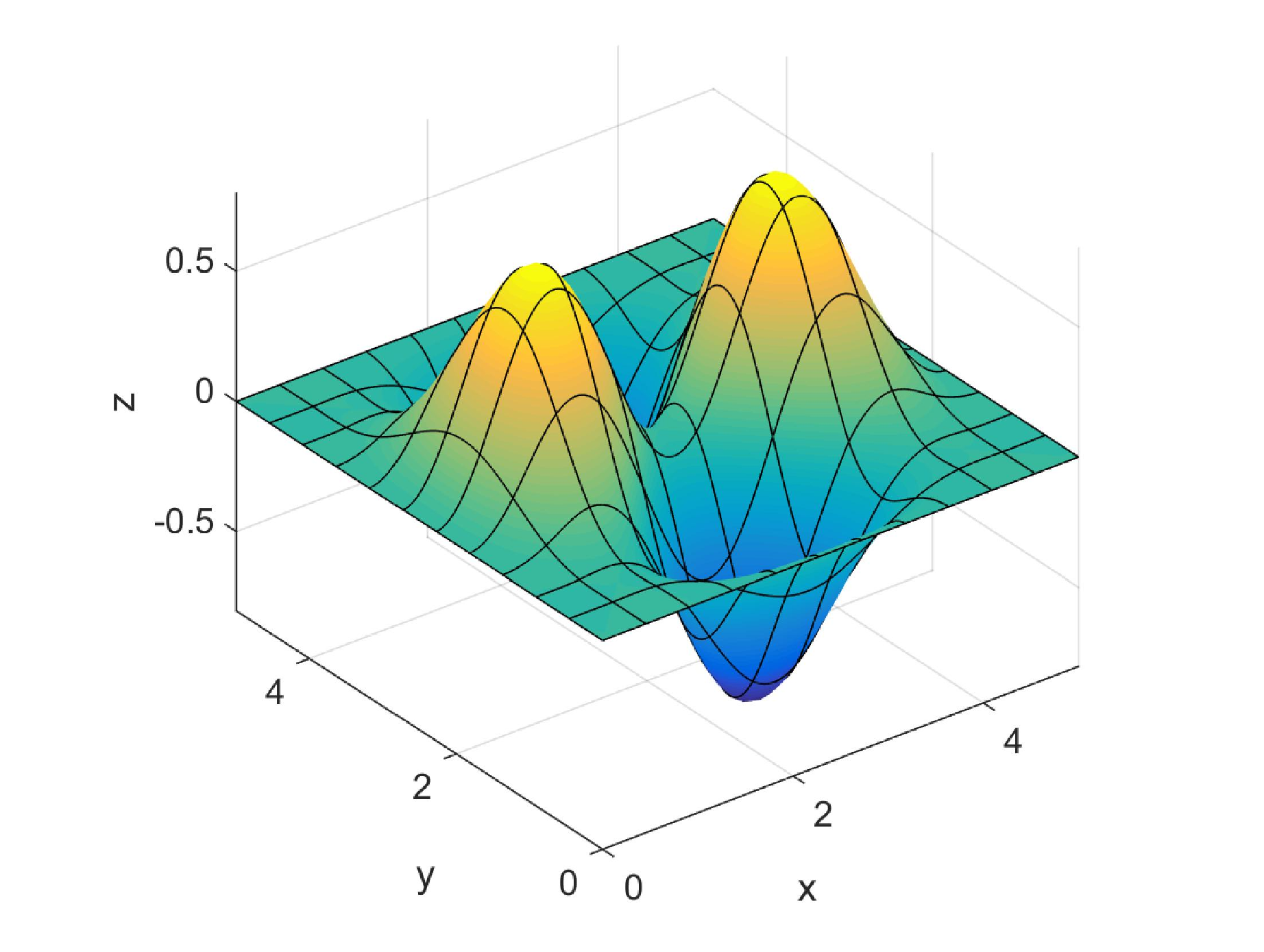}
		\caption{$\bar{P}_5 = 78.9774$.}
	\end{subfigure}
	\begin{subfigure}{0.49\textwidth}
		\includegraphics[width=\linewidth]{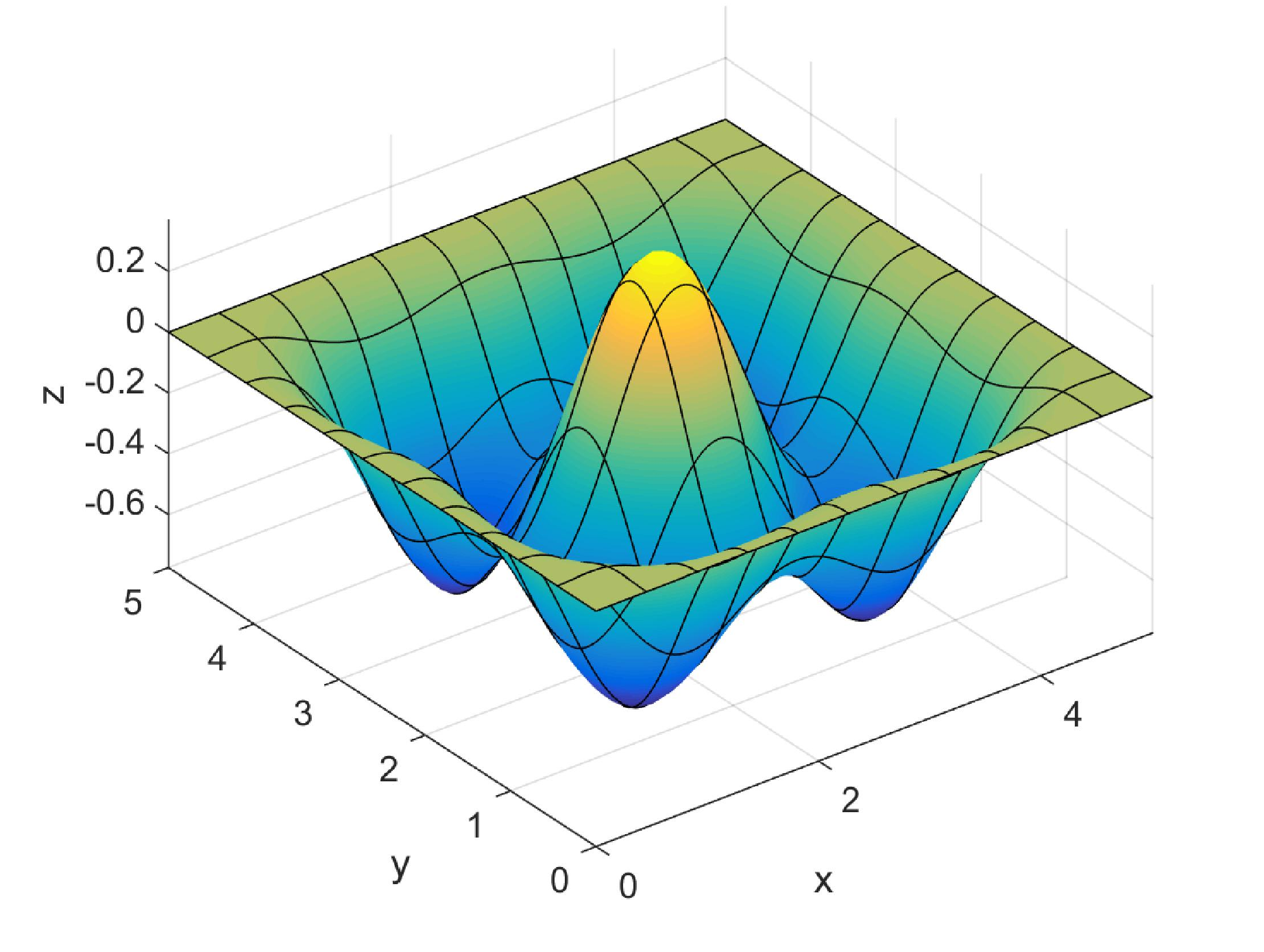}
		\caption{$\bar{P}_6 = 84.1043$.}
	\end{subfigure}
	\caption{The first six buckling mode shapes of Al/Al$_2$O$_3$ CCCC square microplates. \label{fig:exp:buck:sq:6modes}}
\end{figure}

Table \ref{tab:exp:buck:sq:Al2O3} presents the biaxial buckling analysis results of Al/Al$_2$O$_3$ square microplates. The results are calculated based on the proposed RPT and quasi-3D theories which could be served as benchmark examples for future references since no report in the similar case is found. The first six non-dimensional biaxial buckling loads of Al/Al$_2$O$_3$ square plates are reported in Table \ref{tab:exp:buckling:sq:6modes} for $n=10$ and $l/h=0.2$. While the results generated from the proposed quasi-3D and IGA-based Zenkour's quasi-3D theories \cite{Zenkour20139041} are relatively close to each other for both type of boundary conditions, the RPT's show remarkable discrepancy to the other theories, especially for the CCCC plates. This is due to the consideration of normal deformation of the quasi-3D theories. For $a/h=5$, six buckling mode shapes of CCCC plates based on the proposed quasi-3D theory which are scaled up for illustration purposes are presented in Fig. \ref{fig:exp:buck:sq:6modes}.

% Table generated by Excel2LaTeX from sheet 'Sheet1'
\begin{table}[htbp]
  \centering
  \caption{Non-dimensional critical buckling load $\bar{P}_{cr} = \dfrac{P_{cr} a^2}{D_m^\dagger} $ of Al/Al$_2$O$_3$ square microplates (Mori-Tanaka scheme)}
    \begin{tabular}{lllllllllll}
    \hline
    BC    & $a/h$   & $l/h$ & $n=0$   &       &       & $n=1$   &       &       & $n=10$  &  \\
\cline{4-5}\cline{7-8}\cline{10-11}          &       &       & RPT   & Quasi-3D   &       & RPT   & Quasi-3D   &       & RPT   & Quasi-3D\\
\hline
    SSSS  & 5     & 0     & 87.4747 & 86.5475 &       & 35.0795 & 35.6610 &       & 21.7236 & 21.7189 \\
          &       & 0.2   & 103.6359 & 101.8797 &       & 42.3816 & 42.6185 &       & 25.4909 & 25.2307 \\
          &       & 0.4   & 152.0215 & 147.6312 &       & 64.2668 & 63.3902 &       & 36.7873 & 35.7417 \\
          &       & 0.6   & 232.4351 & 223.1633 &       & 100.6957 & 97.7189 &       & 55.6034 & 53.1891 \\
          &       & 0.8   & 344.7280 & 327.6683 &       & 151.6424 & 145.2904 &       & 81.9336 & 77.4961 \\
          &       & 1     & 488.8378 & 460.3907 &       & 217.0988 & 205.8260 &       & 115.7764 & 108.5941 \\
          & 20    & 0     & 105.6668 & 105.6221 &       & 42.0033 & 43.4065 &       & 27.5182 & 27.8578 \\
          &       & 0.2   & 123.5339 & 123.4102 &       & 49.9504 & 51.3272 &       & 31.5521 & 31.8705 \\
          &       & 0.4   & 177.1295 & 176.7715 &       & 73.7907 & 75.0881 &       & 43.6520 & 43.9085 \\
          &       & 0.6   & 266.4425 & 265.6972 &       & 113.5226 & 114.6856 &       & 63.8152 & 63.9707 \\
          &       & 0.8   & 391.4648 & 390.1733 &       & 169.1451 & 170.1141 &       & 92.0401 & 92.0558 \\
          &       & 1     & 552.1931 & 550.1813 &       & 240.6580 & 241.3659 &       & 128.3267 & 128.1618 \\
          & 100   & 0     & 107.0958 & 107.1271 &       & 42.5423 & 44.0165 &       & 27.9978 & 28.3566 \\
          &       & 0.2   & 125.0926 & 125.1206 &       & 50.5370 & 52.0102 &       & 32.0486 & 32.4065 \\
          &       & 0.4   & 179.0825 & 179.1009 &       & 74.5212 & 75.9913 &       & 44.2009 & 44.5563 \\
          &       & 0.6   & 269.0653 & 269.0682 &       & 114.4947 & 115.9596 &       & 64.4546 & 64.8061 \\
          &       & 0.8   & 395.0406 & 395.0223 &       & 170.4576 & 171.9154 &       & 92.8097 & 93.1557 \\
          &       & 1     & 557.0082 & 556.9633 &       & 242.4098 & 243.8584 &       & 129.2661 & 129.6052 \\ \\
    CCCC  & 5     & 0     & 178.2578 & 188.3478 &       & 72.2150 & 78.0421 &       & 42.1220 & 45.4860 \\
          &       & 0.2   & 206.9297 & 215.7331 &       & 85.4028 & 90.6731 &       & 49.1101 & 52.1032 \\
          &       & 0.4   & 292.6287 & 295.5433 &       & 124.8630 & 127.4320 &       & 69.9984 & 71.1001 \\
          &       & 0.6   & 435.0030 & 424.8837 &       & 190.5274 & 187.0650 &       & 104.7011 & 101.8084 \\
          &       & 0.8   & 633.8611 & 601.6084 &       & 282.3839 & 268.7548 &       & 153.1709 & 144.0664 \\
          &       & 1     & 889.1171 & 823.8983 &       & 400.4310 & 371.8457 &       & 215.3906 & 197.7752 \\
          & 20    & 0     & 273.9507 & 288.6976 &       & 109.0237 & 117.3906 &       & 70.8926 & 75.2177 \\
          &       & 0.2   & 308.7803 & 323.9291 &       & 124.5481 & 133.0578 &       & 78.8262 & 83.2085 \\
          &       & 0.4   & 413.0833 & 429.0109 &       & 171.0330 & 179.8054 &       & 102.5575 & 107.0378 \\
          &       & 0.6   & 586.6304 & 603.0356 &       & 248.3764 & 257.2793 &       & 141.9833 & 146.4815 \\
          &       & 0.8   & 829.3607 & 845.5938 &       & 356.5558 & 365.3387 &       & 197.0586 & 201.4295 \\
          &       & 1     & 1141.2885 & 1156.5929 &       & 495.5793 & 503.9582 &       & 267.7793 & 271.8584 \\
          & 100   & 0     & 283.7646 & 292.2008 &       & 112.7294 & 119.3117 &       & 74.1644 & 77.0743 \\
          &       & 0.2   & 319.1326 & 327.9091 &       & 128.4414 & 135.1344 &       & 82.1289 & 85.1055 \\
          &       & 0.4   & 425.0550 & 434.6081 &       & 175.4917 & 182.4322 &       & 105.9838 & 109.1182 \\
          &       & 0.6   & 601.2918 & 611.6685 &       & 253.7742 & 260.9696 &       & 145.6726 & 148.9836 \\
          &       & 0.8   & 847.7694 & 858.8291 &       & 363.2623 & 370.6627 &       & 201.1735 & 204.6383 \\
          &       & 1     & 1164.4998 & 1176.0787 &       & 503.9643 & 511.5167 &       & 272.4865 & 276.0730 \\
    \hline
    \multicolumn{11}{l}{$^\dagger D_m = \dfrac{E_m h^3}{12 (1-\nu_m^2)}$}
    \end{tabular}
  \label{tab:exp:buck:sq:Al2O3}
\end{table}

% Table generated by Excel2LaTeX from sheet 'Buckling'
\begin{table}[htbp]
  \centering
  \caption{The first six non-dimensional buckling loads $\bar{P} = \dfrac{P a^2}{D_m} $ of Al/Al$_2$O$_3$ square microplates, $n=10$, $l/h=0.2$ (Mori-Tanaka scheme)}
    \begin{tabular}{lllllllll}
    \hline
    BC    & $a/h$   & Theory & \multicolumn{6}{l}{Mode} \\
\cline{4-9}          &       &       & 1     & 2     & 3     & 4     & 5     & 6 \\
    \hline
    SSSS  & 5     & IGA-Zenkour& 25.4064 & 47.0158 & 47.0158 & 62.7947 & 66.4706 & 66.4706 \\
          &       & Quasi-3D (Present) & 25.2307 & 46.2828 & 46.2828 & 61.4689 & 64.8256 & 64.8256 \\
          &       & RPT (Present) & 25.4909 & 47.9306 & 47.9306 & 64.7895 & 68.7456 & 68.7456 \\
          & 10    & IGA-Zenkour& 30.3554 & 67.5835 & 67.5835 & 101.6321 & 115.5866 & 115.5866 \\
          &       & Quasi-3D (Present) & 30.2890 & 67.2254 & 67.2254 & 100.9300 & 114.4706 & 114.4706 \\
          &       & RPT (Present) & 30.1012 & 67.3570 & 67.3570 & 101.9693 & 116.1512 & 116.1512 \\
          & 20    & IGA-Zenkour & 31.8815 & 76.0319 & 76.0319 & 121.4313 & 143.5095 & 143.5095 \\
          &       & Quasi-3D (Present) & 31.8705 & 75.8988 & 75.8988 & 121.1662 & 143.0021 & 143.0021 \\
          &       & RPT (Present) & 31.5521 & 75.2628 & 75.2628 & 120.4131 & 142.2468 & 142.2468 \\
          & 100   & IGA-Zenkour & 32.3844 & 79.0722 & 79.0722 & 129.2976 & 155.1773 & 155.1773 \\
          &       & Quasi-3D (Present) & 32.4065 & 79.1198 & 79.1198 & 129.3662 & 155.2457 & 155.2457 \\
          &       & RPT (Present) & 32.0486 & 78.2296 & 78.2296 & 127.9519 & 153.4678 & 153.4678 \\
    CCCC  & 5     & IGA-Zenkour & 50.6557 & 66.0350 & 66.0350 & 78.0114 & 80.2597 & 84.0819 \\
          &       & Quasi-3D (Present) & 52.1032 & 66.3156 & 66.3156 & 77.2179 & 78.9774 & 84.1043 \\
          &       & RPT (Present) & 49.1101 & 66.7907 & 66.7907 & 80.3088 & 82.9464 & 85.5400 \\
          & 10    & IGA-Zenkour& 74.8665 & 114.2196 & 114.2196 & 146.4170 & 160.1781 & 172.6816 \\
          &       & Quasi-3D (Present) & 75.4661 & 115.7598 & 115.7598 & 147.8696 & 161.3725 & 174.2442 \\
          &       & RPT (Present) & 70.1001 & 110.6293 & 110.6293 & 144.4919 & 158.8413 & 167.4729 \\
          & 20    & IGA-Zenkour& 83.8952 & 140.5084 & 140.5084 & 191.9601 & 219.4306 & 239.0973 \\
          &       & Quasi-3D (Present) & 83.2085 & 140.4610 & 140.4610 & 192.6237 & 220.3776 & 239.1104 \\
          &       & RPT (Present) & 78.8262 & 133.9531 & 133.9531 & 185.0964 & 211.8925 & 227.4324 \\
          & 100   & IGA-Zenkour& 85.2475 & 149.0723 & 149.0723 & 210.8794 & 246.4373 & 267.8293 \\
          &       & Quasi-3D (Present) & 85.1055 & 148.8861 & 148.8861 & 210.7133 & 246.2408 & 267.4391 \\
          &       & RPT (Present) & 82.1289 & 143.8128 & 143.8128 & 203.9109 & 237.9639 & 257.7571 \\
    \hline
    \end{tabular}
  \label{tab:exp:buckling:sq:6modes}
\end{table}

Finally, this section ends up with a number of investigations on buckling behaviours of circular FG microplates. Table \ref{tab:exp:buckling:cir:comparison} presents  the results of critical buckling load of CCCC Al/ZrO$_2$-2 circular plates without considering couple stress effects. It should be noted that, for this particular attempt of comparison purpose, the material volumes are defined as $V_m = \left(0.5-z/h \right)^n$ and $V_c = 1 - V_m $ \cite{Ma200485, Saidi2009110, NguyenXuan2014222, Tran2013368}. The comparison reveals that the results generated from proposed RPT are in good agreement with those of other shear deformation theories even with relatively thick and thick plates. The proposed quasi-3D yields slightly different results in all cases.

\begin{figure}
	\centering
	\begin{subfigure}{0.8\textwidth}
		\includegraphics[width=\linewidth]{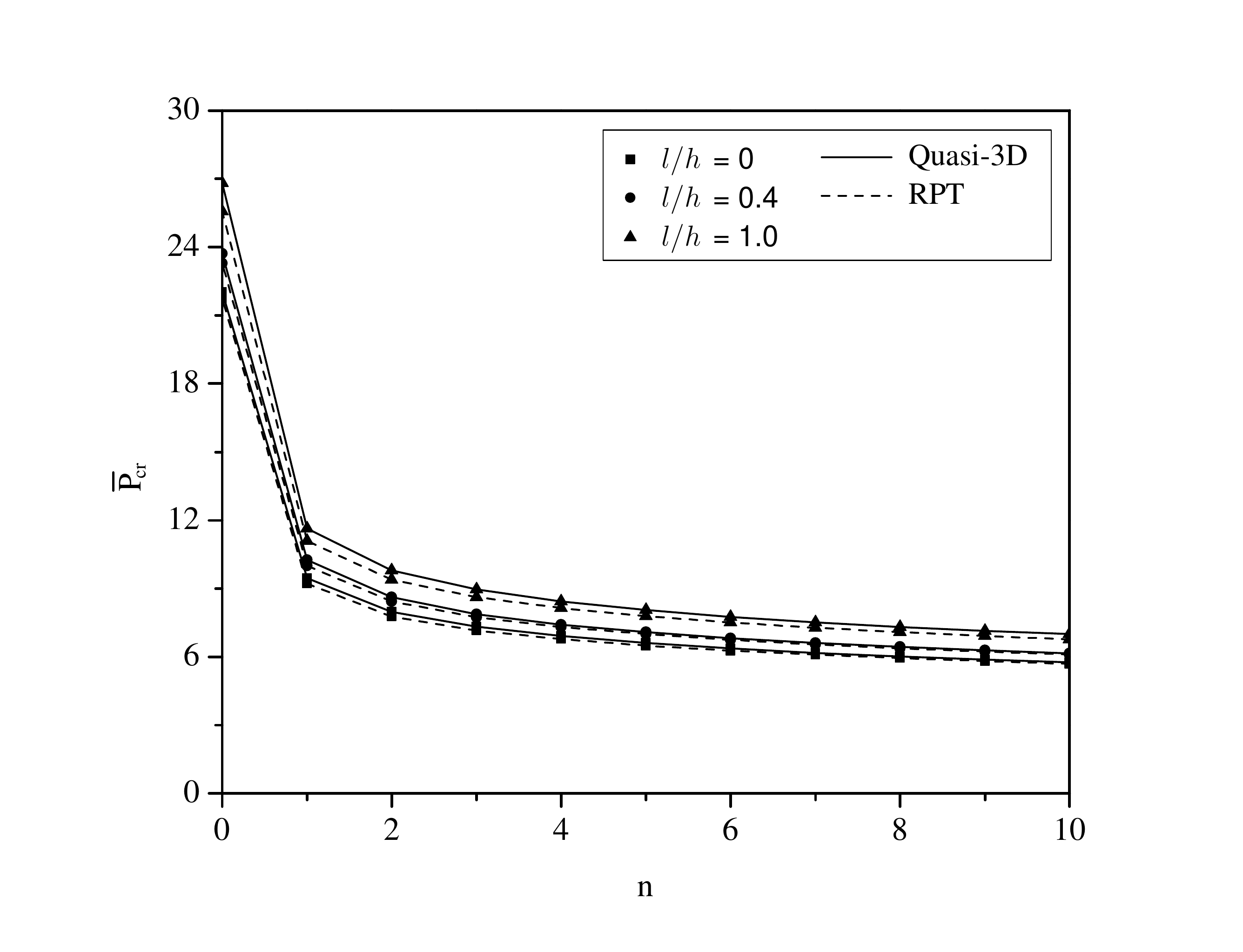}
		\caption{Simply support.}
	\end{subfigure}
	%\hspace*{\fill}
	\begin{subfigure}{0.8\textwidth}
		\includegraphics[width=\linewidth]{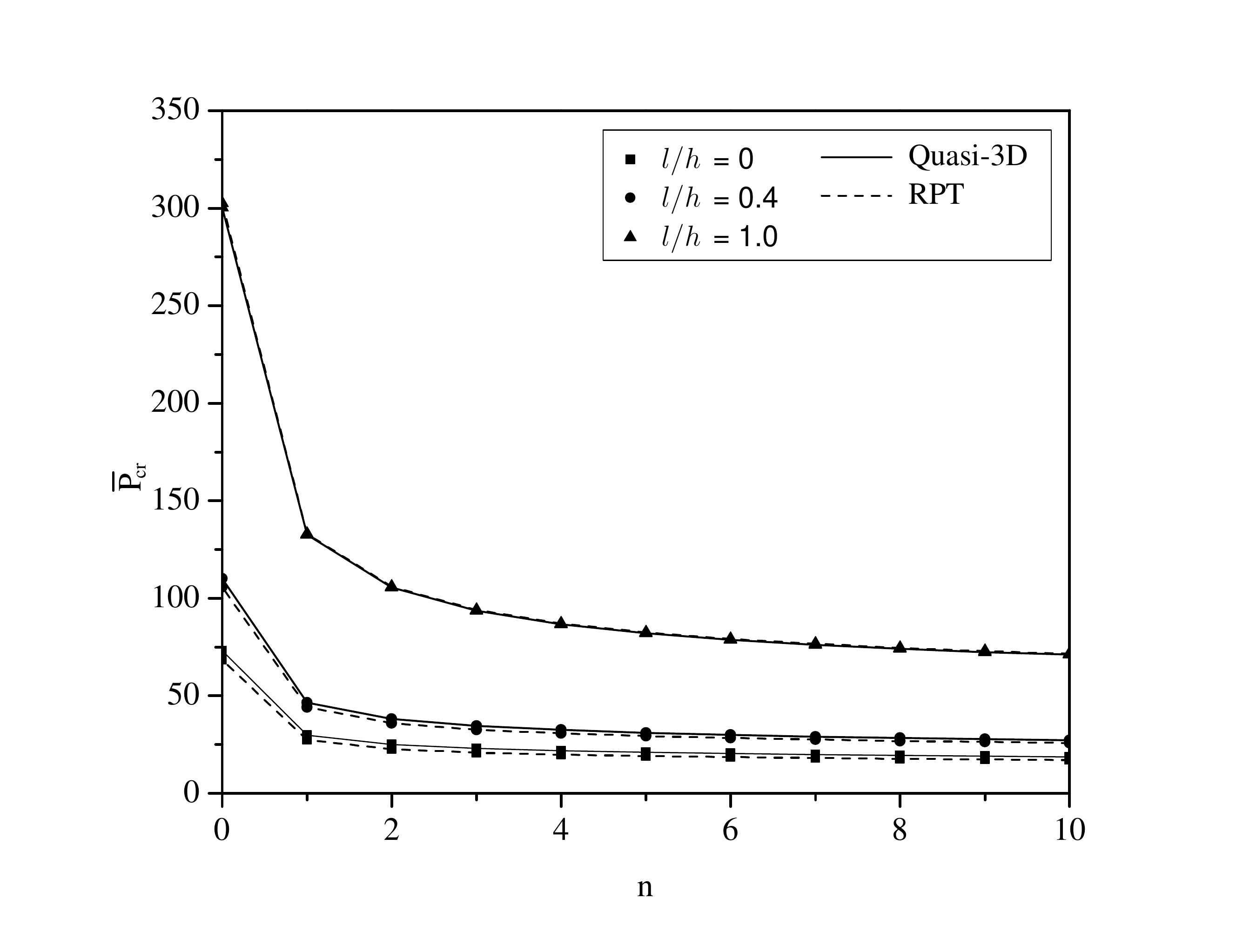}
		\caption{Clamped support.}
	\end{subfigure}
	\caption{Effects of material index $n$ and material length scale ratio $l/h$ on the critical buckling load of Al/Al$_2$O$_3$ circular microplates, $h/R=0.2$ (Mori-Tanaka scheme). \label{fig:exp:buck:cir:effects}}
\end{figure}

\begin{figure}
	\centering
	\begin{subfigure}{0.49\textwidth}
		\includegraphics[width=\linewidth]{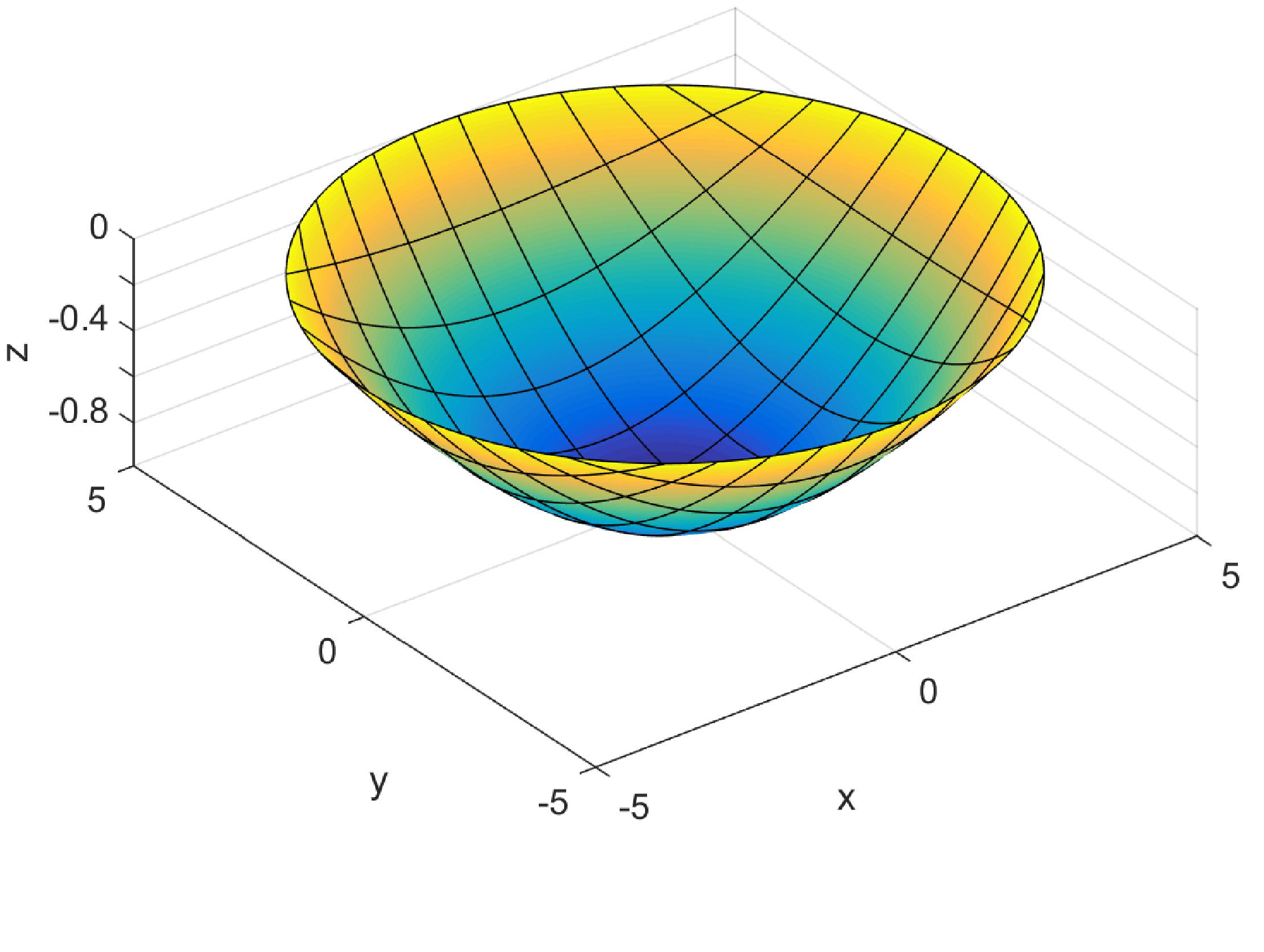}
		\caption{$\bar{P}_1 = 10.7545$.}
	\end{subfigure}
	%\hspace*{\fill}
	\begin{subfigure}{0.49\textwidth}
		\includegraphics[width=\linewidth]{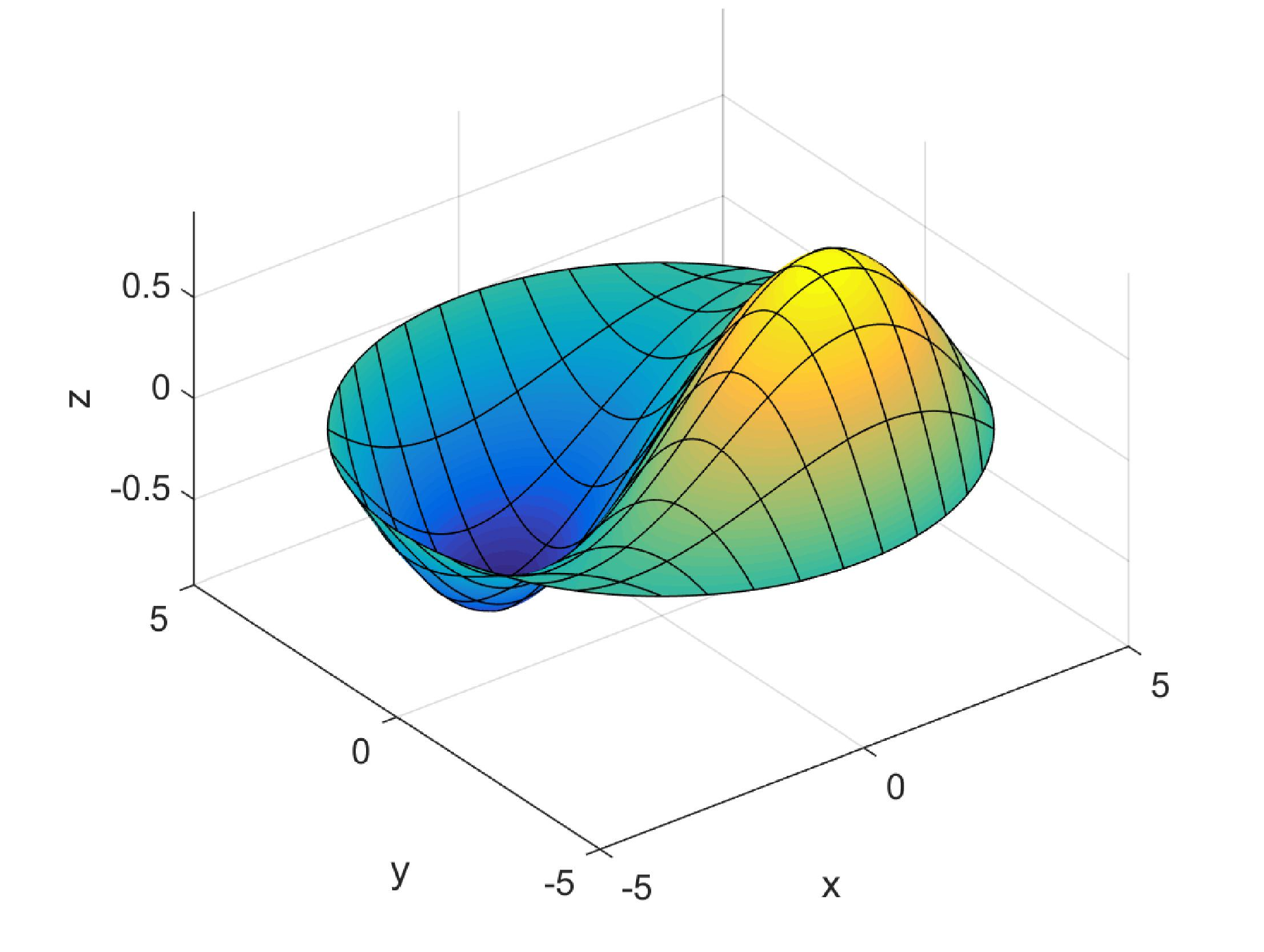}
		\caption{$\bar{P}_2 = 49.5358$.}
	\end{subfigure}
	\begin{subfigure}{0.49\textwidth}
		\includegraphics[width=\linewidth]{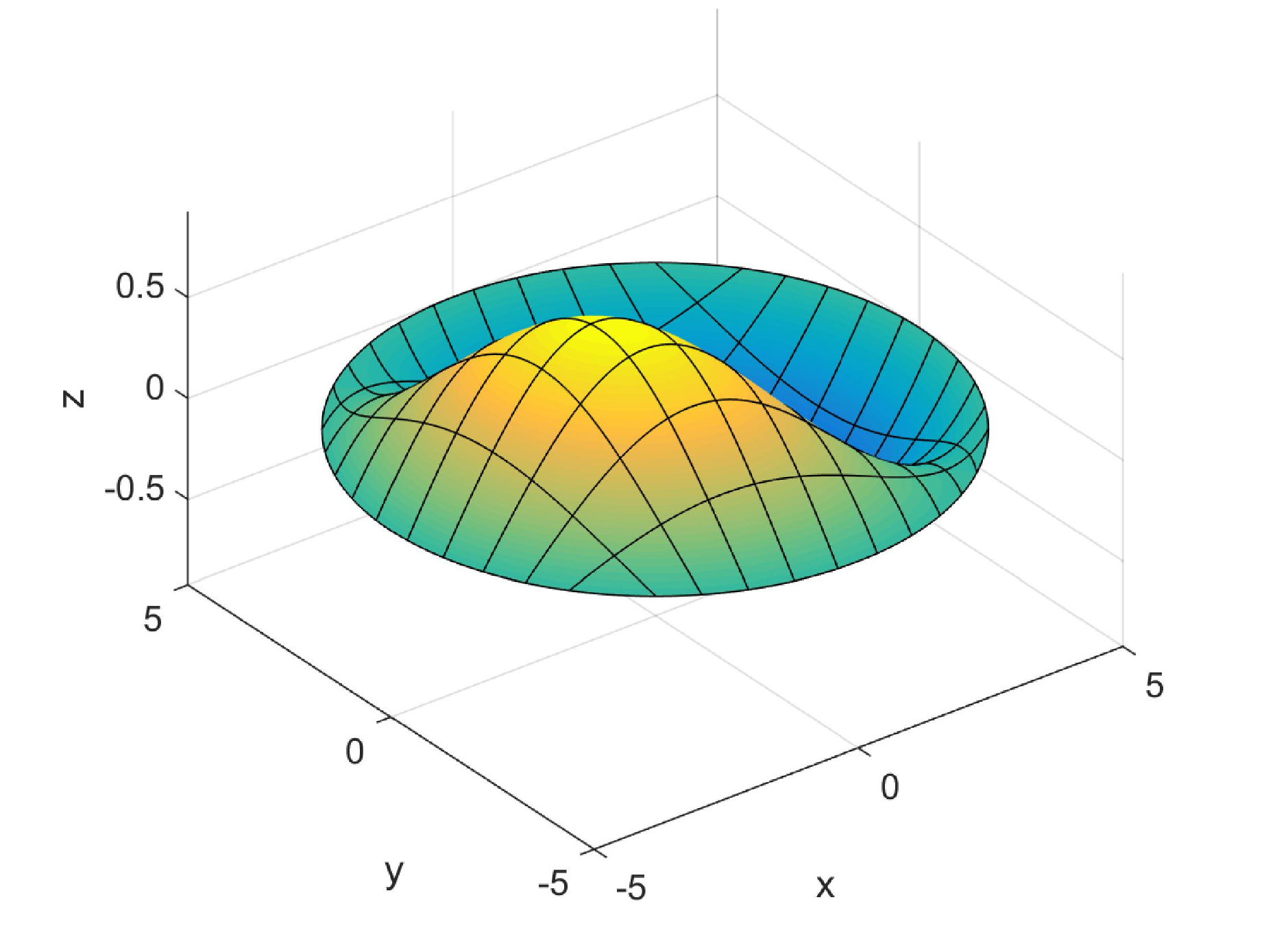}
		\caption{$\bar{P}_3 = 49.5358$.}
	\end{subfigure}
	\begin{subfigure}{0.49\textwidth}
		\includegraphics[width=\linewidth]{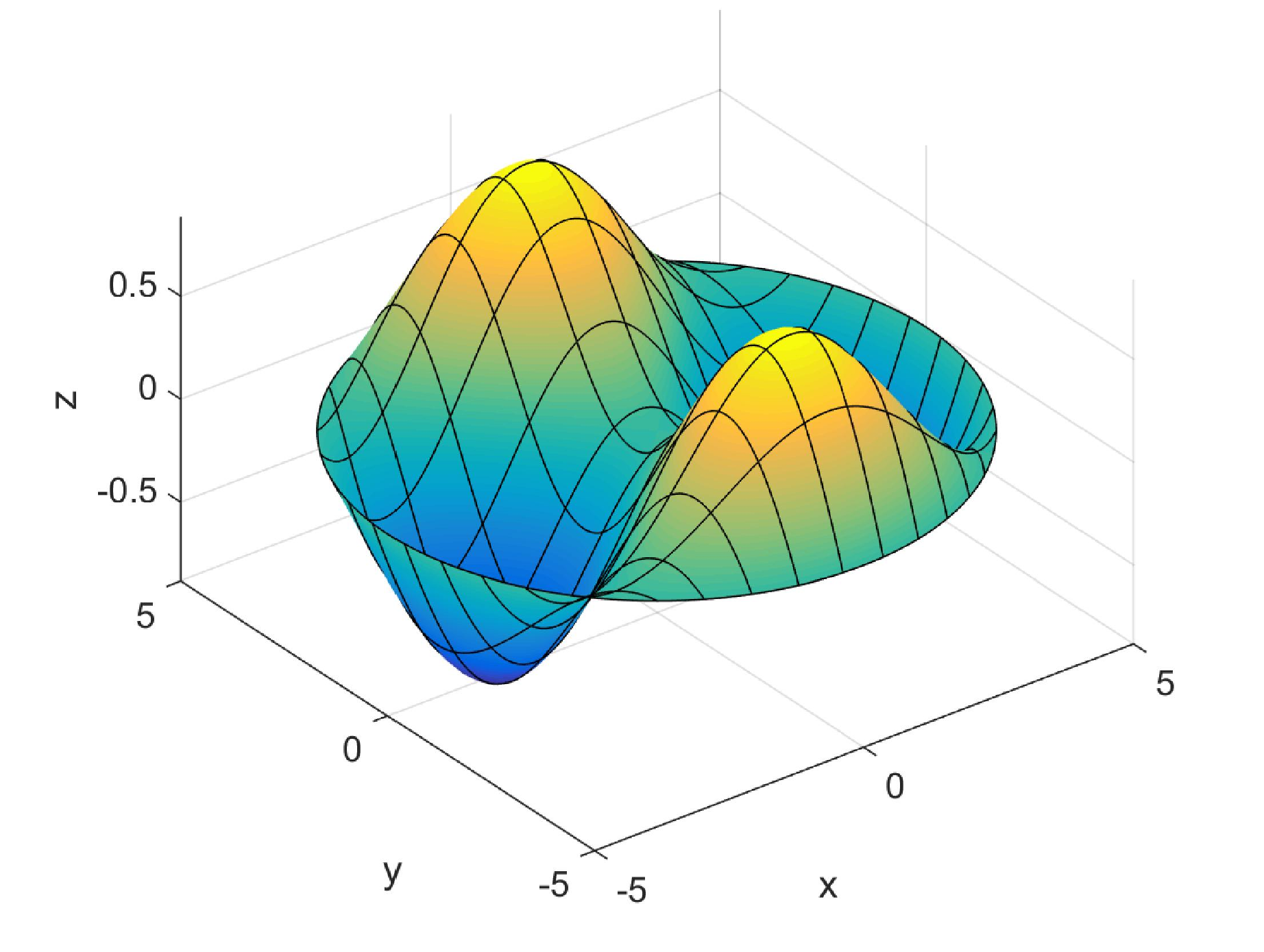}
		\caption{$\bar{P}_4 = 80.6606$.}
	\end{subfigure}
	\begin{subfigure}{0.49\textwidth}
		\includegraphics[width=\linewidth]{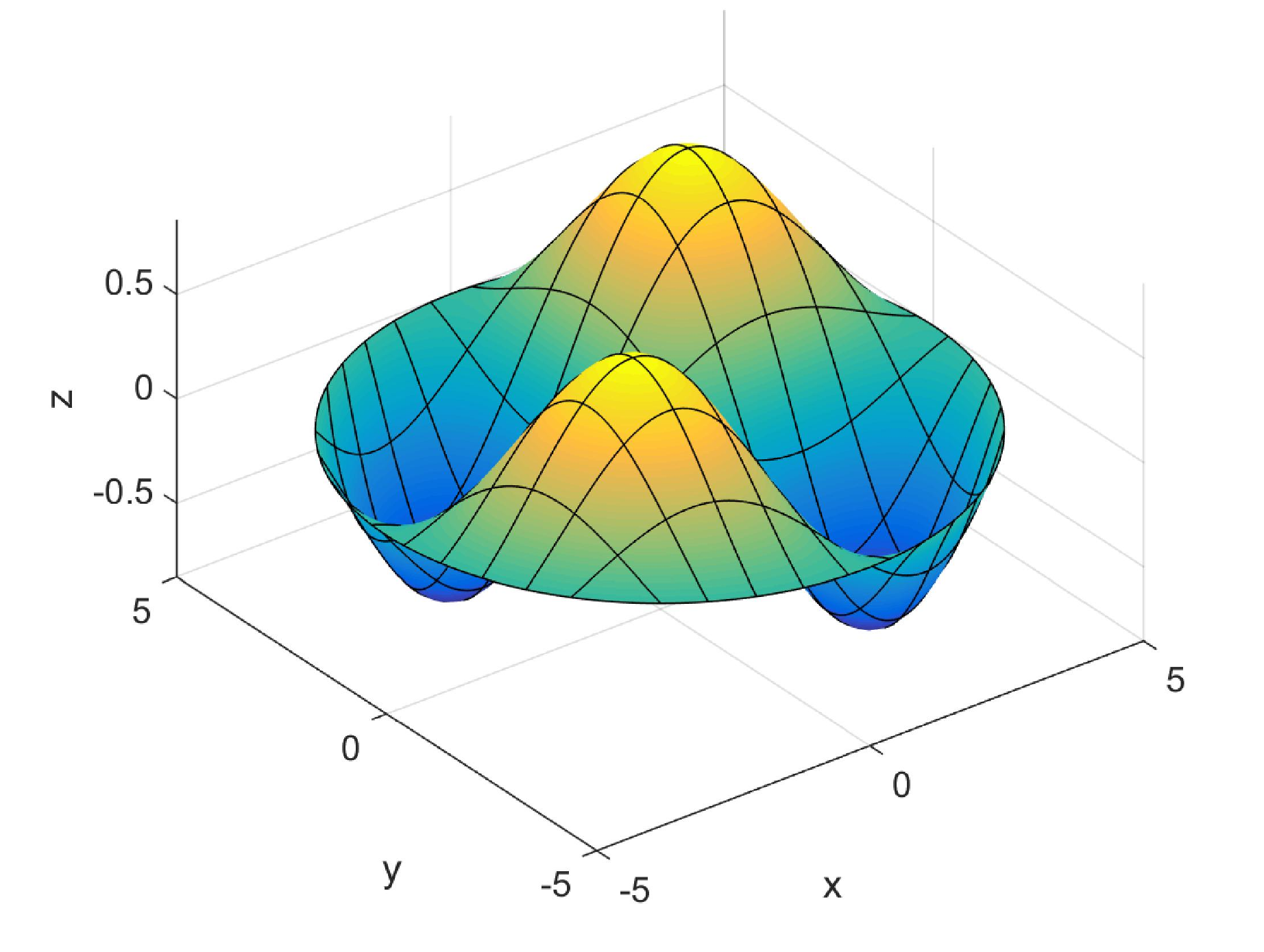}
		\caption{$\bar{P}_5 = 104.8026$.}
	\end{subfigure}
	\begin{subfigure}{0.49\textwidth}
		\includegraphics[width=\linewidth]{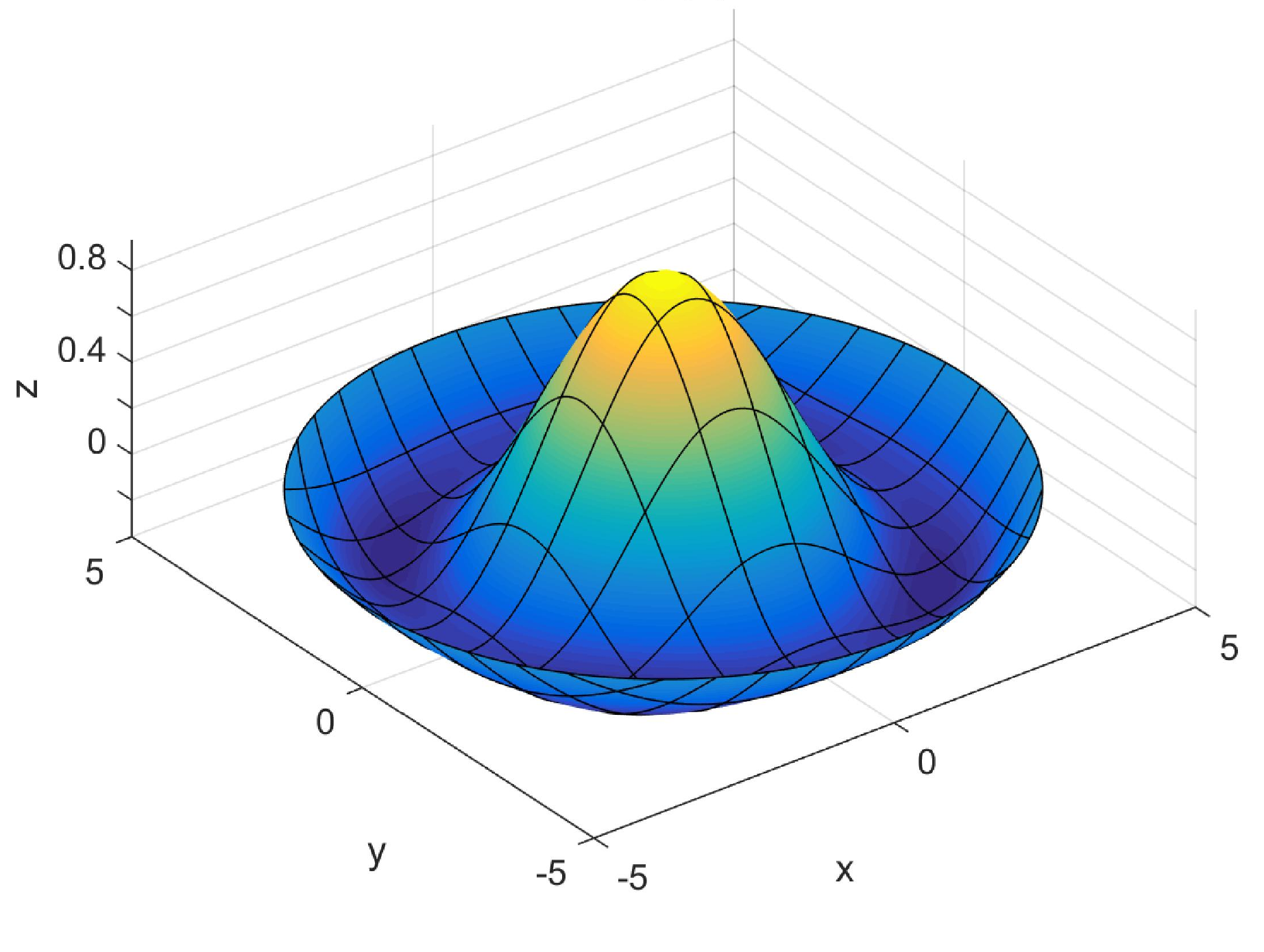}
		\caption{$\bar{P}_6 = 108.1436$.}
	\end{subfigure}
	\caption{The first six buckling mode shapes of simply-supported Al/Al$_2$O$_3$ circular microplates. \label{fig:exp:buck:cir:6modes}}
\end{figure}

Table \ref{tab:exp:buckling:cir:Al2O3} presents the non-dimensional critical buckling load of Al/Al$_2$O$_3$ circular microplates with various boundary conditions based on the proposed RPT and quasi-3D theories. The difference between the two theories is relatively small for both case of boundary conditions. The effects of material index $n$ and material length scale ratio $l/h$ on the critical buckling load of simply and clamped support of Al/Al$_2$O$_3$ circular microplates with $h/R=0.2$ are illustrated in Fig. \ref{fig:exp:buck:cir:effects}. The first six buckling loads of Al/Al$_2$O$_3$ circular microplates with various aspects ratios for $n=1$ and $l/h=0.6$ along with the results generated from Zenkour's quasi-3D theory \cite{Zenkour20139041} using proposed IGA approach are reported in Table \ref{tab:exp:buckling:6modes}. Six scaled buckling mode shapes corresponding to simply-supported circular microplates for $h/R=0.2$ based on proposed quasi-3D theory are presented in Fig. \ref{fig:exp:buck:cir:6modes}.

% Table generated by Excel2LaTeX from sheet 'Sheet1'
\begin{table}[htbp]
  \centering
  \caption{Comparison of non-dimensional critical buckling load $\bar{P_{cr}} = \dfrac{P_{cr}R^2}{D_m}$ of CCCC Al/ZrO$_2$-2 circular plates (rule of mixtures scheme)}
    \begin{tabular}{llllll}
    \hline
    $n$     & Theory & \multicolumn{4}{l}{$h/R$}\\
\cline{3-6}          &       & 0.1   & 0.2   & 0.25  & 0.3 \\
    \hline
    0     & TSDT \cite{Ma200485} & 14.089 & 12.574 & 11.638 & 10.670 \\
          & UTSDT \cite{Saidi2009110} & 14.089 & 12.575 & 11.639 & 10.670 \\
          & TSDT \cite{Tran2013368} & 14.1089 & 12.5914 & 11.6540 & 10.6842 \\
          & RPT \cite{NguyenXuan2014222} & 14.2023 & 12.7281 & 11.8143 & 10.8666 \\
          & RPT (Present) & 14.0932 & 12.5776 & 11.6409 & 10.6719 \\
          & Quasi-3D (Present) & 14.8264 & 13.4557 & 12.4564 & 11.3775 \\
    0.5   & TSDT \cite{Ma200485} & 19.411 & 17.311 & 16.013 & 14.672 \\
          & UTSDT \cite{Saidi2009110} & 19.413 & 17.310 & 16.012 & 14.672 \\
          & TSDT \cite{Tran2013368} & 19.4391 & 17.3327 & 16.0334 & 14.6910 \\
          & RPT \cite{NguyenXuan2014222} & 19.5663 & 17.5180 & 16.2506 & 14.9381 \\
          & RPT (Present) & 19.4169 & 17.3133 & 16.0153 & 14.6740 \\
          & Quasi-3D (Present) & 20.5166 & 18.6074 & 17.2206 & 15.7222 \\
    2     & TSDT \cite{Ma200485} & 23.074 & 20.803 & 19.377 & 17.882 \\
          & UTSDT \cite{Saidi2009110} & 23.075 & 20.805 & 19.378 & 17.881 \\
          & TSDT \cite{Tran2013368} & 23.1062 & 20.8319 & 19.4033 & 17.9060 \\
          & RPT \cite{NguyenXuan2014222} & 23.2592 & 21.0569 & 19.6687 & 18.2099 \\
          & RPT (Present) & 23.0809 & 20.8088 & 19.3812 & 17.8848 \\
          & Quasi-3D (Present) & 24.4332 & 22.3510 & 20.8035 & 19.1161 \\
    5     & TSDT \cite{Ma200485} & 25.439 & 22.971 & 21.414 & 19.780 \\
          & UTSDT \cite{Saidi2009110} & 25.442 & 22.969 & 21.412 & 19.778 \\
          & TSDT \cite{Tran2013368} & 25.4743 & 22.9992 & 21.4407 & 19.8043 \\
          & RPT \cite{NguyenXuan2014222} & 25.6418 & 23.2426 & 21.7268 & 20.1313 \\
          & RPT (Present) & 25.4469 & 22.9742 & 21.4168 & 19.7813 \\
          & Quasi-3D (Present) & 26.8812 & 24.6195 & 22.9303 & 21.0878 \\

    10    & TSDT \cite{Ma200485} & 27.133 & 24.423 & 22.725 & 20.948 \\
          & UTSDT \cite{Saidi2009110} & 27.131 & 24.422 & 22.725 & 20.949 \\
          & TSDT \cite{Tran2013368} & 27.1684 & 24.4542 & 22.7536 & 20.9750 \\
          & RPT \cite{NguyenXuan2014222} & 27.3429 & 24.6994 & 23.0389 & 21.2986 \\
          & RPT (Present) & 27.1395 & 24.4287 & 22.7297 & 20.9524 \\
          & Quasi-3D (Present) & 28.6197 & 26.1483 & 24.3140 & 22.3196 \\
    \hline
    \end{tabular}
  \label{tab:exp:buckling:cir:comparison}
\end{table}

% Table generated by Excel2LaTeX from sheet 'Buckling'
\begin{table}[htbp]
  \centering
  \caption{Non-dimensional critical buckling load $\bar{P_{cr}} = \dfrac{P_{cr} R^2}{D_m} $ of Al/Al$_2$O$_3$ circular microplates (Mori-Tanaka scheme)}
    \begin{tabular}{llllllllll}
    \hline
    $h/R$   & $l/h$ & $n=0$   &       &       & $n=1$   &       &       & $n=10$  & \\
\cline{3-4}\cline{6-7}\cline{9-10}          &       & RPT   & Quasi-3D &       & RPT   & Quasi-3D &       & RPT   & Quasi-3D\\
    \hline
   \multicolumn{10}{l}{Simple support} \\
    0.1   & 0     & 22.5182 & 22.6953 &       & 9.5368 & 9.7960 &       & 5.9574 & 6.0309 \\
          & 0.2   & 23.0489 & 23.2627 &       & 9.8179 & 10.0793 &       & 6.0876 & 6.1617 \\
          & 0.4   & 24.1022 & 24.4172 &       & 10.3474 & 10.6292 &       & 6.3543 & 6.4360 \\
          & 0.6   & 25.0682 & 25.5337 &       & 10.8125 & 11.1391 &       & 6.6130 & 6.7163 \\
          & 0.8   & 25.7985 & 26.4629 &       & 11.1543 & 11.5512 &       & 6.8182 & 6.9619 \\
          & 1     & 26.3291 & 27.2420 &       & 11.3981 & 11.8897 &       & 6.9718 & 7.1752 \\
    0.2   & 0     & 21.7456 & 22.0263 &       & 9.2059 & 9.4510 &       & 5.6942 & 5.7510 \\
          & 0.2   & 22.2719 & 22.5928 &       & 9.4873 & 9.7327 &       & 5.8318 & 5.8833 \\
          & 0.4   & 23.2739 & 23.7059 &       & 9.9893 & 10.2520 &       & 6.1019 & 6.1509 \\
          & 0.6   & 24.2066 & 24.8190 &       & 10.4423 & 10.7545 &       & 6.3676 & 6.4377 \\
          & 0.8   & 24.9581 & 25.8386 &       & 10.8003 & 11.2074 &       & 6.5891 & 6.7170 \\
          & 1     & 25.5480 & 26.7897 &       & 11.0777 & 11.6254 &       & 6.7647 & 6.9866 \\
    0.3   & 0     & 20.5707 & 20.8606 &       & 8.7033 & 8.9036 &       & 5.3041 & 5.3287 \\
          & 0.2   & 21.0993 & 21.4128 &       & 8.9860 & 9.1777 &       & 5.4526 & 5.4621 \\
          & 0.4   & 22.0854 & 22.4669 &       & 9.4810 & 9.6686 &       & 5.7409 & 5.7276 \\
          & 0.6   & 23.0439 & 23.5641 &       & 9.9511 & 10.1693 &       & 6.0352 & 6.0281 \\
          & 0.8   & 23.8746 & 24.6485 &       & 10.3517 & 10.6610 &       & 6.2927 & 6.3431 \\
          & 1     & 24.5707 & 25.7228 &       & 10.6823 & 11.1443 &       & 6.5055 & 6.6623 \\
    \multicolumn{10}{l}{Clamped support} \\
    0.1   & 0     & 76.5059 & 80.4859 &       & 30.4539 & 32.7398 &       & 19.7743 & 20.9413 \\
          & 0.2   & 86.3696 & 90.4095 &       & 34.8524 & 37.1578 &       & 22.0235 & 23.1967 \\
          & 0.4   & 115.9595 & 120.0466 &       & 48.0474 & 50.3612 &       & 28.7616 & 29.9236 \\
          & 0.6   & 165.2737 & 169.2564 &       & 70.0379 & 72.3004 &       & 39.9728 & 41.0821 \\
          & 0.8   & 234.3115 & 237.9906 &       & 100.8235 & 102.9597 &       & 55.6486 & 56.6598 \\
          & 1     & 323.0727 & 326.2293 &       & 140.4042 & 142.3316 &       & 75.7879 & 76.6551 \\
    0.2   & 0     & 68.2782 & 73.0452 &       & 27.3245 & 29.8617 &       & 17.1442 & 18.6240 \\
          & 0.2   & 77.6521 & 82.3859 &       & 31.5460 & 34.0702 &       & 19.3412 & 20.7979 \\
          & 0.4   & 105.7687 & 110.0650 &       & 44.2098 & 46.5462 &       & 25.9122 & 27.1925 \\
          & 0.6   & 152.6199 & 155.7501 &       & 65.3154 & 67.1544 &       & 36.8218 & 37.7110 \\
          & 0.8   & 218.1997 & 219.2330 &       & 94.8629 & 95.8167 &       & 52.0495 & 52.3297 \\
          & 1     & 302.5060 & 300.3159 &       & 132.8523 & 132.4584 &       & 71.5905 & 71.0373 \\
    0.3   & 0     & 57.9331 & 61.7636 &       & 23.3438 & 25.4648 &       & 14.0464 & 15.3047 \\
          & 0.2   & 66.6302 & 70.2136 &       & 27.3083 & 29.3268 &       & 16.1359 & 17.3213 \\
          & 0.4   & 92.7032 & 95.0207 &       & 39.2004 & 40.6571 &       & 22.3875 & 23.1508 \\
          & 0.6   & 136.1160 & 135.6444 &       & 59.0182 & 59.2280 &       & 32.7704 & 32.6634 \\
          & 0.8   & 196.8399 & 191.6493 &       & 86.7603 & 84.8742 &       & 47.2643 & 45.8312 \\
          & 1     & 274.8619 & 262.6208 &       & 122.4264 & 117.4429 &       & 65.8628 & 62.6312 \\
    \hline
    \end{tabular}
  \label{tab:exp:buckling:cir:Al2O3}
\end{table}

% Table generated by Excel2LaTeX from sheet 'Buckling'
\begin{table}[htbp]
\begin{adjustwidth}{-1.8cm}{}
  \centering
  \caption{The first six non-dimensional buckling loads $\bar{P} = \dfrac{P R^2}{D_m} $ of Al/Al$_2$O$_3$ circular microplates, $n=1$, $l/h=0.6$ (Mori-Tanaka scheme)}
    \begin{tabular}{lllllllll}
    \hline
    BC    & $h/R$   & Theory & Mode  &       &       &       &       &  \\
\cline{4-9}          &       &       & 1     & 2     & 3     & 4     & 5     & 6\\
    \hline
    Simple support & 0.1   & IGA-Zenkour & 11.4689 & 54.8031 & 54.8031 & 95.3143 & 123.2488 & 129.6396 \\
          &       & Quasi-3D (Present) & 11.1391 & 54.0884 & 54.0884 & 93.7842 & 122.0854 & 128.1298 \\
          &       & RPT (Present) & 10.8125 & 53.6215 & 53.6215 & 93.3944 & 122.3489 & 128.3281 \\
          & 0.2   & IGA-Zenkour& 11.3113 & 50.7663 & 50.7663 & 83.1176 & 107.1987 & 110.9343 \\
          &       & Quasi-3D (Present) & 10.7545 & 49.5358 & 49.5358 & 80.6606 & 104.8026 & 108.1436 \\
          &       & RPT (Present) & 10.4423 & 50.2378 & 50.2378 & 83.3889 & 109.1766 & 112.9355 \\
          & 0.3   & IGA-Zenkour & 10.8014 & 45.2015 & 45.2015 & 69.1649 & 89.4125 & 90.9795 \\
          &       & Quasi-3D (Present) & 10.1693 & 43.6917 & 43.6917 & 66.3080 & 86.1911 & 87.4685 \\
          &       & RPT (Present) & 9.9511 & 45.6739 & 45.6739 & 71.5215 & 93.8041 & 95.5819 \\
    Clamped support & 0.1   & IGA-Zenkour & 72.6509 & 126.0657 & 126.0657 & 172.3464 & 205.4419 & 222.4846 \\
          &       & Quasi-3D (Present) & 72.3004 & 125.7477 & 125.7477 & 171.5372 & 205.0589 & 221.0537 \\
          &       & RPT (Present) & 70.0379 & 123.1185 & 123.1185 & 169.2944 & 203.3425 & 218.6139 \\
          & 0.2   & IGA-Zenkour& 66.8949 & 108.9270 & 108.9270 & 140.4481 & 166.8569 & 175.2599 \\
          &       & Quasi-3D (Present) & 67.1544 & 108.3685 & 108.3685 & 138.0488 & 164.4947 & 172.3137 \\
          &       & RPT (Present) & 65.3154 & 109.4813 & 109.4813 & 143.4150 & 171.7706 & 179.5962 \\
          & 0.3   & IGA-Zenkour& 58.9576 & 90.1444 & 90.1444 & 110.5155 & 131.5270 & 134.2574 \\
          &       & Quasi-3D (Present) & 59.2280 & 88.4285 & 88.4285 & 106.6289 & 127.0395 & 129.6436 \\
          &       & RPT (Present) & 59.0182 & 93.6789 & 93.6789 & 117.0824 & 140.3857 & 143.1361 \\
    \hline
    \end{tabular}
  \label{tab:exp:buckling:6modes}
  \end{adjustwidth}
\end{table}

\section{Conclusions}
\label{sec:Conclusion}

In this study, a novel computational approach, which employs the modified couple stress theory, the four-variable refined plate theory and quasi-3D, and the NURBS-based isogemetric analysis, has been presented to investigate the static bending, free vibration, and buckling behaviours of functionally graded microplates with various geometric domains and boundary conditions. Within this proposed approach, the mathematical model governing the behaviours of the plates is constructed based on the modified couple stress theory with only one material length scale which efficiently taken into account the size-dependent effects of small-scale structures and the novel seventh-order refined plate theory and quasi-3D theory with only four unknowns. This proposed quasi-3D not only considers shear deformations but also is able to well capture the thickness stretching effect which is neglected by other classical, higher-order shear deformation, and refined plate theories. The NURBS-based isogeometric analysis is incorporated within the computational process to exactly describe the geometry an approximately construct the unknown field in which higher-order derivatives and continuity elements are readily satisfied. A large number of investigations including some benchmark examples have confirmed the validity and efficiency of the proposed approach. The obtained results also reveal that the increase of the material length scale parameter rises the microplates' stiffness which results in the decline in the central displacement and increase in the natural frequency and buckling load. On the contrary, the growth of the material index leads to completely reverse trends in which microplates' stiffness are fallen. Meanwhile, the numerical results that are generated by the proposed refined plate theory and quasi-3D theory are slightly different due to the consideration of the thickness stretching effect. However, these discrepancies become gradually vague as the microplates get thinner.

%\section*{Acknowledgement}
%\label{sec:acknowledgements}

%% References
%%
%% Following citation commands can be used in the body text:
%% Usage of \cite is as follows:
%%   \cite{key}         ==>>  [#]
%%   \cite[chap. 2]{key} ==>> [#, chap. 2]
%%
%% References with bibTeX database:
 \bibliographystyle{elsarticle-num}

\bibliography{Ref_FGM_MCS_IGA}

\end{document}